\documentclass{article}
\usepackage{theorem,amsmath,amssymb}
\newtheorem{Th}{Theorem}[section]
\newtheorem{Le}{Lemma}[section]
\newtheorem{Co}{Corollary}[section]

\newtheorem{Rem}{Remark}[section]
\newcommand{\atops}{\genfrac{}{}{0pt}{} }
\begin{document}
\renewcommand{\theequation}{\arabic{section}.\arabic{equation}}
\title{Pointwise estimates
for Green's kernel of a mixed boundary value problem to the Stokes system in a polyhedral cone}
\date{}
\author{by V.~Maz'ya$^1$ and J.~Rossmann$^2$}

\maketitle

$^1$ University of Link\"oping, Department of Mathematics,\\ \hspace*{2em}
  58183 Link\"oping, Sweden \\ \hspace*{2em} vlmaz@mai.liu.se\\

$^2$ University of Rostock, Department of Mathematics,\\ \hspace*{2em}
  18051 Rostock, Germany \\ \hspace*{2em} juergen.rossmann@mathematik.uni-rostock.de \\

{\small{\bf Key words}: Stokes system, Green's matrix, nonsmooth domains\\

{\bf MSC (2000)}: 35J25, 35J55, 35Q30}

\begin{abstract}
The paper deals with a mixed boundary value problem for the Stokes system in a polyhedral
cone. Here different boundary conditions (in particular, Dirichlet, Neumann,
free surface conditions) are prescribed on the sides of the polyhedron.
The authors obtain regularity results for weak solutions in weighted $L_2$
Sobolev spaces and prove point estimates of Green's matrix.
\end{abstract}

\section{Introduction}
We consider here a mixed boundary value problem for the linear Stokes system
\begin{equation} \label{intro1}
-\Delta u + \nabla p = f, \qquad -\nabla\cdot u = g
\end{equation}
in a polyhedral cone, where on each of the sides $\Gamma_j$ one of
the following boundary conditions is given:
\begin{itemize}
\item[(i)] $u=h$,
\item[(ii)] $u_\tau=h,\quad -p+2\varepsilon_{n,n}(u) = \phi$,
\item[(iii)] $u_n  = h, \quad \varepsilon_{n,\tau}(u)=\phi$,
\item[(iv)] $-p n + 2\varepsilon_n (u) = \phi$.
\end{itemize}
Here $n$ is the outward normal, $u_n=u\cdot n$ is the normal and $u_\tau=u-u_n n$ the tangent component of
the velocity $u$. Furthermore, $\varepsilon(u)$ denotes the matrix with the components
$\frac 12 (\partial_{x_i}u_j+\partial_{x_j}u_i)$,
$\varepsilon_n(u)$ is the vector $\varepsilon(u)n$, $\varepsilon_{n,n} =\varepsilon_n(u)\cdot n$ its normal
component and $\varepsilon_{n,\tau}(u)$ its tangent component.

Conditions (i)--(iv) are frequently used in the study of steady-state flows of
incompressible viscous Newtonian fluids. For example, on solid walls there is the Dirichlet condition (i),
a no-friction condition (Neumann condition) $2\varepsilon(u)\, n -pn =0$ may be useful
on an artificial boundary such as the exit of a canal or a free surface, the slip condition for
uncovered fluid surfaces has the form (iii), and conditions for in/out-stream surfaces can be written
in the form (ii).

Our goal is to obtain estimates for Green's matrix.
For the case of the Dirichlet problem, such estimates were obtained by
Maz'ya and Plamenevski\u{\i} \cite{mp-83}. As in \cite{mp-83} we obtain point estimates of Green's matrix
by means of weighted $L_2$ estimates of the solutions and their derivatives. Here, the weights are powers
of the distances to the edges and corners. However, while the problem with Dirichlet boundary
conditions can be handled in weighted Sobolev spaces with so-called homogeneous norms, the more
general boundary value problem considered in the present paper requires the use of weighted Sobolev
spaces with inhomogeneous norms. This makes the consideration of the boundary value problem more difficult.

We outline the main results of the paper.
In Section 2 we deal with the boundary value problem for the Stokes system in a dihedron ${\cal D}$,
where on both sides $\Gamma^+$, $ \Gamma^-$ one of the boundary conditions (i)--(iv) is given.
We consider weak solutions $(u,p)\in {\cal H}\times L_2({\cal D})$, where ${\cal H}$ is the closure
of $C_0^\infty(\bar{\cal D})^3$ (the set of infinitely differential vector functions on $\bar{\cal D}$
having compact support) with respect to the norm
\[
\| u\|_{\cal H} = \Big( \int\limits_{\cal D} \sum_{j=1}^3 |\partial_{x_j}u|^2\, dx,  \Big)^{1/2}
\]
and obtain regularity assertions for the solutions.
The smoothness of the solutions near the edge depends on the smoothness of the data and on the eigenvalues
of a certain operator pencil $A(\lambda)$ generated by the corresponding problem in a two-dimensional
angle. These eigenvalues can be calculated as the zeros of certain transcendental functions. For example,
in the cases of the Dirichlet and Neumann problems, the spectrum of the pencil $A(\lambda)$ consists
of the solutions of the equation
\[
\sin(\lambda\theta)\, \big( \lambda^2\sin^2\theta -\sin^2(\lambda\theta)\big)=0
\]
($\lambda=0$ for the Dirichlet problem), where $\theta$ is the angle at the edge.
One of our results is the following. Let $\lambda_1$ be the eigenvalue of $A(\lambda)$ with smallest
positiv real part, and let $W_\delta^l({\cal D})$ be the closure of $C_0^\infty(\bar{\cal D})$ with
respect to the norm
\begin{equation} \label{normW}
\| u\|_{W_\delta^l({\cal D})} = \Big( \int_{\cal D}\sum_{|\alpha|\le l}
  r^{2\delta}\, |\partial_x^\alpha u(x)|^2\, dx \Big)^{1/2},
\end{equation}
where $r$ denotes the distance to the edge. We suppose that $f\in W_\delta^{l-2}({\cal D})^3$, $g\in W_\delta^{l-1}
({\cal D})$, the boundary data belong to corresponding trace spaces and satisfy a certain compatibility condition
on the edge, and $\max(l-1-\mbox{Re}\, \lambda_1,0)<\delta<l-1$. Then $\zeta(u,p) \in
W_\delta^l({\cal D})^3\times W_\delta^{l-1}({\cal D})$ for an arbitrary
smooth function $\zeta$ with compact support.
If the spectrum of the pencil $A(\lambda)$ does not contain the value $\lambda=0$ (this is the case, e.g.,
if the Dirichlet condition is given on at least one of the sides $\Gamma^+,\Gamma^-$), then this result is
also true in the class of the weighted spaces $V_\delta^l({\cal D})$ with the norm
\begin{equation} \label{normV}
\| u\|_{V_\delta^l({\cal D})} = \Big( \int_{\cal D}\sum_{|\alpha|\le l}
  r^{2(\delta-l+|\alpha|)}\, |\partial_x^\alpha u(x)|^2\, dx \Big)^{1/2}.
\end{equation}
Furthermore in some cases, when $\lambda_1=1$ (e.g., in the case of the Dirichlet problem, $\theta<\pi$),
the eigenvalue $\lambda_1$ in the $W_\delta^l$ regularity result can be replaced by
the second eigenvalue $\lambda_2$.

The a priori estimates of weak solutions are used in Section 3 for the proof of
point estimates of Green's matrix $(G_{i,j}(x,\xi))_{i,j=1}^4$.
For example in the case $|x-\xi|\ge \min(|x'|,|\xi'|)$, we obtain
\begin{eqnarray}\label{estimateG1}
\big| \partial_{x'}^\alpha \partial_{x_3}^\sigma\partial_{\xi'}^\beta\partial_{\xi_3}^\tau
  G_{i,j}(x,\xi)  \big| & \le & c\, |x-\xi|^{-1-\delta_{i,4}-\delta_{j,4}-|\alpha|-|\beta|-\sigma-\tau}\nonumber \\
&&  \times \Big( \frac{|x'|}{|x-\xi|} \Big)^{\min(0,\mu-|\alpha|-\delta_{i,4}-\varepsilon)} \,
  \Big( \frac{|\xi'|}{|x-\xi|}  \Big)^{\min(0,\mu-|\beta|-\delta_{j,4}-\varepsilon)}
\end{eqnarray}
if the edge of ${\cal D}$ coincides with the $x_3$-axis, where $x'=(x_1,x_2)$, $\xi'=(\xi_1,\xi_2)$,
$\mu=\mbox{Re}\, \lambda_1$, and $\varepsilon$ is an arbitrarily small positive number.
If $\lambda=0$ is not an eigenvalue of $A(\lambda)$, then there is the sharper estimate
\begin{eqnarray*}
\big| \partial_{x'}^\alpha \partial_{x_3}^\sigma\partial_{\xi'}^\beta\partial_{\xi_3}^\tau
  G_{i,j}(x,\xi)  \big| & \le & c\, |x-\xi|^{-1-\delta_{i,4}-\delta_{j,4}-|\alpha|-|\beta|-\sigma-\tau}\nonumber \\
&& \quad  \times \Big( \frac{|x'|}{|x-\xi|} \Big)^{\mu-|\alpha|-\delta_{i,4}-\varepsilon} \,
  \Big( \frac{|\xi'|}{|x-\xi|}  \Big)^{\mu-|\beta|-\delta_{j,4}-\varepsilon} ,
\end{eqnarray*}
Moreover, in some cases when $\lambda_1=1$, this eigenvalue can be replaced in (\ref{estimateG1}) by
$\lambda_2$. This improves the estimates given in \cite{mp-83,mps}.
For the components $G_{i,4}$ we obtain a representation of the form
\begin{equation} \label{Gi4}
G_{i,4}(x,\xi)=-\nabla_\xi\cdot \vec{\cal P}_i(x,\xi)+ {\cal Q}_i(x,\xi),
\end{equation}
where the normal components of $\vec{\cal P}_i(x,\xi)$ vanish if $\xi$ lies on the boundary, and
$\vec{\cal P}_i(x,\xi)$, ${\cal Q}_i(x,\xi)$ satisfy the estimates
\begin{equation} \label{Gi4a}
|\partial_x^\alpha\partial_\xi^\gamma \vec{\cal P}_i(x,\xi)| \le c_{\alpha,\gamma} \,
  |x-\xi|^{-1-\delta_{i,4}-|\alpha|-|\gamma|}, \quad
|\partial_x^\alpha\partial_\xi^\gamma {\cal Q}_i(x,\xi)| \le c_{\alpha,\gamma} \,
  |x'|^{-2-\delta_{i,4}-|\alpha|-|\gamma|}
\end{equation}
for $|x-\xi|<\min(|x'|,|\xi'|)$. The last result is of importance for the estimation of the
integral
\[
\int_{\cal D} g(\xi)\, G_{i,4}(x,\xi)\, d\xi
\]
in the representation of the solution of the boundary value problem to the Stokes system.

Section 4 is concerned with the boundary value problem in a polyhedral cone ${\cal K}$ with vertex at the
origin and edges $M_1,\ldots,M_n$. The smoothness of solutions in a neighborhood of an edge point
$x_0\in M_k$ depends again on the eigenvalues of certain operator pencils $A_k(\lambda)$. These eigenvalues
are, as in the case of a dihedron, zeros of special transcendental functions.
The smoothness of solutions in a neighborhood of the vertex depends additionally on the eigenvalues of a
certain operator pencil ${\mathfrak A}(\lambda)$. Here ${\mathfrak A}(\lambda)$ is the operator of a
parameter-depending boundary value problem on the intersection of the cone ${\cal K}$ with the unit sphere.
Spectral properties of this operator pencil are given in papers by Dauge \cite{Dauge-89},
Kozlov, Maz'ya and Schwab \cite{kms} for the Dirichlet problem,
by Kozlov and Maz'ya \cite{km-88} for the Neumann problem and in the book by Kozlov, Maz'ya and Rossmann
\cite{kmr2} for boundary conditions (i)--(iii). We prove the existence of uniquely determined weak solutions
in $W_{\beta,0}^1({\cal K})^3\times W_{\beta,0}^0({\cal K})$ if the line $\mbox{Re}\, \lambda = -\beta-1/2$
is free of eigenvalues of the pencil ${\mathfrak A}(\lambda)$, where $W_{\beta,0}^l({\cal K})$ is the closure
of $C_0^\infty(\bar{\cal K}\backslash\{ 0\})$ with respect to the norm
\[
\| u\|_{W_{\beta,0}^l({\cal K})} = \Big( \sum_{|\alpha|\le l} |x|^{2(\beta-l+|\alpha|)}\, |\partial_x^\alpha
  u(x)|^2\, dx\Big)^{1/2}.
\]
Furthermore, we obtain regularity results for these solutions.
The absence of eigenvalues of the pencil ${\mathfrak A}(\lambda)$ on the line $\mbox{Re}\, \lambda = -\beta-1/2$
guarantees also the existence of a Green matrix $(G_{i,j}(x,\xi))_{i,j=1}^4$ of the boundary value problem
in the cone ${\cal K}$ such that the functions $x\to \zeta(|x-\xi|/r(\xi))\, G_{i,j}(x,\xi)$ belong to
$W_{\beta,0}^1({\cal K})$ for every $\xi\in {\cal K}$, $i=1,2,3$ and to $W_{\beta,0}^0({\cal K})$ for $i=4$.
Here $\zeta$ is an arbitrary smooth function on $[0,\infty)$ equal to one in $(1,\infty)$ and to zero in $(0,\frac 12 )$.
In the last subsection we derive point estimates of this Green matrix. In the case $|x|/2 < |\xi|< 2|x|$
we obtain analogous estimates to the case of a dihedron, while in the case
$|\xi|<|x|/2$ the following estimate holds:
\begin{eqnarray}\label{estimateG2}
\big| \partial_x^\alpha\partial_\xi^\gamma G_{i,j}(x,\xi)\big| & \le & \!\!c\,
  |x|^{\Lambda_- -\delta_{i,4} -|\alpha|+\varepsilon}\ |\xi|^{-\Lambda_- -1 -\delta_{j,4}
  -|\gamma|-\varepsilon} \nonumber \\
&& \!\!\times \prod_{k=1}^n \Big( \frac{r_k(x)}{|x|}\Big)^{\min(0,\mu_k-|\alpha|-\delta_{i,4}-\varepsilon)}
  \prod_{k=1}^n \Big( \frac{r_k(\xi)}{|\xi|}\Big)^{\min(0,\mu_k-|\gamma|-\delta_{j,4}-\varepsilon)}.
\end{eqnarray}
Here $r_k(x)$ denotes the distance to the edge $M_k$, $\Lambda_- < \mbox{Re}\, \lambda < \Lambda_+$
is the widest strip in the complex plane containing the line $\mbox{Re}\, \lambda = -\beta-1/2$ which is
free of eigenvalues of the pencil ${\mathfrak A}(\lambda)$. Furthermore, $\mu_k = \mbox{Re}\, \lambda_1^{(k)}$,
where $\lambda_1^{(k)}$ is the eigenvalue of the pencil $A_k(\lambda)$ with smallest positive real part.

Note again that the exponent $\min(0,\mu_k-|\alpha|-\delta_{i,4}-\varepsilon)$ in the above estimate
can be replaced by $\mu_k-|\alpha|-\delta_{i,4}-\varepsilon$ if $\lambda=0$ is not an eigenvalue of
the pencil $A_k(\lambda)$. Furthermore in some cases (when $\lambda_1^{(k)}=1$), estimate (\ref{estimateG2})
is valid with $\mu_k=\mbox{Re}\, \lambda_2^{(k)}$, where $\lambda_2^{(k)}$ is the eigenvalue
with smallest real part greater than 1.

The estimates of Green's matrix obtained in the present paper can be used to prove regularity results
for weak solutions of the mixed boundary value problem in (weighted and nonweighted) Sobolev
and H\"older spaces.
As an example, we consider the mixed problem with boundary conditions (i)--(iii). We give some
regularity results for weak solutions of this problem which are proved  in \cite{mr-04}.
Suppose the Dirichlet condition is given on at least one of the adjoining sides of every edge $M_k$.
We denote by $I$ the set of all $k$ such that the Dirichlet condition is prescribed on both adjoining
sides of $M_k$. Then (\ref{estimateG2}) is valid with $\Lambda_- =-1-2\varepsilon$ (see \cite[Th.6.1.5]{kmr2}),
$\mu_k=\frac 12 +2\varepsilon$ for $k\in I$, $\mu_k=\frac 14 +2\varepsilon$ for $k\notin I$. This estimate together
with analogous estimates for the cases $|\xi|>2|x|$ and $|x|/2 < |x-\xi|<2|x|$ allows, for example,
to obtain the following regularity result for the weak solution $(u,p)\in W^{1,2}({\cal G})\times
L_2({\cal G})$ of the mixed problem with homogeneous boundary conditions (i)--(iii) in a bounded
polyhedron ${\cal G}$.
\begin{itemize}
\item If $f\in (W^{1,s'}({\cal G})^*)^3$ and $g\in L_s({\cal G})$, $2<s\le 8/3$, $s'=s/(s-1)$, then
$(u,p)\in W^{1,s}({\cal G})^3\times L_2({\cal G})$.
\end{itemize}
Here $W^{l,s}({\cal G})$ denotes the Sobolev spaces of all functions $u$ such that
$\partial_x^\alpha u\in L_s({\cal G})$ for $|\alpha|\le l$. Under additional assumptions
on the polyhedron ${\cal G}$, this result can be improved. If the angle $\theta_k$ at the edge
$M_k$ is less than $\frac 32 \pi$ for $k\notin I$, then $\mu_k>1/3$, and the above regularity
result is true for $2<s\le 3$.
We assume that the Dirichlet condition is given on all sides of a convex polyhedron ${\cal G}$ except
the side $\Gamma_1$, where condition (ii) is prescribed, and that $\theta_k<\pi/2$ for $k\not\in I$.
Then the following regularity result holds.
\begin{itemize}
\item If $f\in (W^{1,2}({\cal G})^*)^3\cap L_s({\cal G})$ and $g\in L_2({\cal G})\cap W^{1,s}({\cal G})$,
$1<s\le 2$ (in the case $s=2$, $g$ must satisfy a compatibility condition on the edges),
then the weak solution  belongs to $W^{2,s}({\cal G})^3\times W^{1,s}({\cal G})$.
\end{itemize}
This result is based on the estimates $\mu_k>1$, $\Lambda_-= -2$ (see \cite[Th.6.2.7]{kmr2}). If $\theta_k
<\frac 34 \pi$ for $k\in I$ and $\theta_k<\frac 38 \pi$ for $k\not\in I$, then $\mu_k>4/3$ and the last
result holds even for $1<s<3$.

The present paper is an extension of our preprint \cite{mr-03}.
We added here estimates for the elements of the last row of Green's matrix
(see (\ref{Gi4}), (\ref{Gi4a}) and the analogous estimate for the problem in a polyhedral cone)
and sharper estimates for the case when $\lambda=0$ is not an eigenvalue of the pencil $A_k(\lambda)$.
The estimates in \cite{mr-03} so completed become sufficient to obtain various regularity results
for solutions to the Stokes system as mentioned above.

\setcounter{equation}{0}
\section{The problem in a dihedron (\boldmath $L_2$-theory)\unboldmath}

In the following, let $K$ be an infinite angle in the $(x_1,x_2)$-plane given in
polar coordinates $r,\varphi$ by the inequalities $0<r<\infty,\ -\theta/2<\varphi<\theta/2$.
Furthermore, let
\[
{\cal D} = \{ x=(x',x_3):\, x'=(x_1,x_2)\in K,\, x_3\in {\Bbb R}\}
\]
be a dihedron. The sides $\{x:\, \varphi=\pm \theta/2\}$ of ${\cal D}$ are denoted by $\Gamma^\pm$,
the edge $\bar{\Gamma}^+\cap \bar{\Gamma}^-$ is denoted by $M$.

We consider a boundary value problem for the Stokes system, where
on each of the sides $\Gamma^\pm$ one of the boundary conditions
(i)--(iv) is given. Let $n^\pm=(n_1^\pm,n_2^\pm,0)$ be the exterior normal to
$\Gamma^\pm$, $\varepsilon_n^\pm(u)=\varepsilon(u)\, n^\pm$ and $\varepsilon_{nn}^\pm(u)
=\varepsilon_n^\pm(u)\cdot n^\pm$. Furthermore, let $d^\pm\in \{ 0,1,2,3\}$ be integer numbers
characterizing the boundary conditions on $\Gamma^+$ and $\Gamma^-$, respectively. We put
\begin{itemize}
\item $S^\pm u = u \quad \mbox{for }d^\pm=0$,
\item $S^\pm u = u-(u\cdot n^\pm)n^\pm, \quad N^\pm(u,p)= -p+2 \varepsilon_{nn}^\pm(u)
  \quad\mbox{for }d^\pm=1,$
\item $S^\pm u = u\cdot n^\pm,\quad N^\pm (u,p)=\varepsilon_n^\pm(u)
  -\varepsilon_{nn}^\pm(u)\, n^\pm \quad \mbox{for } d^\pm=2$
\item $N^\pm(u,p)=-pn^\pm +2\varepsilon_n^\pm(u)  \quad \mbox{for }d^\pm=3$
\end{itemize}
and consider the boundary value problem
\begin{eqnarray} \label{StokesD}
&& -\Delta u + \nabla p = f,\quad -\nabla\cdot u = g\quad\mbox{in }{\cal D}, \\
&& S^\pm u = h^\pm,\quad N^\pm (u,p)=\phi^\pm \quad \mbox{on }\Gamma^\pm. \label{bcD}
\end{eqnarray}
Here the condition $N^\pm(u,p)=\phi^\pm$ is absent in the case $d^\pm=0$, while the condition
$S^\pm u=h^\pm$ is absent in the case $d^\pm=3$.
The boundary conditions (\ref{bcD}) arise from the bilinear form
\begin{equation} \label{blf}
b(u,v)= 2 \int_{\cal D} \sum_{i,j=1}^3 \varepsilon_{i,j}(u)\ \varepsilon_{i,j}(v)\, dx
\end{equation}
(see formula (\ref{GreenD}) below).

\subsection{Green's formula for Stokes system}

Let $b(\cdot,\cdot)$ be the bilinear form (\ref{blf}). Partial integration yields
\begin{equation} \label{GreenD}
b(u,v)-\int_{\cal D} p\, \nabla\cdot v\, dx = \int_{\cal D} (-\Delta u - \nabla\nabla\cdot u
  + \nabla p)\cdot v\, dx + \sum_\pm \int_{\Gamma^\pm} (-pn^\pm + 2\varepsilon(u)n^\pm)\cdot v\, dx
\end{equation}
for arbitrary $u,v\in C_0^\infty(\bar{\cal D})^3$, $p\in C_0^\infty(\bar{\cal D})$.
As a consequence of (\ref{GreenD}), the following Green formula holds for $u,v\in C_0^\infty(\bar{\cal D})^3$,
$p,q\in C_0^\infty(\bar{\cal D})$.
\begin{eqnarray} \label{Green2}
&& \hspace{-3em}\int_{\cal D} (-\Delta u-\nabla\nabla\cdot u+\nabla p)\cdot v\, dx - \int_{\cal D}(\nabla\cdot u)\, q\, dx
  + \sum_{\pm} \int_{\Gamma^\pm} \big(-pn^\pm+2\varepsilon(u) n^\pm\big)\cdot v\, dx \nonumber\\
&& \hspace{-3em}= \int_{\cal D} u\cdot (-\Delta v-\nabla\nabla\cdot v+\nabla q)\, dx - \int_{\cal D}p\, \nabla\cdot v\, dx
  + \sum_{\pm} \int_{\Gamma^\pm} u\cdot \big(-qn^\pm+2\varepsilon(v) n^\pm\big)\, dx.
\end{eqnarray}

\subsection{Weighted Sobolev spaces}

For arbitrary real $\delta$ we denote by $V_\delta^l({\cal D})$
the closure of $C_0^\infty(\bar{\cal D}\backslash M)$ (the set of
all infinitely differentiable functions with compact support in
$\bar{\cal D}\backslash M)$ with respect to the norm (\ref{normV}),
where $r=|x'|$ denotes the distance to the edge. For real
$\delta>-1$ let $W_\delta^l({\cal D})$ be the closure of
$C_0^\infty(\overline{\cal D})$ with respect to the norm (\ref{normW}).
Furthermore, let $V_\delta^{l-1/2}(\Gamma^\pm)$ be the space of
traces of functions from $V_\delta^l({\cal D})$ on $\Gamma^\pm$.
The norm in this space is
\[
\| u\|_{V_\delta^{l-1/}(\Gamma^\pm)} = \inf\{ \|
v\|_{V_\delta^l({\cal D})}\, : \
  v\in V_\delta^l({\cal D}), \ v = u\mbox{ on }\Gamma^\pm\}.
\]
Analogously, the norm in trace space
$W_\delta^{l-1/2}(\Gamma^\pm)$ for $W_\delta^l({\cal D})$ is
defined. Note that the norm in $V_\delta^{l-1/2}(\Gamma^\pm)$ is
equivalent (see \cite[Le.1.4]{mp78a}) to
\begin{eqnarray} \label{eqnorm}
\| u\| & = & \Big( \int_{\gamma^\pm} \int_{\Bbb R}\int_{\Bbb R} r^{2\delta}\,
 \big| \partial_{x_3}^{l-1}u(r,x_3)-\partial_{y_3}^{l-1}(r,y_3)\big|^2\,
 \frac{dx_3\, dy_3}{|x_3-y_3|^2} \, dr \nonumber \\
& & + \int_{\Bbb R} \int_{\gamma^\pm}\int_{\gamma^\pm}
  \big| r_1^\delta (\partial_r^{l-1}u)(r_1,x_3)-(\partial_r^{l-1}u)(r_2,x_3)\big|^2
  \, \frac{dr_1\, dr_2}{|r_1-r_2|^2}\, dx_3 \nonumber \\
& & + \int_{\Gamma^\pm} \sum_{j=0}^{l-1} r^{2(\delta-l+j)+1} \, |\partial_r^j
  u(r,x_3)|^p\, dr\, dx_3\Big)^{1/2} .
\end{eqnarray}
If $\delta$ is not integer, $-1< \delta<l-1$, then the trace of an
arbitrary function $u\in W_\delta^l({\cal D})$ on the edge $M$
belongs to the Sobolev space $W^{l-\delta-1}(M)$ (see, e.g.,
\cite[Le.1.1]{mr-88}). Furthermore, the following lemma holds (see
\cite[Le.1.3]{mr-88}).

\begin{Le} \label{l0}
{\em 1)} The space $W_\delta^l({\cal D})$ is continuously imbedded
into $W_{\delta-1}^{l-1}({\cal D})$ if $\delta>0$ and into
$V_\delta^l({\cal D})$ if $\delta>l-1$.

{\em 2)} Let $u\in W_\delta^l({\cal D})$, where $\delta$ is not
integer, $\delta>-1$. Then for the inclusion $u\in
V_\delta^l({\cal D})$ it is necessary and sufficient that
$\partial_x^\alpha u|_M=0$ for $|\alpha|<l-\delta-1$.
\end{Le}

Analogously to $V_\delta^l({\cal D})$ and $W_\delta^l({\cal D})$, we define the weighted
Sobolev spaces $V_\delta^l(K)$ and $W_\delta^l(K)$ on the two-dimensional angle $K$.
Here in the definition of the norms one has only to replace ${\cal D}$ and $x$ by $K$
and $x'$, respectively. For the space $W_\delta^l(K)$ a result analogous to Lemma \ref{l0} holds.
Furthermore, we will use the following lemma.

\begin{Le} \label{l0a}
Let $u\in V_\delta^l(K)$, $l\ge 2$. Then there exists a constant $c$ independent of $u$ such that
\[
\sup_{x'\in{\cal K}} |x'|^{\delta-l+1}\, |u(x')| \le c\, \| u\|_{V_\delta^l(K)}.
\]
\end{Le}

\noindent P r o o f.
Let $K_0 = \{ x'\in K: \, 1/2< |x'|<2\}.$ By Sobolev's lemma, we have
\[
|v(x')| \le c\, \| v\|_{W^l(K_0)} \quad\mbox{for each } v\in W^l(K_0),\ |x'|=1.
\]
Now let $u\in V_\delta^l(K)$ and $|x'|=\rho$. We introduce the function $v(x')=u(\rho x')$ and
obtain
\begin{eqnarray*}
\hspace{-1.5em}\rho^{\delta-l+1} \, |u(x')| & = & \rho^{\delta-l+1}\, \big|v(x'/\rho)\big| \le
  c\, \rho^{\delta-l+1}\, \| v\|_{W^l(K_0)} = c\, \rho^{\delta-l+1} \Big( \sum_{|\alpha|\le l} \int_{K_0}
  \big| \partial_{x'}^\alpha u(\rho x')\big|^2\, dx'\Big)^{1/2} \\
& = & c\, \Big( \sum_{|\alpha|\le l} \int\limits_{\substack{K \\ \rho/2 <|x'|<2\rho}}
   \rho^{2(\delta-l+|\alpha|)} \big|\partial_{x'}^\alpha u(x')\big|^2\, dx'\Big)^{1/2}.
\end{eqnarray*}
This proves the lemma. \hfill$\Box$

\subsection{Weak solutions of the boundary value problem}

Let $L_2^1({\cal D})$ be the closure of the set $C_0^\infty(\bar{\cal D})$ with respect to the norm
\begin{equation} \label{normH}
\| u\|_{L_2^1({\cal D})} = \Big( \int_{\cal D} \sum_{j=1}^3 |\partial_{x_j}u|^2\, dx\Big)^{1/2}
\end{equation}
The closure of the set $C_0^\infty({\cal D})$ with respect to this norm is denoted by
$\stackrel{\circ}{L}\!\!{}_2^1({\cal D})$.

Furthermore, let ${\cal H}=L_2^1({\cal D})^3$ and $V=\{ u\in {\cal H}:\, S^\pm u =0 \mbox{ on }\Gamma^\pm\}$.
In the case of the Dirichlet problem ($S^\pm u=u$), we have $V=\stackrel{\circ}{L}\!\!{}_2^1({\cal D})^3$.
Note that, by Hardy's inequality,
\[
\int_{\cal D} |x|^{-2} |u|^2\, dx \le 4 \, \| u\|^2_{L_2^1({\cal D})} \quad \mbox{for }
  u\in C_0^\infty(\overline{\cal D}).
\]
Therefore, the norm (\ref{normH}) is equivalent to the norm
\begin{equation} \label{normH2}
\| u\| = \Big( \int\limits_{\cal D} \big( |x|^{-2} \, |u|^2 + \sum_{j=1}^3
  |\partial_{x_j}u|^2\, dx \Big)^{1/2},
\end{equation}
and ${\cal H}$ can be also defined as the closure of $C_0^\infty(\overline{\cal D})^3$ with respect to the norm
(\ref{normH2}). Obviously, every $u\in {\cal H}$ is quadratically
summable on each compact subset of ${\cal D}$. From Hardy's
inequality it follows that
\[
\| u\|^2_{L_2({\cal D})} \le \int_{\cal D} (r^{2\delta}+r^{2\delta-2})\,
  |u|^2\, dx \le \int_{\cal D} r^{2\delta}\big( |u|^2+   c\, |\nabla u|^2\big)\, dx
\]
for $u\in C_0^\infty(\overline{\cal D})$ and $0<\delta<1$, where $c$ depends only on $\delta$. Consequently,
there are the continuous imbeddings $W_\delta^1({\cal D})\subset L_2({\cal D})$ and
$W_\delta^2({\cal D})^3 \subset {\cal H}$ if $0<\delta<1$.

By (\ref{GreenD}), every solution $(u,p)\in W_\delta^2({\cal D})\times W_\delta^1({\cal D})$ of problem
(\ref{StokesD}), (\ref{bcD}) satisfies the integral equality
\[
b(u,v)-\int_{\cal D} p\, \nabla\cdot v\, dx = \int_{\cal D} (f + \nabla g)\cdot v\, dx
  + \sum_\pm \int_{\Gamma^\pm} \phi^\pm\cdot v\, dx.
\]
for all $v\in C_0^\infty({\cal D})^3$ (in the case $S^\pm v=v_n$, the function $\phi^\pm$ has to be
replaced by the vector $\phi^\pm n^\pm$).
By a weak solution of problem (\ref{StokesD}), (\ref{bcD}) we mean
a pair $(u,p) \in {\cal H}\times L_2({\cal D})$ satisfying
\begin{eqnarray} \label{bvpD1}
&& b(u,v) - \int_{\cal D} p\, \nabla\cdot v\, dx = F(v) \ \mbox{ for all }v\in V,\\
&& - \nabla\cdot u=g \ \mbox{ in }{\cal D},\quad S^\pm u = h^\pm \ \mbox{ on } \Gamma^\pm , \label{bvpD2}
\end{eqnarray}
where
\begin{equation} \label{funct}
F(v) =\int_{\cal D} (f +\nabla g)\cdot v\, dx + \sum_\pm
\int_{\Gamma^\pm}
  \phi^\pm\cdot  v\, dx,
\end{equation}
provided the functional (\ref{funct}) belongs to the dual space $V^*$ of $V$. For example,
$F\in V^*$ if $f\in W_\delta^0({\cal D})^3$, $g\in W_\delta^1 ({\cal D})$, $\phi^\pm \in
W_\delta^{1/2}(\Gamma^\pm)$, $\delta<1$, and the supports of $f,\ g$ and $\phi^\pm$ are compact.

\subsection{A property of the operator div}

The goal of this subsection is to prove that the operator $\mbox{div}:\, \stackrel{\circ}{L}\!\!{}_2^1
({\cal D})^3 \to L_2({\cal D})$ is surjective. For this end, we show that its dual operator is injective
and has closed range. We start with an assertion in the two-dimensional angle $K$. Here
$\stackrel{\circ}{W}\!\!{}^1(K)$ denotes the closure of $C_0^\infty(K)$ with respect to the norm
\[
\| u\|_{W^1(K)} = \Big( \int_K \big( |u|^2 + |\partial_{x_1}u|^2+ |\partial_{x_2}u|^2 \big)\, dx\Big)^{1/2}
\]
and $W^{-1}(K)$ its dual space (with respect to the $L_2$ scalar product in $K$).

\begin{Le} \label{X}
For arbitrary $f\in L_2(K)$ there is the estimate
\begin{equation} \label{x1}
\| f\|_{L_2(K)} \le c\, \Big( \|\partial_{x_1}f \|_{W^{-1}(K)} + \|\partial_{x_2}f \|_ {W^{-1}(K)}
  + \| f \|_{W^{-1}(K)}\Big)
\end{equation}
with a constant $c$ independent of $f$.
\end{Le}

\noindent P r o o f. For bounded Lipschitz domains the assertion of the lemma can be found e.g. in
\cite[Ch.2,{\S}2]{Girault}. Let ${\cal U}_j$, $j=1,2,\ldots$, be pairwise disjoint congruent parallelograms
such that $\bar{K} = \bar{\cal U}_1\cup\bar{\cal U}_2\cup\cdots$. Then
\begin{equation} \label{x2}
\| f\|^2_{L_2(K)} = \sum_{j=1}^\infty \| f\|^2_{L_2({\cal U}_j)} \le c\, \sum_{j=1}^\infty \Big(
  \|\partial_{x_1}f \|^2_{W^{-1}({\cal U}_j)} + \|\partial_{x_2}f \|^2_{W^{-1}({\cal U}_j)}
  + \| f \|^2_{W^{-1}({\cal U}_j)} \Big).
\end{equation}
By Riesz' representation theorem, there exist functions $g\in \stackrel{\circ}{W}\!\!{}^1(K)$,
$g_j \in \stackrel{\circ}{W}\!\!{}^1({\cal U}_j)$ such that
\begin{eqnarray*}
&&  \| f\|_{W^{-1}(K)} = \| g  \|_{W^1(K)} ,\quad \int_K f\, \bar{v}\, dx = (g,v)_{W^1(K)}
  \quad\mbox{for all }v\in \stackrel{\circ}{W}\!\!{}^1(K),\\
&&  \| f\|_{W^{-1}({\cal U}_j)} = \| g_j  \|_{W^1({\cal U}_j)},\quad \int_{{\cal U}_j} f\, \bar{v}\, dx
  = (g_j,v)_{W^1({\cal U}_j)} \quad\mbox{for all }  v\in \stackrel{\circ}{W}\!\!{}^1({\cal U}_j),\ j=1,2,\ldots.
\end{eqnarray*}
Let $g_{j,0}$ be the extension of $g_j$ by zero. Then
\[
\| g_j\|^2_{W^1({\cal U}_j)} = \int_{{\cal U}_j} f\, \bar{g}_j\, dx = \int_K f\, \bar{g}_{j,0}\, dx
  = (g,g_{j,0})_{W^1(K)} = (g,g_j)_{W^1({\cal U}_j)}
\]
and, therefore
\[
\| g_j\|_{W^1({\cal U}_j)} \le \| g\|_{W^1({\cal U}_j)}.
\]
Consequently,
\[
\sum_{j=1}^\infty \| f \|^2_{W^{-1}({\cal U}_j)} = \sum_{j=1}^\infty \| g_j\|^2_{W^1({\cal U}_j)}
  \le \sum_{j=1}^\infty \| g\|^2_{W^1({\cal U}_j)} = \| g\|^2_{W^1(K)} = \| f\|^2_{W^{-1}(K)}
\]
The same inequality holds for $\partial_{x_j}f$. This together with (\ref{x2}) implies (\ref{x1}).
\hfill $\Box$\\

Next we give equivalent norms in $L_2({\cal D})$ and in the dual space $L_2^{-1}({\cal D})$ of
$\stackrel{\circ}{L}\!\!{}_2^1({\cal D})$.

\begin{Le} \label{Y}
Let $F(y,\xi)=\tilde{f}(|\xi|^{-1}y,\xi)$, where $\tilde{f}(x',\xi)$ denotes the Fourier transform
of $f(x',x_3)$ with respect to the last variable. Then
\begin{eqnarray*}
&& \| f\|^2_{L_2({\cal D})} = \int_{\Bbb R} \xi^{-2} \| F(\cdot,\xi)\|^2_{L_2(K)}\, d\xi
  \quad\mbox{if } f\in L_2({\cal D}), \\
&& \| f\|^2_{L_2^{-1}({\cal D})} = \int_{\Bbb R} \xi^{-4} \| F(\cdot,\xi)\|^2_{W^{-1}(K)}
  \, d\xi   \quad\mbox{if } f\in L_2^{-1}({\cal D}).
\end{eqnarray*}
\end{Le}

\noindent P r o o f.
The first equality follows immediately from Parseval's equality. We prove the second one.
By Riesz' representation theorem, there exists a function $u\in \stackrel{\circ}{L}\!\!{}_2^1({\cal D})$ such that
\begin{equation} \label{y1}
\| f\|_{L_2^{-1}({\cal D})} = \| u\|_{L_2^1({\cal D})},\quad
  \int_{\cal D} f\, \bar{v}\, dx = \int_{\cal D} \nabla u \cdot \nabla\bar{v}\, dx \ \mbox{ for all }
  v \in \stackrel{\circ}{L}\!\!{}_2^1({\cal D}).
\end{equation}
Furthermore, for arbitrary $\xi \in {\Bbb R}$ there exists a function $W(\cdot,\xi) \in
\stackrel{\circ}{W}\!\!{}^1(K)$ such that
\begin{eqnarray*}
&& \| F(\cdot,\xi)\|_{W^{-1}(K)} = \| W(\cdot,\xi)\|_{W^1(K)} ,\quad \\
&& \int_K  F(y,\xi)\, \overline{V(y,\xi)}\, dy = \int_K \big( \nabla_y W(y,\xi)\cdot \nabla_y \overline{V(y,\xi)} +
  W(y,\xi)\, \overline{V(y,\xi)} \big)\, dy,
\end{eqnarray*}
where $V(y,\xi)=\tilde{v}(|\xi|^{-1}y,\xi)$ and $\tilde{v}$ is the Fourier transform with respect to $x_3$
of an arbitrary function $v\in  \stackrel{\circ}{L}\!\!{}_2^1({\cal D})$. From the last equality it follows
that
\[
\int_K \tilde{f}(x',\xi)\, \overline{\tilde{v}(x',\xi)}\, dx' = \xi^{-2}\int_K \nabla_{x'} W(|\xi|x',\xi)\cdot
  \nabla_{x'} \overline{\tilde{v}(x',\xi)} + \xi^2 W(|\xi|x',\xi)\, \overline{\tilde{v}(x',\xi)} \big)\, dx',
\]
Comparing this with (\ref{y1}), we conclude that $W(|\xi|x',\xi) = \xi^2 \tilde{u}(x',\xi)$.
Consequently,
\begin{eqnarray*}
\hspace{-3em} \int_{\Bbb R} \| F(\cdot,\xi)\|^2_{W_2^{-1}(K)}\, d\xi & = &\!\!\! \int_{\Bbb R} \xi^{-4}\,
  \| W(\cdot,\xi)\|^2_{W^1(K)} d\xi= \int_{\Bbb R} \int_K \big( |\nabla_{x'}\tilde{u}(x',\xi)|^2+
  \xi^2\, |\tilde{u}(x',\xi)|^2\big)\, dx'\, d\xi \\
& = & \| u\|^2_{L_2^1({\cal D})} = \| f\|^2_{L_2^{-1}({\cal D})}.
\end{eqnarray*}
The proof is complete. \hfill $\Box$

\begin{Th} \label{t1}
{\em 1)} There exists a constant $c$ such that
\[
\| f \|_{L_2({\cal D})} \le c\, \sum_{j=1}^3 \| \partial_{x_j} f\|_{L_2^{-1}({\cal D})}
  \quad \mbox{for all }f\in L_2({\cal D}).
\]

{\em 2)} For arbitrary $f\in L_2({\cal D})$ there exists a vector function $u\in
\stackrel{\circ}{L}\!\!{}_2^1({\cal D})^3$ such that $\nabla\cdot u =f$ and
\[
\| u\|_{L_2^1({\cal D})^3} \le c \, \| f\|_{L_2({\cal D})},
\]
where the constant $c$ is independent of $f$.
\end{Th}

\noindent P r o o f.
Let $f$ be an arbitrary function in $L_2({\cal D})$ and
$F(y,\xi)=\tilde{f}(|\xi|^{-1}y,\xi)$, where $\tilde{f}$ denotes the Fourier transform of $f$
with respect to $x_3$. Then, by Lemmas \ref{X}, \ref{Y}, we have
\begin{eqnarray*}
\hspace{-2em} \| f \|^2_{L_2({\cal D})} & = & \int_{\Bbb R} \xi^{-2}\, \| F(\cdot,\xi)\|^2_{L_2(K)}\, d\xi
  \le c\int_{\Bbb R} \Big( \sum_{j=1}^2 \|\partial_{y_j}F(\cdot,\xi)\|^2_{W^{-1}(K)}
  + \|F(\cdot,\xi)\|^2_{W^{-1}(K)} \big)\, d\xi \\
& = & c\, \sum_{j=1}^3 \| \partial_{x_j} f\|^2_{L_2^{-1}({\cal D})} .
\end{eqnarray*}
From this it follows, in particular, that the range of the mapping
\[
L_2({\cal D}) \ni f \to \nabla f \in L_2^{-1}({\cal D})^3
\]
is closed. Moreover, the kernel of this operator is obviously trivial. Consequently, by
the closed range theorem, its dual operator $u \to -\nabla\cdot u$
maps $\stackrel{\circ}{L}\!\!{}_2^1({\cal D})^3$ onto $L_2({\cal D})$.
This proves the theorem. \hfill $\Box$

\subsection{Existence and uniqueness of weak solutions}

\begin{Le} \label{l2}
There exists a positive constant $c$ such that $\displaystyle
b(u,\bar{u})\ge c\, \| u\|^2_{\cal H}$ for all $u\in {\cal H}$.
\end{Le}

\noindent P r o o f. We have
\[
b(u,\bar{u})+\| u\|^2_{L_2({\cal D})^3} \ge c\, \| u\|^2_{\cal H}
\]
for all $u\in C_0^\infty(\overline{\cal D})^3,$ $u(x)=0$ for
$|x|>1$ (see, e.g., \cite{Girault}). We consider the set of all $u
\in C_0^\infty (\overline{\cal D})^3$ with support in the ball
$|x|\le \varepsilon$. For such $u$ Hardy's inequality implies
\[
\int_{\cal D} |u(x)|^2\, dx \le c_1\, \int_{\cal D} |x|^2\,
\sum_{j=1}^3
  |\partial_{x_j} u(x)|^2\, dx \le c_1\, \varepsilon^2\,
  \| u\|^2_{\cal H}
\]
and, therefore,
\[
b(u,\bar{u}) \ge \frac c2 \,  \|  u\|^2_{\cal H}
\]
if $\varepsilon$ is sufficiently small. Applying the similarity
transformation $x=\alpha y$, we obtain the same inequality for
arbitrary $u\in C_0^\infty( \overline{\cal D})^3$. The result
follows. \hfill $\Box$

\begin{Th} \label{t2}
Let $F\in V^*$, $g\in L_2({\cal D})$, and let $h^\pm$ be such that
there exists a vector function $w\in {\cal H}$ satisfying the
equalities $S^\pm w=h^\pm$ on $\Gamma^\pm$. Then there exists a
unique solution $(u,p)\in {\cal H}\times L_2 ({\cal D})$ of the
problem {\em (\ref{bvpD1}), (\ref{bvpD2})}. Furthermore, $(u,p)$
satisfies the estimate
\[
\| u\|_{{\cal H}} + \| p\|_{L_2({\cal D})} \le c\, \Big( \|
F\|_{V^*} + \| g\|_{L_2({\cal D})}
  + \| w\|_{{\cal H}}\Big)
\]
with a constant $c$ independent of $F$, $g$ and $w$.
\end{Th}

\noindent P r o o f. 1) We prove the existence of a solution. By our
assumption on $h^\pm$ and by Theorem \ref{t1}, we may restrict
ourselves to the case $g=0$, $h^\pm=0$. Let $V_0=\{ u\in V:\, \nabla\cdot u=0\}$,
and let $V_0^\perp$ be the orthogonal complement of $V_0$ in $V$. Then, by Lax-Milgram's lemma, there
exists a vector function $u\in V_0$ such that
\[
b(u,v)=F(v)\quad \mbox{for all }v\in V_0, \quad \| u\|_{\cal H} \le c\, \| F\|_{V_0^*} \le c\, \| F\|_{V^*}.
\]
By Theorem \ref{t1}, the operator $B:\, u\to -\nabla\cdot u$ is
an isomorphism from $V_0^\perp$ onto $L_2({\cal D})$. Hence, the mapping
\[
L_2({\cal D}) \ni q \to \ell(q) \stackrel{def}{=} F(B^{-1} q)- b(u,B^{-1} q)
\]
defines a linear and continuous functional on $L_2({\cal D})$. By
Riesz representation theorem, there exists a function $p\in
L_2({\cal D})^3$ satisfying
\[
\int_{\cal D} p\, q\, dx = \ell(q)\quad\mbox{for all } q\in
L_2({\cal D}), \quad
   \| p\|_{L_2({\cal D})} \le c\, \Big( \| F\|_{V^*}+ \| u\|_{\cal H}\Big).
\]
If we set $q=- \nabla\cdot v$, where $v$ is an arbitrary element of $V_0^\perp$, we obtain
\[
-\int_{\cal D}p\, \nabla\cdot v\, dx=F(v)-b(u,v)\quad\mbox{for all
}v\in V_0^\perp .
\]
Since both sides of the last equality vanish for $v\in V_0$, we
get (\ref{bvpD1}). Furthermore, $-\nabla\cdot u=0$ and $S^\pm
u|_{\Gamma^\pm}=0$.

2) We prove the uniqueness. Suppose $(u,p) \in {\cal H}\times L_2({\cal D})$ is a solution of problem
(\ref{bvpD1}), (\ref{bvpD2}) with $F=0$, $g=0$, $h^\pm=0$. Then, in particular,
$u\in V_0$ and $b(u,u)=0$ what, by Lemma \ref{l2}, implies $u=0$. Consequently,
\[
\int_{\cal D} p\, \nabla\cdot v\, dx =0 \quad\mbox{for all }v\in
V.
\]
Since $v$ can be chosen such that $\nabla\cdot v=p$, we obtain
$p=0$. The proof is complete. \hfill $\Box$

\begin{Rem} \label{r1}
{\em The assumption on $h^\pm$ in Theorem \ref{t2} is satisfied e.g. for $h^\pm \in V_0^{1/2}
(\Gamma^\pm)^{3-d^\pm}$. Then, by \cite[Le.1.2]{mp78a}, there exists a vector function $w\in
V_0^1({\cal D})^3\subset {\cal H}$ satisfying $S^\pm w =h^\pm$ on $\Gamma^\pm$.
(Note that $h^\pm$ is a vector-function if $d^\pm \le 1$ and a scalar function if $d^\pm=2$.
In the case $d^\pm=1$ the vector $h^\pm$ is tangential to $\Gamma^\pm$, therefore, the corresponding
function space can be identified with $V_0^{1/2}(\Gamma^\pm)^2$.)

Furthermore, for $h^\pm \in W_\delta^{3/2}(\Gamma^\pm)^{3-d^\pm}$, $0<\delta<1$, satisfying a compatibility
condition on $M$  there exists a vector function $w \in W_\delta^2({\cal D})^3 \subset {\cal H}$ such that
$S^\pm w =h^\pm$ on $\Gamma^\pm$ (see Lemma \ref{l11}) below.}
\end{Rem}

\subsection{Reduction to homogeneous boundary conditions}

In the sequel, we will consider weak solutions of problem (\ref{StokesD}), (\ref{bcD})
with data $f \in W_\delta^1({\cal D})^3$, $g\in W_\delta^1({\cal D})$, $h^\pm \in
W_\delta^{3/2}(\Gamma^\pm)^{3-d^\pm}$ and $\phi^\pm \in W_\delta^{1/2}(\Gamma^\pm)^{d^\pm}$,
$0<\delta<1$.

\begin{Le} \label{l10}
{\em 1)} For arbitrary $h^\pm \in V_\delta^{1/2}(\Gamma^\pm)^{3-d^\pm}$ there exists a vector
function $u\in V_\delta^1({\cal D})^3$ such that $S^\pm u = h^\pm$ on $\Gamma^\pm$ and
\[
\| u\|_{V_\delta^1({\cal D})^3} \le c\sum_\pm \| h^\pm\|_{V_\delta^{1/2}(\Gamma^\pm)^{3-d^\pm}}
\]
with a constant $c$ independent of $h^+$ and $h^-$.

{\em 2)} For arbitrary $h^\pm \in V_\delta^{l-1/2}(\Gamma^\pm)^{3-d^\pm}$ and
$\phi^\pm \in V_\delta^{l-3/2}(\Gamma^\pm)^{d^\pm}$, $l\ge 2$, there exists a
vector function $u \in V_\delta^l({\cal D})^3$ satisfying the
boundary conditions
\begin{equation} \label{1l10}
S^\pm u = h^\pm,\quad N^\pm(u,0)=\phi^\pm \quad \mbox{ on }\Gamma^\pm
\end{equation}
and the estimate
\begin{equation} \label{3l10}
\| u\|_{V_\delta^l({\cal D})^3}\le c\, \sum_\pm \Big( \| h^\pm
  \|_{V_\delta^{l-1/2}(\Gamma^\pm)^{3-d^\pm}}
  + \|\phi^\pm\|_{V_\delta^{l-3/2}(\Gamma^\pm)^{3-d^\pm}}\Big).
\end{equation}
Moreover, if $\mbox{\em supp}\, h^\pm \in \Gamma^\pm \cap {\cal
U}$ and $\mbox{\em supp}\, \phi^\pm \in \Gamma^\pm \cap {\cal U}$,
where ${\cal U}$ is an arbitrary domain in ${\Bbb R}^3$, then $u$
can be chosen such that $\mbox{\em supp}\, u \in {\cal U}$.
\end{Le}

\noindent P r o o f. The first part of the lemma follows immediately from \cite[Le.1.2]{mp78a}.
For the proof of the second part, we assume for simplicity that $\Gamma^-$ coincides
with the half-plane $\varphi=0$ (i.e., $x_1>0$, $x_2=0$) and
$\Gamma^+$ with the half-plane $\varphi=\theta$. Then the boundary conditions
\begin{equation} \label{2l10}
S^- u = h^-,\quad N^-(u,0)=\phi^\pm \quad \mbox{ on }\Gamma^-
\end{equation}
have the form $u=h^-$ on $\Gamma^-$ if $d^-=0$,
\begin{eqnarray*}
&& u_1=h_1^-,\ \ u_3=h_3^-, \ \ 2\partial_{x_2}u=\phi^- \ \mbox{ on }\Gamma^-\  \mbox{ if } d^-=1, \\
&& u_2=h^-,\ \ -\varepsilon_{1,2}(u)=\phi_1^-, \ \ -\varepsilon_{3,2}(u)=\phi_3^- \
  \mbox{ on }\Gamma^-\  \mbox{ if } d^-=2, \\
&& -2\varepsilon_{j,2}(u)=\phi_j^-\ \mbox{ on }\Gamma^-\  \mbox{ for } j=1,2,3\  \mbox{ if } d^-=3.
\end{eqnarray*}
In all these cases, the existence of a vector function $u\in V_\delta^l({\cal D})^3$ satisfying
(\ref{2l10}) can be easily deduced from \cite[Le.3.1]{mp78a}. Analogously, there exists a
function $v\in V_\delta^2({\cal D})^3$ satisfying $S^+v=h^+$ and $N^+(v,0)v=\phi^+$ on $\Gamma^+$.
Let $\zeta=\zeta(\varphi)$ be a smooth function on $[0,\theta]$ equal to 1 for $\varphi<\theta/2$
and to zero for $\varphi>3\theta/4$. Then the function $w(x)=\zeta(\varphi)\, u(x) + (1-\zeta(\varphi))\, v(x)$
satisfies (\ref{1l10}). \hfill $\Box$\\

The analogous result in the space $W_\delta^l$ holds only under additional assumptions on the
boundary data. If $u\in W_\delta^l({\cal D})^3$, $\delta<l-1$, then there exists the trace
$u|_M \in W^{l-1-\delta}(M)^3$, and from the boundary conditions (\ref{bcD}) it follows that
$S^\pm u|_M = h^\pm|_M$. Here $S^+$ and $S^-$ are considered as operators on $W^{1-\delta}(M)^3$.
Consequently, the boundary data $h^+$ and $h^-$ must satisfy the compatibility condition
\begin{equation} \label{cc}
\big( h^+|_M\, ,h^-|_M \big) \in R(T),
\end{equation}
where $R(T)$ is the range of the operator $T=(S^+,S^-)$. For example, in the case of the Dirichlet
problem ($d^+=d^-=0$), condition (\ref{cc}) is satisfied if and only if $h^+|_M =h^-|_M$,
while in the case $d^- = 0,\ d^+=2$ condition (\ref{cc}) is equivalent to $h^-|_M\cdot n^+ = h^+|_M$.

\begin{Le} \label{l11}
Let $h^\pm \in W_\delta^{l-1/2}(\Gamma^\pm)^{3-d^\pm}$ and $\phi^\pm \in W_\delta^{l-3/2}(\Gamma^\pm)^{d^\pm}$,
$0\le l-2<\delta<l-1$, be functions vanishing for $r(x)>C$. Suppose that $h^+$ and $h^-$
satisfy the compatibility condition {\em (\ref{cc})} on $M$. Then there exists a vector function
$u\in W_\delta^l({\cal D})^3$ satisfying {\em (\ref{1l10})} and an estimate analogous to {\em
(\ref{3l10})}. Moreover, if $\mbox{\em supp}\, h^\pm \in \Gamma^\pm \cap {\cal U}$ and
$\mbox{\em supp}\, \phi^\pm \in \Gamma^\pm \cap {\cal U}$, where ${\cal U}$ is an arbitrary domain
in ${\Bbb R}^3$, then $u$ can be chosen such that $\mbox{\em supp}\, u \in {\cal U}$.
\end{Le}

\noindent P r o o f. By (\ref{cc}), there exists a vector function $\psi\in W^{l-1-\delta}(M)^3$ such that
$S^\pm \psi =h^\pm|_M$. Let $v \in W_\delta^l({\cal D})^3$ be an extension of $\psi$. Then the
trace of $h^\pm -S^\pm v|_{\Gamma^\pm}$ on $M$ is equal to zero and, consequently, $h^\pm -S^\pm
v|_{\Gamma^\pm}\in V_\delta^{l-1/2}(\Gamma^\pm)^{3-d^\pm}$ (cf. Lemma \ref{l0}).
Furthermore, $\phi^\pm -N^\pm (v,0)|_{\Gamma^\pm} \in W_\delta^{l-3/2}(\Gamma^\pm)^{d^\pm}\subset
V_\delta^{l-3/2}(\Gamma^\pm)^{d^\pm}$. Thus, according to Lemma \ref{l10}, there exists a function
$w\in V_\delta^l({\cal D})^3$ such that $S^\pm w =h^\pm -S^\pm v$ and $N^\pm (w,0)=\phi^\pm
-N^\pm (v,0)$ on $\Gamma^\pm$. Then $u=v+w$ satisfies (\ref{1l10}). \hfill $\Box$

\subsection{A priori estimates for the solutions}

The proofs of the following lemmas are essentially based on local estimates for
solutions of elliptic boundary value problems in smooth domains. In the sequel,
$W_{loc}^l(\overline{\cal D}\backslash M)$ denotes the set of all functions $u$ on ${\cal D}$ such that
$\zeta u$ belongs to the Sobolev space $W^l({\cal D})$ for arbitrary $\zeta\in
C_0^\infty(\overline{\cal D}\backslash M)$.

\begin{Le} \label{l6}
Let $u\in W^1_{loc}(\overline{\cal D}\backslash M)^3\cap V_{\delta-1}^0({\cal D})^3$ and $p\in W_{loc}^0
(\overline{\cal D}\backslash M)\cap V_{\delta-1}^{-1}({\cal D})$ satisfy
\begin{equation} \label{ve}
b(u,v)-\int_{\cal D} p\, \nabla\cdot v\, dx = F(v)\quad \mbox{for all } v\in
  C_0^\infty(\overline{\cal D}\backslash M)^3,
\end{equation}
$-\nabla u= g$ in ${\cal D}$ and $S^\pm u =h^\pm$ on $\Gamma^\pm$, where $F\in V_\delta^{-1}({\cal D})^3$,
$g\in V_\delta^0({\cal D})$, $h^\pm \in V_\delta^{1/2}(\Gamma^\pm)^{3-d^\pm}$, $0<\delta<1$.
Then $u\in V_\delta^1({\cal D})^3$, $p\in V_\delta^0({\cal D})$, and
\[
\| u\|_{V_\delta^1({\cal D})^3}+ \|p\|_{V_\delta^0({\cal D})} \le c\, \Big(
  \|F\|_{V_\delta^{-1}({\cal D})^3} +\| g\|_{V_\delta^0({\cal D})}
  + \sum_\pm \|h^\pm\|_{V_{\delta}^{1/2}(\Gamma^\pm)}
  + \|u\|_{V_{\delta-1}^0({\cal D})^3} + \| p\|_{V_{\delta-1}^{-1}({\cal D})} \Big).
\]
\end{Le}

\noindent P r o o f. Due to Lemma \ref{l10}, we may assume without loss of generality that $h^\pm =0$
Let $\zeta_{j,k}$ be infinitely differentiable functions such that
\[
\hspace{-2em}\mbox{supp}\, \zeta_{j,k}\subset\{ x:\ 2^{k-1}< |x'|<2^{k+1}, \ j-1< 2^{-k}x_3< j+1\}, \ \
  \sum_{j,k=-\infty}^{+\infty} \zeta_{j,k}=1,\ \ \big|\partial_x^\alpha\zeta_{j,k}(x)\big|\le c\, 2^{-k|\alpha|}.
\]
Furthermore, let $\displaystyle \eta_{j,k}=  \sum_{i=j-1}^{j+1} \,
\sum_{l=k-1}^{k+1} \zeta_{i,l}$. Obviously, $\eta_{j,k}=1$ on supp$\, \zeta_{j,k}$.
We introduce the functions $\tilde{\zeta}_{j,k}(x)=\zeta_{j,k}(2^k x)$, $\tilde{\eta}_{j,k}(x)
=\eta_{j,k}(2^k x)$, $\tilde{g}(x)=g({2^k}x)$, $\tilde{p}(x)=p(2^kx)$,
$\tilde{u}(x)=2^{-k}u(2^kx)$, and $\tilde{v}(x)=2^{-k}v(2^kx)$, where $v$ is an arbitrary
vector function from $C_0^\infty(\overline{\cal D}\backslash M)^3$.
Then the support of $\zeta_{j,k}$ is contained in the set $\{x:\, 1/2< r(x)<2,\ j-1< x_3<j+1\}$
and the derivatives of $\zeta_{j,k}$ are bounded by constants independent of $j$ and $k$.
Furthermore, we have $-\nabla\cdot \tilde{u}
=\tilde{g}$ in ${\cal D}$, $S^\pm \tilde{u}=0$ on $\Gamma^\pm$, and
\[
b(\tilde{u},\tilde{v})-\int_{\cal D} \tilde{p}\, \nabla\cdot \tilde{v}\, dx = \tilde{F}(\tilde{v}),
\]
where $\tilde{F}(\tilde{v}) =2^{-3k}F(v)$. Using local estimates of weak solutions of elliptic
boundary value problems in smooth domains (see e.g. \cite{adn}, \cite[Sec.3.2]{kmr1}), we obtain
\[
\hspace{-1em}\|\tilde{\zeta}_{j,k}\tilde{u}\|^2_{V_\delta^1({\cal D})^3}+\|\tilde{\zeta}_{j,k}\tilde{p}
  \|^2_{V_\delta^0({\cal D})}
\le c\, \Big( \|\tilde{\eta}_{j,k}\tilde{F}\|^2_{V_\delta^{-1}({\cal D})^3} +
  \|\tilde{\eta}_{j,k}\tilde{g} \|^2_{V_\delta^0({\cal D})}+ \| \tilde{\eta}_{j,k} \tilde{u}
  \|^2_{V_{\delta-1}^0({\cal D})^3} + \| \tilde{\eta}_{j,k}\tilde{p}\|^2_{V_{\delta-1}^{-1}({\cal D})}\Big)
\]
with a constant $c$ independent of $j$ and $k$. This implies
\[
\hspace{-1em}\|\zeta_{j,k}u\|^2_{V_\delta^1({\cal D})^3}+\|\zeta_{j,k}p\|^2_{V_\delta^0({\cal D})}
 \le c\, \Big( \|\eta_{j,k}F\|^2_{V_\delta^{-1}({\cal D})^3} +
  \| \eta_{j,k}g  \|^2_{V_\delta^0({\cal D})}+ \| \eta_{j,k} u\|^2_{V_{\delta-1}^0({\cal D})^3}
  + \| \eta_{j,k} p\|^2_{V_{\delta-1}^{-1}({\cal D})}\Big)
\]
with the same constant $c$. It can be proved analogously to \cite[Le.6.1.1]{kmr1} that the norm in
$V_\delta^l({\cal D})$ is equivalent to
\[
\| u\| = \Big( \sum_{j,k=-\infty}^{+\infty} \| \zeta_{j,k} u\|^2_{V_\delta^l({\cal D})} \Big)^{1/2}.
\]
The same is true (cf. \cite[Sec.6.1]{kmr1}) for the norm of the dual space $V_{-\delta}^{-l}({\cal D})$.
Using this and the last inequality, we conclude that $u\in V_\delta^1({\cal D})^3$, $p\in
V_\delta^0({\cal D})$. Moreover, the desired estimate for $u$ and $p$ holds. \hfill $\Box$\\

Analogously, the following lemma can be proved (see \cite[Th.4.1]{mp78a}).

\begin{Le} \label{l4}
Let $(u,p)$ be a solution of problem {\em (\ref{StokesD}), (\ref{bcD})}, $u\in W_{loc}^2
(\overline{\cal D}\backslash M)^3 \cap V_{\delta-1}^{l-1}({\cal D})^3$, \linebreak $p\in W_{loc}^1
(\overline{\cal D}\backslash M)\cap V_{\delta-1}^{l-2}({\cal D})$ $l\ge 2$.
Suppose  that $f\in V_\delta^{l-2}({\cal D})^3$, $g\in V_\delta^{l-1}({\cal D})$, and the components
of $h^\pm$ and $\phi^\pm$ are from $V_\delta^{l-1/2}(\Gamma^\pm)$ and
$V_\delta^{l-3/2}(\Gamma^\pm)$, respectively. Then $u\in V_\delta^l({\cal D})^3$ and $p\in V_\delta^{l-1}({\cal D})$.
\end{Le}

Furthermore, there is the following result (cf. \cite[Le.2.3]{mr-01}).

\begin{Le} \label{l4a}
Let $(u,p)$ be a solution of problem {\em (\ref{StokesD}), (\ref{bcD})}, and let $\zeta,\eta$
be infinitely differentiable functions on $\bar{\cal D}$ with compact supports such that
$\eta=1$ in a neighborhood of $\mbox{\em supp}\,\zeta$. Suppose that $\eta u\in W_{loc}^2(\overline{\cal D}
\backslash M)^3 \cap W_{\delta-1}^{l-1}({\cal D})^3$, $\eta p\in W_{loc}^1(\overline{\cal D}\backslash M)
\cap W_{\delta-1}^{l-2}({\cal D})$ $l\ge 2$, $\eta f\in W_\delta^{l-2}({\cal D})^3$, $\eta g\in
W_\delta^{l-1}({\cal D})$, and the components of $\eta h^\pm$ and $\eta \phi^\pm$ are
from $W_\delta^{l-1/2}(\Gamma^\pm)$ and $W_\delta^{l-3/2}(\Gamma^\pm)$, respectively. Then $\zeta u\in
W_\delta^l({\cal D})^3$ and $\zeta p\in W_\delta^{l-1}({\cal D})$.
\end{Le}

\subsection{Smoothness of \boldmath$x_3$-derivatives\unboldmath}

Our goal is to show that the solution $(u,p)$ of problem (\ref{StokesD}), (\ref{bcD}) belongs to
$W_\delta^2({\cal D})^3\times W_\delta^1({\cal D})$ if $f \in W_\delta^1({\cal D})^3$, $g\in
W_\delta^1({\cal D})$, $h^\pm \in W_\delta^{3/2}(\Gamma^\pm)^{3-d^\pm}$ and $\phi^\pm
W_\delta^{3/2}(\Gamma^\pm)^{d^\pm}$, $0<\delta<1$. For this end, we show in this subsection that
$\partial_{x_3}u \in V_\delta^1({\cal D})^3$ and $\partial_{x_3}p \in V_\delta^0({\cal D})$ under
the above assumptions on the data. Due to Lemma \ref{l11}, we may restrict ourselves to
homogeneous boundary conditions ($h^\pm=0$, $\phi^\pm=0$).

\begin{Le} \label{l7}
Let $(u,p)\in {\cal H} \times L_2({\cal D})$ be a solution of problem {\em (\ref{bvpD1}), (\ref{bvpD2})}.
Suppose that $g\in V_\delta^1({\cal D})$, $0<\delta<1$, $h^\pm =0$, and the functional $F$ has the form
\[
F(v) = \int_{\cal D} f\cdot v\, dx,
\]
where $f\in V_\delta^0({\cal D})^3$. We assume further that $f$ and $g$ have compact support. Then
$\partial_{x_3} u \in V_{\delta-1}^0({\cal D})^3$, $\partial_{x_3} p \in V_{\delta-1}^{-1}({\cal D})$ and
\[
\| \partial_{x_3} u \|_{V_{\delta-1}^0({\cal D})^3} + \|\partial_{x_3} p
  \|_{V_{\delta-1}^{-1}({\cal D})} \le c\, \Big( \| f\|_{V_\delta^0({\cal D})^3}
  + \| g\|_{V_\delta^1({\cal D})} \Big).
\]
\end{Le}

\noindent P r o o f. First note that $F\in {\cal H}^*\subset V^*$ and $g \in L_2({\cal D})$ under the assumptions
of the lemma. For arbitrary real $h$ let $u_h(x)=h^{-1}\big( u(x',x_3+h)-u(x',x_3)\big)$. Obviously,
\[
b(u_h,v)-\int_{\cal D} p_h\, \nabla\cdot v\, dx = b(u,v_{-h})-\int_{\cal D} p\,
  \nabla\cdot v_{-h} \, dx = F(v_{-h}) = \int_{\cal D} f_h\cdot v\, dx
\]
for all $v\in V$, $-\nabla\cdot u_h = g_h$ in ${\cal D}$, and $S^\pm u_h=0$ on $\Gamma^\pm$. Consequently,
by Theorem \ref{t2}, there exists a constant $c$ independent of $u,p$ and $h$ such that
\begin{equation} \label{1l7}
\| u_h\|^2_{\cal H} + \| p_h\|^2_{L_2({\cal D})} \le c\, \Big(
  \| f_h\|^2_{V^*} + \| g_h\|^2_{L_2({\cal D})}\Big).
\end{equation}
We prove that
\begin{equation} \label{2l7}
\int_0^\infty h^{2\delta-1} \Big( \| f_h\|^2_{V^*} +
  \| g_h\|^2_{L_2({\cal D})} \Big)\, dh \le c\, \Big( \| f\|^2_{V_\delta^0({\cal D})^3}
  + \| g\|^2_{V_\delta^1({\cal D})} \Big).
\end{equation}
Indeed, let $\tilde{g}(x',\xi)$ be the Fourier transform of $g(x',x_3)$ with respect to the variable $x_3$. Then
\begin{eqnarray*}
&& \hspace{-2em} \int_0^\infty h^{2\delta-1} \| g_h\|^2_{L_2({\cal D})}\, dh = \int_0^\infty
  \int_{\Bbb R} \int_K h^{2\delta-3}\, |e^{i\xi h}-1|^2\, |\tilde{g}(x',\xi)|^2\,   dx'\, d\xi\, dh \\
&& \hspace{-2em} = c \int_{\Bbb R} \int_K |\xi|^{2-2\delta}\, |\tilde{g}(x',\xi)|^2\,
  dx'\, d\xi \le \int_{\Bbb R} \int_K r^{2\delta-2}\, (1+r^2\xi^2)\,
  |\tilde{g}(x',\xi)|^2\,  dx'\, d\xi \le c\, \| g\|^2_{V_\delta^1({\cal D})}
\end{eqnarray*}
Furthermore, if $\chi$ is a smooth function, $0\le \chi\le 1$, $\chi=1$ for $r>h$, $\chi=0$ for $r<h/2$,
$|\nabla \chi|\le c\, h^{-1}$, and $\varepsilon$ is an arbitrary positive number, then
\begin{eqnarray*}
&& \Big| \int_{\cal D} f_h\cdot v\, dx\Big|  =  \Big| \int_{\cal D} f\cdot v_{-h}\, dx\Big| \\
&& \le  \| \chi f\|_{L_2({\cal D})^3} \, \| v_{-h}\|_{L_2({\cal D})^3} +
  \| r^{1-\varepsilon}h^{\varepsilon-1}f\|_{L_2({\cal D}_h)^3}\,
  \| r^{\varepsilon-1}h^{1-\varepsilon}(1-\chi)v_{-h}\|_{L_2({\cal D}_h)^3},
\end{eqnarray*}
where ${\cal D}_h=\{ x\in {\cal D}:\, r(x)<h\}$. Here
\[
\| v_{-h}\|^2_{L_2({\cal D})^3} = \int_{\cal D}\Big|\int_0^1
  \partial_{x_3}v(x',x_3-th)\, dt\Big|^2\, dx\le \| \partial_{x_3} v
  \|^2_{L_2({\cal D})^3} \le \| v\|^2_{\cal H}.
\]
Using Hardy's inequality, we further obtain
\begin{eqnarray*}
\hspace{-1em} \| (r/h)^{\varepsilon-1}v_{-h}\|^2_{L_2({\cal D}_h)^3}
  \le  c\int\limits_{\cal D} r^{2\varepsilon} h^{2-2\varepsilon} \big| \partial_r(1-\chi)
  v_{-h}\big|^2\, dx \le c\int\limits_{\atops{{\cal D}} {r<h}} \big( |v_{-h}|^2+ |\partial_r v_{-h}|^2\big)\, dx
\le  c\, \| v\|^2_{\cal H}.
\end{eqnarray*}
Consequently, for $0<\varepsilon<1-\delta$ we obtain
\begin{eqnarray*}
&& \int_0^\infty h^{2\delta-1}\, \| f_h\|^2_{V^*}  \le
  c \int_0^\infty \big( \| \chi f\|_{L_2({\cal D})^3} + \| r^{1-\varepsilon}
  h^{\varepsilon-1}f\|^2_{L_2({\cal D}_h)^3}\big)\, dh \\
&& \le \int_{\cal D} |f(x)|^2 \, \Big( \int_0^{2r} h^{2\delta-1}\, dh +
  r^{2-2\varepsilon} \int_r^\infty h^{2\delta-3+2\varepsilon}\, dh\Big)\, dx
  \le c\, \| f\|^2_{V_\delta^0({\cal D})^3}.
\end{eqnarray*}
This proves (\ref{2l7}). Next we prove that
\begin{equation} \label{3l7}
\| u_z\|^2_{V_{\delta-1}^0({\cal D})^3} + \| p_z\|^2_{V_{\delta-1}^{-1}({\cal D})}
  \le c\, \int_0^\infty h^{2\delta-1}\, \Big( \| u_h\|^2_{\cal H}
  + \| p_h\|^2_{L_2({\cal D})}\Big)\, dh.
\end{equation}
It can be easily shown (see \cite[Le.3]{r90}) that
\begin{eqnarray*}
&& \hspace{-2em} \| \partial_{x_3}u\|^2_{V_{\delta-1}^0({\cal D})^3} = \int_{\Bbb R}\int_K
  |x'|^{2\delta-2}\, |\xi|^2\, |\tilde{u}(x',\xi)|^2\, dx'\, d\xi\\
&& \hspace{-2em} \le  c\,  \int_{\Bbb R}\int_K  |\xi|^{2-2\delta}\, \Big( |\xi|^2\, |\tilde{u}(x',\xi)|^2
  + \sum_{j=1}^2 |\partial_{x_j}\tilde{u}(x',\xi)|^2\Big)\, dx'\, d\xi\\
&& \hspace{-2em} =  c\int\limits_{\Bbb R} \int\limits_K \Big( \int\limits_0^\infty h^{2\delta-3}
  |e^{i\xi h}-1|^2\, dh\Big)\, \Big( |\xi|^2\, |\tilde{u}(x',\xi)|^2
  + \sum_{j=1}^2 |\partial_{x_j}\tilde{u}(x',\xi)|^2\Big)\, dx'\, d\xi =
  \int\limits_0^\infty h^{2\delta-1}\, \| u_h\|^2_{\cal H}\, .
\end{eqnarray*}
Furthermore, since
\begin{eqnarray*}
&& \Big| \int_{\cal D} \partial_{x_3}p \,\, v\, dx\Big|^2 \le
  \int_{\Bbb R}\int_K |\xi|^{2\delta}\, |\tilde{v}(x',\xi)|^2\, dx'\, d\xi \
  \int_{\Bbb R}\int_K |\xi|^{2-2\delta}\, |\tilde{p}(x',\xi)|^2\, dx'\, d\xi \\
&& \le \int_{\Bbb R}\int_K r^{-2\delta}\, (1+r^2|\xi|^2)\, |\tilde{v}(x',\xi)|^2\, dx'\,
  d\xi\  \int_{\Bbb R}\int_K |\xi|^{2-2\delta}\, |\tilde{p}(x',\xi)|^2\, dx'\, d\xi \\
&& \le \| v\|^2_{V_{1-\delta}^1({\cal D})} \ \int_{\Bbb R}\int_K |\xi|^{2-2\delta}\,
  |\tilde{p}(x',\xi)|^2\, dx'\, d\xi
\end{eqnarray*}
and
\begin{eqnarray*}
\int\limits_{\Bbb R} \int\limits_K |\xi|^{2-2\delta} \, |\tilde{p}(x',\xi)|^2\, dx'\, d\xi
  & = & c\, \int\limits_0^\infty \int\limits_{\Bbb R}\int\limits_K h^{2\delta-3}|e^{i\xi h}-1|^2 \,
  |\tilde{p}(x',\xi)|^2\, dx'\, d\xi\, dh\\
& = & c\, \int\limits_0^\infty h^{2\delta-1} \| p_h\|^2_{L_2({\cal D})} \, dh,
\end{eqnarray*}
we obtain
\[
\| \partial_{x_3}p\|^2_{V_{\delta-1}^{-1}({\cal D})} \le
\int_{\Bbb R}\int_K
  |\xi|^{2-2\delta}\, |\tilde{p}(x',\xi)|^2\, dx'\, d\xi \le c\,
  \int_0^\infty h^{2\delta-1} \| p_h\|^2_{L_2({\cal D})} \, dh.
\]
This proves (\ref{3l7}). Now the assertion of the lemma follows immediately from (\ref{1l7})--(\ref{3l7}).
\hfill $\Box$

\begin{Co} \label{c2}
Let the assumptions of Lemma {\em \ref{l7}} be satisfied. Then $\partial_{x_3}u \in V_\delta^1({\cal D})^3$,
$\partial_{x_3}p\in V_\delta^0({\cal D})$ and
\[
\|\partial_{x_3}u\|_{V_\delta^1({\cal D})^3}+\|\partial_{x_3}p\|_{V_\delta^0({\cal D})}
  \le c\, \Big( \| f\|_{V_\delta^0({\cal D})^3} + \| g\|_{V_\delta^1({\cal D})} \Big).
\]
with a constant $c$ independent of $u$ and $p$.
\end{Co}

\noindent P r o o f.
From Lemma \ref{l7} and well-known local regularity results for solutions of elliptic boundary
value problems (see e.g. \cite{adn},\cite[Sect.3.2]{kmr1}) we conclude that $\partial_{x_3}u \in
W_{loc}^1(\overline{\cal D}\backslash M)^3\cap V_{\delta-1}^0({\cal D})^3$ and $\partial_{x_3}p \in
L_{2,loc}(\overline{\cal D}\backslash M)\cap V_{\delta-1}^{-1}({\cal D})$. Obviously,
\[
b(\partial_{x_3}u,v) - \int_{\cal D} \partial_{x_3}p \, \nabla\cdot v\, dx =
  \int_{\cal D} \partial_{x_3}f \cdot v\, dx \quad\mbox{for all }v\in
  C_0^\infty(\overline{\cal D}\backslash M)^3,
\]
where $\partial_{x_3}f\in V_\delta^{-1}({\cal D})^3)$, $-\nabla \cdot \partial_{x_3}u =\partial_{x_3}g
\in V_\delta^0({\cal D})$, and $S^\pm \partial_{x_3}u=0$ on $\Gamma^\pm$. Applying Lemma \ref{l6},
we obtain $\partial_{x_3}u \in V_{\delta}^1({\cal D})^3$, $\partial_{x_3}p \in V_{\delta}^0({\cal D})$
and the desired inequality. \hfill $\Box$

\subsection{Auxiliary problems in the angle \boldmath$K$\unboldmath, operator pencils}

Suppose $(u,p)$ is a solution of the Stokes system (\ref{StokesD}) with homogeneous boundary conditions
(\ref{bcD}) which is independent of $x_3$. Then $u_3$ is a solution of the problem
\begin{equation} \label{Laplace}
-\Delta_{x'} u_3 = f_3\ \mbox{ in }K, \quad u_3=h_3^\pm \ \mbox{on }\Gamma^\pm\ \mbox{ for }
  d^\pm\le 1,\quad \frac{\partial u_3}{\partial n^\pm}=\phi_3^\pm \mbox{ on }\Gamma^\pm
  \ \mbox{ for } d^\pm\ge 2,
\end{equation}
where $\Delta_{x'}$ denotes the Laplace operator in the coordinates $x'=(x_1,x_2)$, whereas the vector
$(u',p)=(u_1,u_2,p)$ is a solution of the two-dimensional Stokes system
\begin{equation} \label{StokesK}
-\Delta_{x'} u' + \nabla_{x'} p= f',\quad -\nabla_{x'}\cdot u'=g\quad\mbox{in }K,
\end{equation}
with the corresponding boundary conditions
\begin{equation} \label{bcK}
\tilde{S}^\pm u'={h'}^\pm,\ \ \tilde{N}^\pm(u',p)={\phi'}^\pm \ \ \mbox{on }\gamma^\pm.
\end{equation}
Here
\begin{eqnarray*}
&& \tilde{S}^\pm u'=u' \ \mbox{ if }d^\pm=0,\quad \tilde{S}^\pm u'=u'\cdot\tau^\pm  \ \mbox{ if }d^\pm=1,
  \quad \tilde{S}^\pm u'=u'\cdot n^\pm \ \mbox{ if }d^\pm=2, \\
&& \tilde{N}^\pm(u',p)=-p+2(\varepsilon(u')n^\pm)\cdot n^\pm \ \mbox{ if }d^\pm=1,\quad
  \tilde{N}^\pm (u',p)=(\varepsilon(u')n^\pm)\cdot \tau^\pm \ \mbox{ if }d^\pm=2,\\
&&  \tilde{N}^\pm (u',p)=-pn^\pm+2\varepsilon(u')n^\pm \ \mbox{ if }d^\pm=3,
\end{eqnarray*}
by $\varepsilon(u')$ we denote the matrix with the components $\varepsilon_{i,j}(u)$, $i,j=1,2$,
and $\tau^\pm$ is the unit vector on $\gamma^\pm$.

Setting $u=r^\lambda U(\varphi)$, $p=r^{\lambda-1}P(\varphi)$ we obtain a boundary value problem
for the vector function $(U,P)$ on the interval $(-\theta/2,+\theta/2)$ quadratically depending on
the parameter $\lambda\in {\Bbb C}$. The operator $A(\lambda)$ of this problem is a continuous mapping
\[
\mbox{$W^2(-\frac\theta 2,+\frac\theta 2) \times W^1(-\frac\theta 2,+\frac\theta 2)
 \to L_2(-\frac\theta 2,+\frac\theta 2)\times W^1(-\frac\theta 2,+\frac\theta 2)\times {\Bbb C}^3$}
\]
for arbitrary $\lambda$. As is known, the spectrum of this pencil $A(\lambda)$ consists only of
eigenvalues with finite geometric and algebraic multiplicities. We give here a description of the
spectrum for different $d^-$ and $d^+$. Without loss of generality, we may assume that $d^-\le d^+$.
\begin{enumerate}
\item In the case of the Dirichlet problem ($d^-=d^+=0$), the spectrum of the pencil
  $A(\lambda)$ consists of the numbers $\frac{j\pi}\theta$, where $j$ is an arbitrary nonzero
  integer, and of the nonzero solutions of the equation
  \begin{equation} \label{eq00}
  \lambda\sin \theta \pm \sin(\lambda\theta) =0.
  \end{equation}
\item $d^-=0,\ d^+=1$: Then the spectrum consists of the numbers $\frac{j\pi}{\theta}$, where
  $j=\pm 1,\pm 2,\ldots$, and of the nonzero solutions of the equation
  \begin{equation} \label{eq01}
  \lambda\sin(2\theta)+\sin(2\lambda\theta)=0.
  \end{equation}
\item $d^-=0,\ d^+=2$: Then the spectrum consists of the numbers
  $\frac{j\pi}{2\theta}$, $j=\pm 1,\pm 2,\ldots$ and of the nonzero solutions of the
  equation
  \begin{equation} \label{eq02}
  \lambda\sin(2\theta) -\sin(2\lambda\theta)=0.
  \end{equation}
\item $d^-=0,\ d^+=3$: Then additionally to the numbers
  $\frac{j\pi}{2\theta}$, $j=\pm 1,\pm 2,\ldots$, the solutions of the equation
  \begin{equation} \label{eq03}
  \lambda\sin \theta \pm \cos(\lambda\theta) =0
  \end{equation}
  are eigenvalues of the pencil $A(\lambda)$.
\item $d^-=d^+=1$: Then the spectrum consists of the numbers $\frac{j\pi}{\theta}$,
  and $\frac{k\pi}\theta \pm 1$, where $j,k$ are arbitrary integers, $j\not=0$.
\item $d^-=1,\ d^+=2$: Then the spectrum consists of the numbers $\frac{j\pi}{2\theta}$,
  and $\frac{k\pi}{2\theta} \pm 1$, where $j$ is an arbitrary integer and $k$ is an
  arbitrary odd integer.
\item $d^-=1,\ d^+=3$: Then the numbers $\frac{j\pi}{2\theta}$, $j=0,\pm 1,\pm
  2,\ldots$, and all solutions of (\ref{eq02}) belong to the spectrum.
\item $d^-=d^+=2$: Then the spectrum consists of the numbers $\frac{j\pi}{\theta}$,
  and $\frac{k\pi}\theta \pm 1$, where $j,k$ are arbitrary integers.
\item $d^-=2,\ d^+=3$: Then additionally to the numbers $\frac{j\pi}\theta$, $j=0,\pm
  1,\ldots$, the solutions of (\ref{eq01}) belong to the spectrum.
\item In the case of the Neumann problem ($d^-=d^+=3$) the spectrum of the pencil $A(\lambda)$
  consists of the numbers $\frac{j\pi}\theta$, where $j$ is an arbitrary integer, and of all
  solutions of (\ref{eq00}).
\end{enumerate}
We refer to \cite{Kalex,orlt} and for the cases 1 and 3 also to \cite{kmr2,mps}.

Note that the line $\mbox{Re}\, \lambda=0$ may contain only the eigenvalue $\lambda=0$ and
that $\lambda=0$ is an eigenvalue only if one of the following conditions is satisfied:
\begin{enumerate}
\item $d^+=3$ and $d^-\ge 1$ (or $d^-=3$ and $d^+\ge 1$),
\item $1\le d^+=d^-\le 2$ and $\theta\in\{\pi,2\pi\}$,
\item $d^+=1,\ d^-=2$ (or $d^-=1,\ d^+=2$) and $\theta\in \{\frac\pi 2 ,\frac{3\pi}2\}$.
\end{enumerate}
The eigenvectors corresponding to the eigenvalue $\lambda=0$ are of the form $(U,P)=(C,0)$, where $C$
is constant, and have rank 2 (i.e., have a generalized eigenvector).

\begin{Le} \label{l8}
Let $(u,p) \in W_0^1(K)^3\times L_2(K)$ be a solution of problem {\em (\ref{Laplace})--(\ref{bcK})}
vanishing outside the unit ball. Suppose that $f\in W_\delta^0(K)^3$, $g\in W_\delta^1(K)$,
$0<\delta<1$, $h^\pm=0$, $\phi^\pm=0$, and that the strip $0<\mbox{\em Re}\, \lambda\le 1-\delta$
does not contain eigenvalues of the pencil $A(\lambda)$. Then $u\in W_\delta^2(K)^3$,
$p\in W_\delta^1(K)$, and
\[
\|u\|_{W_\delta^2(K)^3} + \|p\|_{W_\delta^1(K)} \le c\, \Big(
  \|f\|_{W_\delta^0(K)^3}\ + \|g\|_{W_\delta^1(K)}\Big) .
\]
\end{Le}

\noindent P r o o f. Since the support of $(u,p)$ is compact, we have
$(u,p)\in V_\varepsilon^1(K)^3 \times V_\varepsilon^0(K)$ with
arbitrary $\varepsilon>0$. Consequently, $u$ admits the
representation $u= c + d\, \log r +v$ with constant vectors $c,d$
and $v\in V_\delta^2(K)^3$. Furthermore, $p\in V_\delta^1(K)$ and
\[
\|v\|_{V_\delta^2(K)^3} + \|p\|_{V_\delta^1(K)} \le c\, \Big(
  \|f\|_{V_\delta^0(K)^3}\ + \|g\|_{V_\delta^1(K)}\Big)
\]
(see, e.g., \cite[Th.8.2.2]{kmr1}). Since $u\in W_0^1(K)^3$, we conclude that $d=0$.
The result follows. \hfill $\Box$\\

For the proof of higher order regularity results, we need the following lemma.

\begin{Le} \label{l8a}
Let $l$ be a integer, $l\ge 1$. Furthermore, let $f$ be a homogenous vector polynomial of degree $l-2$,
$f=0$ if $l=1$, $g$ a homogeneous polynomial of degree $l-1$, $h^\pm = c^\pm r^l$, and
$\phi^\pm= d^\pm r^{l-1}$, where $c^\pm \in {\Bbb C}^{3-d^\pm}$,
and $d^\pm \in {\Bbb C}^{d^\pm}$. We suppose that $\lambda = l$ is not an
eigenvalue of the pencil $A(\lambda)$. Then there exist unique homogeneous polynomials
\begin{equation} \label{1l8a}
u = \sum_{i+j=l} c_{i,j}\, x_1^i\, x_2^j,\qquad p=\sum_{i+j=l-1} d_{i,j}\, x_1^i\, x_2^j,
\end{equation}
where $c_{i,j}\in {\Bbb C}^3$, $d_{i,j} \in {\Bbb C}$, such that $(u,p)$ is a solution of
{\em (\ref{Laplace})--(\ref{bcK})}.
\end{Le}

\noindent P r o o f.
Inserting (\ref{1l8a}) into (\ref{Laplace})--(\ref{bcK}), we obtain a linear system of $4l+3$ equations
for $4l+3$ unknowns $c_{i,j},d_{i,j}$. Since $\lambda=l$ is not an eigenvalue of $A(\lambda)$,
the corresponding homogenous system has only the trivial solution. Therefore, the inhomogeneous system
is uniquely solvable. This proves the lemma. \hfill $\Box$

\begin{Le} \label{l9}
{\em 1)} Let $(u,p) \in V_\delta^{l-1}(K)^3 \times V_\delta^{l-2}(K)$ be a solution of
{\em (\ref{Laplace})--(\ref{bcK})}. Suppose that
\[
f\in V_\delta^{l-2}(K)^3,  \ g\in V_\delta^{l-1}(K), \ h^\pm \in V_\delta^{l-1/2}(\gamma^\pm)^{3-d^\pm},\
  \phi^\pm \in V_\delta^{l-3/2}(\gamma^\pm)^{d^\pm},
\]
$l\ge 3$, and that the strip $l-2-\delta\le \mbox{\em Re}\, \lambda
\le l-1-\delta$ does not contain eigenvalues of the pencil $A(\lambda)$. Then $(u,p)\in V_\delta^l(K)^3
\times V_\delta^{l-1}(K)$.

{\em 2)} Let $(u,p) \in W_\delta^{l-1}(K)^3 \times W_\delta^{l-2}(K)$ be a solution of {\em (\ref{Laplace})--(\ref{bcK})}.
Suppose that
\[
f\in W_\delta^{l-2}(K)^3,\ g\in W_\delta^{l-1}(K),\ h^\pm \in W_\delta^{l-1/2}(\gamma^\pm)^{3-d^\pm},\
 \phi^\pm \in W_\delta^{l-3/2}(\gamma^\pm)^{d^\pm},
\]
$l\ge 3$, $0<\delta<1$, $\delta$ is not integer, and that the strip $l-2-\delta\le \mbox{\em Re}\, \lambda
\le l-1-\delta$ does not contain eigenvalues of the pencil $A(\lambda)$. Then $u\in W_\delta^l(K)^3$ and
$p\in W_\delta^{l-1}(K)$.
\end{Le}

\noindent P r o o f. For the first assertion we refer to \cite[Th.8.2.2]{kmr1}. We prove the second one.
Let $\zeta$ be a smooth cut-off function in ${\Bbb R}^2$ equal to one for $|x'|<1$ and to zero
for $|x'|>2$. Furthermore let $u^{(1)}$ be the Taylor polynomial of degree $l-3$ of $u$, and $p^{(1)}$
the Taylor polynomial of degree $l-4$ of $p$, $p^{(1)}=0$ if $l=3$. Then there are the representations
\[
u= u^{(0)} + \zeta u^{(1)},\ \ p = p^{(0)} + \zeta p^{(1)},
\]
where $u^{(0)} \in V_\delta^{l-1}(K)^3$ and $p^{(0)} \in V_\delta^{l-2}(K)$ (see \cite[Th.7.1.1]{kmr1}).
From (\ref{Laplace})--(\ref{bcK}) it follows that
\[
-\Delta_{x'}\big( u_1^{(0)} ,u_2^{(0)} \big) + \nabla_{x'}p' = F',
\quad -\nabla_{x'}\cdot \big( u_1^{(0)} ,u_2^{(0)} \big) = G,\quad
-\Delta_{x'} u_3^{(0)} =F_3 \ \mbox{in }K,
\]
where $F=(F',F_3) \in V_\delta^{l-3}(K)^3 \cap W_\delta^{l-2}(K)^3$, $G\in V_\delta^{l-2}(K)\cap
W_\delta^{l-1}(K)$. Furthermore,
\[
\tilde{S}^\pm (u_1^{(0)},u_2^{(0)},p^{(0)})= H^\pm \in V_\delta^{l-3/2}(\gamma^\pm)
  \cap W_\delta^{l-1/2}(\gamma^\pm)
\]
and
\[
\tilde{N}^\pm (u_1^{(0)},u_2^{(0)},p^{(0)})=\Phi^\pm \in V_\delta^{l-5/2}(\gamma^\pm)
  \cap W_\delta^{l-3/2}(\gamma^\pm).
\]
An analogous inclusion holds for the traces of $u_3^{(0)}$ and $\partial u_3^{(0)}/\partial n^\pm$, respectively.
By \cite[Th.7.1.1,7.1.3]{kmr1}, there are the representations
\[
F = \zeta\, \sum_{i+j=l-4} F_{i,j}\, x_1^i\, x_2^j + \tilde{F},\quad
  G = \zeta\, \sum_{i+j=l-3} G_{i,j}\, x_1^i\, x_2^j + \tilde{G},
\]
$H^\pm = \zeta\, c^\pm\, r^{l-2} + \tilde{H}^\pm,$ and $\Phi^\pm = \zeta\, d^\pm\, r^{l-3} + \tilde{\Phi}^\pm$,
where $\tilde{F} \in V_\delta^{l-2}(K)^3$, $\tilde{G} \in V_\delta^{l-1}(K)$, $\tilde{H}^\pm \in
V_\delta^{l-1/2}(\gamma^\pm)$, $\tilde{\Phi}^\pm \in V_\delta^{l-1/3}(\gamma^\pm)$.
Now the desired result can be easily deduced from assertion 1) and Lemma \ref{l8a}. \hfill $\Box$

\subsection{Regularity assertions for weak solutions}

\begin{Le} \label{l12}
Let $(u,p)\in {\cal H}\times L_2({\cal D})$ be a solution of {\em (\ref{bvpD1}), (\ref{bvpD2})}.
We assume that the support of $(u,p)$ is compact, the functional $F$ has the form {\em (\ref{funct})},
where
\[
f\in W_\delta^0({\cal D})^3, \ \ g \in W_\delta^1({\cal D}),  \ \ \phi^\pm \in
  W_\delta^{1/2}(\Gamma^\pm)^{d^\pm},
\]
$0<\delta<1$, and that $h^\pm \in W_\delta^{3/2}(\Gamma^\pm)^{3-d^\pm}$ satisfy the compatibility
condition {\em (\ref{cc})}. If there are no eigenvalues of the pencil $A(\lambda)$ in the strip
$0<\mbox{\em Re}\, \lambda \le 1-\delta$, then $(u,p)\in W_\delta^2({\cal D})^3\times
W_\delta^1 ({\cal D})$ and
\begin{eqnarray} \label{1l12}
&& \| u\|_{W_\delta^2({\cal D})^3} + \| p\|_{W_\delta^1({\cal D})} \nonumber \\
&& \le c\, \Big(
  \| f\|_{W_\delta^0({\cal D})^3} + \| g\|_{W_\delta^1({\cal D})}
  + \sum_\pm \| h^\pm\|_{W_\delta^{3/2}(\Gamma^\pm)^{3-d^\pm}}
  + \sum_\pm \| \phi^\pm\|_{W_\delta^{1/2}(\Gamma^\pm)^{3-d^\pm}}\Big),
\end{eqnarray}
where the constant $c$ is independent of $f$, $g$, $h^\pm$, and $\phi^\pm$.
\end{Le}

\noindent P r o o f. Due to Lemma \ref{l11}, we may assume without loss of
generality that $h^\pm=0$ and $\phi^\pm=0$. By our assumptions on
$u$ and $p$, we have $u(\cdot,x_3) \in W_0^1(K)^3$ and
$p(\cdot,x_3)\in L_2(K)$ for almost all $x_3$. Furthermore, by
Corollary \ref{c2}, $(\partial_{x_3}u)(\cdot,x_3) \in
W_\delta^1(K)^3$ and $(\partial_{x_3}p)(\cdot,x_3) \in
W_\delta^0(K)$. Consequently, for almost all $x_3$, the function
$u_3(\cdot,x_3)$ is a solution of problem (\ref{Laplace}) with
$f_3+\partial^2_{x_3}u_3-\partial_{x_3}p \in W_\delta^0(K)$ on the
right-hand side of the differential equation, while
$u'(\cdot,x_3)=(u_1(\cdot,x_3),u_2(\cdot,x_3))$ is a solution of
problem (\ref{StokesK}), where $f'$ und $g$ have to be replaced by
$f'+\partial_{x_3}^2 u'$ and $g+\partial_{x_3}u_3$, respectively.
Applying Lemma \ref{l8}, we obtain $u(\cdot,x_3) \in
W_\delta^2(K)^3$, $p(\cdot,x_3)\in W_\delta^1(K)$, and
\begin{eqnarray*}
&& \| u(\cdot,x_3) \|^2_{W_\delta^2(K)^3} + \| p(\cdot,x_3)\|^2_{W_\delta^1(K)} \\
&& \le c\, \Big( \|
  f(\cdot,x_3)\|^2_{W_\delta^0(K)^3}+ \|   g(\cdot,x_3)\|^2_{W_\delta^1(K)}
 +\ \| (\partial_{x_3}u)(\cdot,x_3)\|^2_{W_\delta^1(K)^3} +
  \| (\partial_{x_3}p)(\cdot,x_3)\|^2_{W_\delta^0(K)}\Big)
\end{eqnarray*}
with a constant $c$ independent of $x_3$. Integrating this
inequality with respect to $x_3$ und using Corollary \ref{c2}, we
obtain the assertion of the lemma. \hfill $\Box$\\

The following lemma can be proved in the same way using Lemma \ref{l9}.

\begin{Le} \label{l12a}
{\em 1)} Let $(u,p)$ be a solution of {\em (\ref{bvpD1}), (\ref{bvpD2})} such that
$\partial_{x_3}^j (u,p) \in W_\delta^{l-1}({\cal D})^3 \times W_\delta^{l-2}({\cal D})$ for $j=0$ and $j=1$.
Suppose that
\[
f\in W_\delta^{l-2}({\cal D})^3,\ g\in W_\delta^{l-1}({\cal D}),\ h^\pm \in W_\delta^{l-1/2}(\Gamma^\pm)^{3-d^\pm},\
 \phi^\pm \in W_\delta^{l-3/2}(\Gamma^\pm)^{d^\pm},
\]
$l\ge 3$, $0<\delta<1$, $\delta$ is not integer, and that the strip $l-2-\delta\le \mbox{\em Re}\, \lambda
\le l-1-\delta$ does not contain eigenvalues of the pencil $A(\lambda)$. Then $u\in W_\delta^l({\cal D})^3$ and
$p\in W_\delta^{l-1}({\cal D})$.

{\em 2)} The same result is true in the class of the spaces $V_\delta^l$.
\end{Le}

\begin{Th} \label{t3}
Let $\zeta,\ \eta$ be smooth cut-off functions with support in the unit ball such that $\eta=1$
in a neighborhood of $\mbox{\em supp}\, \zeta$, and let $(u,p)\in {\cal H}\times L_2({\cal D})$ be
a solution of {\em (\ref{bvpD1}), (\ref{bvpD2})}, where $F$ is a functional on $V$ which has
the form {\em (\ref{funct})} with
\[
\eta \partial_{x_3}^j f\in W_\delta^0({\cal D})^3, \ \ \eta \partial_{x_3}^j g \in W_\delta^1({\cal D}),
  \ \ \eta\partial_{x_3}^j \phi^\pm \in W_\delta^{1/2}(\Gamma^\pm)^{d^\pm} \ \ for \ j=0,1,\ldots,k.
\]
We assume further that $\eta \partial_{x_3}^j h^\pm \in W_\delta^{3/2}(\Gamma^\pm)^{3-d^\pm}$
for $j=0,1,\ldots,k$, $h^\pm$ satisfy the compatibility condition {\em (\ref{cc})},
$0<\delta<1$, and the strip $0<\mbox{\em Re}\, \lambda \le 1-\delta$
does not contain eigenvalues of the pencil $A(\lambda)$. Then $\zeta\partial_{x_3}^k (u,p)\in
W_\delta^2({\cal D})^3\times W_\delta^1({\cal D})$ and there is the estimate
\begin{eqnarray} \label{1t3}
&& \hspace{-4.5em} \| \zeta \partial_{x_3}^k u\|_{W_\delta^2({\cal D})^3} + \|\zeta \partial_{x_3}^k p
  \|_{W_\delta^1({\cal D})} \le c\, \Big( \sum_{j=0}^k \Big( \|\eta \partial_{x_3}^j
  f\|_{W_\delta^0({\cal D})^3} + \|\eta \partial_{x_3}^j g\|_{W_\delta^1({\cal D})} \nonumber\\
&& \hspace{-3.5em}+ \sum_\pm \| \eta \partial_{x_3}^j h^\pm\|_{W_\delta^{3/2}(\Gamma^\pm)^{3-d^\pm}}
  + \sum_\pm \| \eta \partial_{x_3}^j \phi^\pm\|_{W_\delta^{1/2}(\Gamma^\pm)^{3-d^\pm}}\Big)
  +\| \eta u\|_{W_\delta^1({\cal D})^3}+ \|\eta p\|_{W_\delta^0({\cal D})}\Big).
\end{eqnarray}
\end{Th}

\noindent P r o o f. 1) First we prove the theorem for $k=0$. From
(\ref{bvpD1}), (\ref{bvpD2}) it follows that $(\zeta u,\zeta p)$
satisfies the equations
\begin{eqnarray*}
&& b(\zeta u,v)-\int_{\cal D} \zeta p\, \nabla\cdot v\, dx =
\int_{\cal D} \tilde{f}\,
  v + \sum_\pm \int_{\Gamma^\pm} \tilde{g}\, v\, dx \ \mbox{ for all }v\in V,\\
&& -\nabla\cdot (\zeta u)= \zeta g -(\nabla\zeta)\cdot u \ \mbox{
in }{\cal D},\quad
  S^\pm (\zeta u) =\zeta h^\pm \ \mbox{ on }\Gamma^\pm,
\end{eqnarray*}
where
\begin{eqnarray*}
\tilde{f}_i  & = & \zeta f_i - 2\sum_{j=1}^3 (\partial_{x_j}
\zeta)\, \varepsilon_{i,j}(u) - \sum_{j=1}^3 \partial_{x_j}\,
\big( u_i\partial_{x_j}\zeta
  +u_j\partial_{x_i}\zeta\big) -p\partial_{x_i}\zeta \in W_\delta^0({\cal D})\\
\tilde{g}_i & = & \zeta g_i + \frac{\partial\zeta}{\partial n}\,
u_i + u_n\,
\partial_{x_i}\zeta \in W_\delta^{1/2}(\Gamma^\pm).
\end{eqnarray*}
Applying Lemma \ref{l12}, we obtain the assertion of the theorem
for $k=0$.

2) Let the conditions of the theorem on $F$, $g$ and $h^\pm$ with $k=1$ be satisfied.
Moreover, we suppose that $\eta\partial_{x_3}^j h^\pm \in
V_\delta^{3/2}(\Gamma^\pm)^{3-d^\pm}$ for $j=0$ and $j=1$. By $\chi$ and $\chi_1$ we
denote smooth function such that $\chi=1$ in a neighborhood of
supp$\, \zeta$, $\chi_1=1$ in a neighborhood of supp$\, \chi$, and
$\eta=1$ in a neighborhood of supp$\, \chi_1$. Furthermore, for an
arbitrary function $v$ on ${\cal D}$ or $\Gamma^\pm$ we set
$v_h(x',x_3)=h^{-1}\big(v(x',x_3+h)-v(x',x_3)\big)$. Obviously,
$(u_h,p_h)\in {\cal H}\times L_2({\cal D})$ for arbitrary real
$h$. Consequently, by Theorem \ref{t3}, we have
\begin{eqnarray} \label{1c3}
&& \hspace{-2em}\| \zeta u_h\|_{W_\delta^2({\cal D})^3} + \|\zeta p_h\|_{W_\delta^1({\cal D})}  \le
  c\, \Big( \|\chi f_h\|_{W_\delta^0({\cal D})^3} + \|\chi g_h\|_{W_\delta^1({\cal D})}
  + \sum_\pm \| \chi h_h^\pm\|_{V_\delta^{3/2}(\Gamma^\pm)^{3-d^\pm}} \nonumber \\
&& \hspace{4em} + \sum_\pm \| \chi
\phi_h^\pm\|_{V_\delta^{1/2}(\Gamma^\pm)^{3-d^\pm}}
  +\| \chi u_h\|_{W_\delta^1({\cal D})^3}+ \|\chi p_h\|_{W_\delta^0({\cal D})}\Big)
\end{eqnarray}
with a constant $c$ independent of $h$, where $\chi$ is a smooth
function, $\chi=1$ in a neighborhood of supp$\, \zeta$, $\eta=1$
in a neighborhood of supp$\, \chi$. Here $\chi f_h=(\chi
f)_h-\chi_h f$ and, for sufficiently small $|h|$ we have
\begin{eqnarray*}
&& \| (\chi f)_h\|^2_{W_\delta^0({\cal D})^3} =\int_{\cal D}
r^{2\delta}\, \Big| \int_0^1
  \frac{\partial (\chi f)}{\partial x_3}(x',x_3+th)\, dt\Big|^2\, dx
  \le  \| \partial_{x_3}(\chi f)\|^2_{W_\delta^0({\cal D})}, \\
&& \|\chi_h f \|^2_{W_\delta^0({\cal D})^3}\le c\, \|\eta
f\|_{W_\delta^0({\cal D})^3}.
\end{eqnarray*}
Analogous estimates hold for the norms of $\chi g_h$, $\chi
h_h^\pm$ and $\chi \phi_h^\pm$, $\chi u_h$, and $\chi p_h$ on the
right-hand side of (\ref{1c3}). Here one can use the equivalence
of the norm in $V_\delta^{l-1/2}(\Gamma\pm)$ with (\ref{eqnorm}).
Hence the right-hand side and, therefore, also the limit (as $h\to
0$) of the left-hand of (\ref{1c3}) are majorized by
\begin{eqnarray*}
&& c \sum_{j=0}^1\Big( \|\eta \partial_{x_3}^j f\|_{W_\delta^0({\cal D})^3}
  + \|\eta\partial_{x_3}^j g\|_{W_\delta^1({\cal D})}
  + \sum_\pm \| \eta\partial_{x_3}^j h^\pm\|_{V_\delta^{3/2}(\Gamma^\pm)^{3-d^\pm}} \\
&&  \qquad + \sum_\pm \| \eta \partial_{x_3}^j  \phi^\pm\|_{V_\delta^{1/2}(\Gamma^\pm)^{3-d^\pm}}
 +\| \chi_1\partial_{x_3}^j u\|_{W_\delta^1({\cal D})^3}+ \|\chi_1 \partial_{x_3}^j p
  \|_{W_\delta^0({\cal D})}\Big).
\end{eqnarray*}
By the first part of the proof, the norm of
$\chi_1\partial_{x_3}(u,p)$ in $W_\delta^1({\cal D})^3\times
W_\delta^0$ is majorized by the right-hand side of (\ref{1t3})
with $k=0$. This implies (\ref{1t3}) for $k=1$.

Suppose now that $\eta \partial_{x_3}^j h^\pm \in
W_\delta^{3/2}(\Gamma^\pm)^{3-d^\pm}$ for $j=0,1$ and the
compatibility condition (\ref{cc}) is satisfied. Then there exists
a vector function $\psi \in W^{k+1-\delta}(M)^3$ such that $S^\pm
\psi=(\eta h^\pm)|_M$. Let $v\in W_\delta^{k+2}({\cal D})^3$ be an
extension of $\psi$. Then $\chi\partial_{x_3}^j (h^\pm-S^\pm
v)|_M=0$ for $j=0,\ldots,k$ and, consequently,
$\chi\partial_{x_3}^j (h^\pm-S^\pm v|_{\Gamma^\pm}) \in
V_\delta^{3/2} (\Gamma^\pm)^{3-d^\pm}$. Now we can apply the
result proved above to the vector function $(u-v,p)$ and obtain
$\zeta\partial_{x_3} (u-v,p) \in W_\delta^2({\cal D})^3\times W_\delta^1({\cal D})$.

3) For $k>1$ the theorem can be easily proved by induction. \hfill $\Box$\\

Moreover, the following generalization of Theorem \ref{t3} holds.

\begin{Th} \label{t4}
Let $\zeta,\ \eta$ be the same cut-off functions as in Theorem {\em \ref{t3}}, and let $(u,p)\in
{\cal H}\times L_2({\cal D})$ be a solutions of {\em (\ref{bvpD1}), (\ref{bvpD2})}. Suppose that
$F$ is a functional on $V$ which has the form {\em (\ref{funct})} with
\[
\eta \partial_{x_3}^j f\in W_\delta^{l-2}({\cal D})^3,\ \ \eta\partial_{x_3}^j g\in W_\delta^{l-1}({\cal D}),
  \ \ \eta\partial_{x_3}^j \phi^\pm \in W_\delta^{l-3/2}(\Gamma^\pm)^{d^\pm} \ \ for \ j=0,1,\ldots,k.
\]
Furthermore, we assume that $\eta \partial_{x_3}^j h^\pm \in W_\delta^{l-1/2}(\Gamma^\pm)^{3-d^\pm}$
for $j=0,1,\ldots,k$, $h^\pm$ satisfy the compatibility condition {\em (\ref{cc})}, $0<\delta<l-1$,
$\delta$ is not integer, and that the strip $0<\mbox{\em Re}\, \lambda \le l-1-\delta$ does not contain
eigenvalues of the pencil $A(\lambda)$. Then $\zeta\partial_{x_3}^j(u,p)\in W_\delta^l({\cal D})^3 \times
W_\delta^{l-1}({\cal D})$ for $j=0,1,\ldots,k$.
\end{Th}

\noindent P r o o f. 1) We consider first the case $k=0$, $0<\delta<1$
and prove the theorem for this case by induction in $l$. For $l=2$ we can refer to Theorem \ref{t3}.
Suppose the assertion is proved for a certain integer $l=s\ge 2$ and the conditions
of the theorem are satisfied for $l=s+1$. We denote by $\chi$ and $\chi_1$ the same
cut-off functions as in the proof of theorem \ref{t3}. Then, by the induction hypothesis,
we have $\chi_1 (u,p) \in W_\delta^s({\cal D})^3\times W_\delta^{s-1}({\cal D})$
and $\chi_1 \partial_{x_3}(u,p) \in W_\delta^{s-1}({\cal D})^3 \times
W_\delta^{s-2}({\cal D})$. If $s\ge 3$, then this implies that
$\chi_1 \partial_{x_3}(u,p) \in {\cal H}\times L_2({\cal D})$ and,
by the induction hypothesis, we obtain $\chi \partial_{x_3}(u,p)
\in W_\delta^{s}({\cal D})^3 \times W_\delta^{s-1}({\cal D})$. For
$s=2$ the last inclusion follows from Theorem \ref{t3}.
Consequently, we have $\chi \partial_{x_3}^j (u,p)\in
W_\delta^{s}({\cal D})^3\times W_\delta^{s-1}({\cal D})$ for $j=0$
and $j=1$. Using this result and Lemma \ref{l12a}, we obtain $\zeta(u,p) \in
W_\delta^{s+1}({\cal D})^3 \times W_\delta^{s}({\cal D})$.

2) Now let $k=0$ and $\sigma<\delta<\sigma+1$, where $\sigma$ is an integer, $1\le \sigma \le l-2$.
Since $0<\delta-\sigma<1$, $\eta f \in W_{\delta-\sigma}^{l-\sigma-2}({\cal D})^3$, $\eta g\in
W_{\delta-\sigma}^{l-\sigma-1}({\cal D})$, $\eta h^\pm \in
W_{\delta-s}^{l-s-1/2} (\Gamma^\pm)^{3-d^\pm}$, and $\eta \phi^\pm\in
W_{\delta-\sigma}^{l-\sigma-3/2}(\Gamma^\pm)^{d^\pm}$, it follows from the first part of the proof
that $\chi(u,p) \in W_{\delta-\sigma}^{l-\sigma} ({\cal D})^3\times W_{\delta-\sigma}^{l-\sigma-1}
({\cal D})$. Using Lemma \ref{l4a}, we obtain the assertion of the theorem for $k=0$.

3) We prove the assertion for $k>0$, $l-2<\delta<l-1$. Then, according to Theorem \ref{t3},
we have $\chi \partial_{x_3}^j (u,p) \in W_{\delta-l+2}^2({\cal D})^3\times W_{\delta-l+2}^1({\cal D})$
for $j=0,1,\ldots,k$. Using Lemma \ref{l4a}, we obtain $\chi \partial_{x_3}^j (u,p) \in W_{\delta}^l
({\cal D})^3\times W_{\delta}^{l-1}({\cal D})$ for $j=0,1,\ldots,k$.

4) In the case $k>0$, $0<\delta<l-2$ we prove the assertion by induction in $k$. Suppose the theorem is proved
for $k-1$. From parts 1) and 2) of the proof  we conclude that $\chi (u,p) \in W_\delta^l({\cal D})^3
\times W_\delta^{l-1}({\cal D})$ and $\chi\partial_{x_3}(u,p) \in W_\delta^{l-1}({\cal D})^3\times
W_\delta^{l-2}({\cal D})$. The last inclusion implies $\chi\partial_{x_3} (u,p) \in
{\cal H}\times L_2({\cal D})$. Consequently, by the induction hypothesis, we obtain $\zeta\partial_{x_3}^j
\partial_{x_3}(u,p) \in W_\delta^l({\cal D})^3\times W_\delta^{l-1}({\cal D})$ for
$j=0,\ldots,k-1$. The proof of the theorem is complete. \hfill $\Box$

\subsection{The case when \boldmath$\lambda=0$ is not an eigenvalue of $A(\lambda)$\unboldmath}

The number $\lambda=0$ does not belong to the spectrum of the pencil $A(\lambda)$ if either $d^+\cdot d^-=0$ or
\begin{equation} \label{erg1}
d^+,d^-\in \{ 1,2\}  \ \mbox{ and } \ \theta\notin \{ \frac \pi k \, ,\frac\pi k + \pi\},
\end{equation}
where $k=1$ if $d^+=d^-$, $k=2$ if $d^+\not= d^-$. In these cases, we have
\[
u|_M= 0 \quad\mbox{if } u\in W_\delta^2({\cal D})^3, \ 0<\delta<1,\ S^+u|_M=0,\ S^- u|_M=0.
\]
Using this fact, we can easily deduce the following result from Theorem \ref{t3}.

\begin{Le} \label{l12b}
Let $\zeta,\ \eta$ be the same cut-off functions as in Theorem {\em \ref{t3}}, and let
$(u,p)\in {\cal H}\times L_2({\cal D})$ be a solution of {\em (\ref{bvpD1}), (\ref{bvpD2})}.
We suppose that
\[
\eta \partial_{x_3}^j f\in V_\delta^0({\cal D})^3, \ \ \eta \partial_{x_3}^j g \in V_\delta^1({\cal D}),\ \
  \eta \partial_{x_3}^j h^\pm \in W_\delta^{3/2}(\Gamma^\pm)^{3-d^\pm}, \ \
  \eta\partial_{x_3}^j \phi^\pm \in W_\delta^{1/2}(\Gamma^\pm)^{d^\pm},
\]
for $j=0,1,\ldots,k$, where $0<\delta<1$, and that the functional $F$ on $V$ has the form {\em (\ref{funct})}.
We assume further that the strip $0\le \mbox{\em Re}\, \lambda \le 1-\delta$
does not contain eigenvalues of the pencil $A(\lambda)$. Then $\zeta\partial_{x_3}^k (u,p)\in
V_\delta^2({\cal D})^3\times V_\delta^1({\cal D})$ and there is an estimate analogous to {\em (\ref{1t3})}.
\end{Le}

\noindent P r o o f.
From Theorem \ref{t3} it follows that $\zeta\partial_{x_3}^k (u,p)\in W_\delta^2({\cal D})^3\times
V_\delta^1({\cal D})$. Furthermore, we have $S^\pm \zeta\partial_{x_3}^k u|_M = \zeta\partial_{x_3}^k h^\pm|_M=0$
and, therefore, $\zeta\partial_{x_3}^ku|_M=0$. From this and from Lemma \ref{l0} we conclude that
$\zeta\partial_{x_3}^k u \in V_\delta^2({\cal D})^3$. This proves the lemma. \hfill $\Box$ \\

Furthermore, the following generalization holds. Using the second part of Lemma \ref{l12a}, it can be proved
analogously to Theorem \ref{t4}.

\begin{Th} \label{t4a}
Let $\zeta,\ \eta$ be the same cut-off functions as in Theorem {\em \ref{t3}}, and let $(u,p)\in
{\cal H}\times L_2({\cal D})$ be a solution of {\em (\ref{bvpD1}), (\ref{bvpD2})}. Suppose that
\[
\eta \partial_{x_3}^j f\in V_\delta^{l-2}({\cal D})^3,\ \ \eta\partial_{x_3}^j g\in V_\delta^{l-1}({\cal D}),\ \
  \eta \partial_{x_3}^j h^\pm \in V_\delta^{l-1/2}(\Gamma^\pm)^{3-d^\pm},\ \
  \eta\partial_{x_3}^j \phi^\pm \in V_\delta^{l-3/2}(\Gamma^\pm)^{d^\pm}
\]
for $j=0,1,\ldots,k$, and that the functional $F$ on $V$ has the form {\em (\ref{funct})}.
If $0<\delta<l-1$, $\delta$ is not integer, and the strip $0\le\mbox{\em Re}\, \lambda \le l-1-\delta$ does not contain
eigenvalues of the pencil $A(\lambda)$, then $\zeta\partial_{x_3}^j(u,p)\in V_\delta^l({\cal D})^3 \times
V_\delta^{l-1}({\cal D})$ for $j=0,1,\ldots,k$.
\end{Th}

\subsection{The case when \boldmath$\lambda=1$ is the smallest positive eigenvalue
of $A(\lambda)$\unboldmath}

The number $\lambda=1$ belongs to the spectrum of the pencil $A(\lambda)$ for all angles $\theta$ if
$d^+ +d^-$ is even. Here, $\lambda=1$ is the eigenvalue with
smallest positive real part if
\begin{equation} \label{evend}
d^+ + d^- \ \mbox{ is even and } \theta < \frac{\pi}{m},\quad \mbox{ where } m=1 \mbox{ if }d^+=d^-, \
  m=2\ \mbox{if }d^+\not= d^-.
\end{equation}
Then the eigenvalue $\lambda=1$ has geometric and algebraic multiplicity 1. For even
$d^+$ and $d^-$ there is the eigenvector $(U,P)=(0,1)$, for odd
$d^+$ and $d^-$ the eigenvector $(U,P)=(\sin\varphi,-\cos\varphi,0,0)$.
Under the above restrictions on $\theta$, generalized eigenvectors corresponding to the eigenvalue
$\lambda=1$ do not exist.

In the case (\ref{evend}), the result of Theorem \ref{t4} can be improved. However, then the boundary data
and the function $g$ must satisfy additional compatibility conditions on the edges.
We restrict ourselves in the proof to the Dirichlet problem
\begin{equation}\label{Dirich}
-\Delta u+\nabla p =f,\ -\nabla\cdot u=g\ \mbox{in }{\cal D},\quad u=h^\pm\ \mbox{on }\Gamma^\pm .
\end{equation}
Suppose that $(u,p) \in W_\delta^3({\cal D})^3 \times W_\delta^2({\cal D})$ is a solution of problem
(\ref{Dirich}), $0<\delta<1$. Then the traces of $\partial_{x_j}u$, $j=1,2,3$ on $M$ exist. From the equations
$-\nabla \cdot u=g$  and $u|_{\Gamma^+}=h^+$ it follows that
\begin{equation} \label{M1}
-\partial_{x_1} u_1|_M  -\partial_{x_2}u_2|_M =g|_M + \partial_{x_3}h_3^+|_M \, .
\end{equation}
Furthermore, the equations $u=h^\pm$ on $\Gamma^\pm$ imply that
\begin{equation} \label{M2}
\cos\frac\theta 2 \, \partial_{x_1}u_j|_M \pm \sin\frac\theta 2 \,
  \partial_{x_2} u_j|_M = \partial_r h_j^\pm|_M \quad\mbox{for }j=1,2,3.
\end{equation}
The algebraic system (\ref{M1}), (\ref{M2}) with the unknowns $\partial_{x_1}u_j|_M$ and
$\partial_{x_2}u_j|_M$, $j=1,2,3$ is solvable if and only if
\begin{equation} \label{cc1}
n^-\cdot \partial_r h^+|_M + n^+\cdot \partial_r h^-|_M = (g|_M + \partial_{x_3}h_3^+|_M)\, \sin\theta.
\end{equation}

\begin{Th} \label{t6}
Let $(u,p)$ be a solution of problem {\em (\ref{Dirich})}, and let $\zeta,\eta$ be the same cut-off
functions as in Theorem {\em \ref{t3}}. Suppose that $\eta(u,p)\in {\cal H}\times L_2({\cal D})$,
$\eta f\in W_\delta^{l-2}({\cal D})^2$, $\eta g\in W_\delta^{l-1}({\cal D})$, $\eta h^\pm
=\eta(h_1^\pm,h_2^\pm,h_3^\pm) \in W_\delta^{l-1/2}(\Gamma^\pm)^3$, $l\ge 3$, $0<\delta<l-2$, $\delta$
is not integer, $\theta<\pi$, $\lambda=1$ is the only eigenvalue of the pencil $A(\lambda)$ in the strip
$0< \mbox{\em Re}\, \lambda \le l-1-\delta$ and that the compatibility conditions $h^+|_M=h^-|_M$
and {\em (\ref{cc1})} are satisfied. Then $\zeta u \in W_\delta^{l}({\cal D})^3$ and
$\zeta p\in W_\delta^{l-1}({\cal D})$.
\end{Th}

\noindent P r o o f. We prove the theorem first for $l=3$. Let $\chi$, $\chi_1$ be smooth functions on $\overline{\cal D}$
such that $\chi=1$ in a neighborhood of supp$\, \zeta$, $\chi_1=1$ in a neighborhood of supp$\, \chi$,
and $\eta=1$ in a neighborhood of supp$\, \chi_1$. Then, by Theorem \ref{t3}, we have
$\chi_1\partial_{x_3}u \in W_\delta^2({\cal D})$ and $\chi_1\partial_{x_3}p \in W_\delta^1({\cal D})$.
Let $c(x_3)$, $d(x_3)$ be vectors satisfying
\[
-c_1(x_3)-d_2(x_3)=g(0,x_3) + (\partial_{x_3}h_3^+)(0,x_3),\qquad \cos\frac\theta 2 \, c(x_3)
  \pm \sin\frac\theta 2 \,  d(x_3) = (\partial_r h^\pm)(0,x_3)
\]
for $x_3\in M\cap \mbox{supp}\, \zeta$ and the estimate
\[
|c(x_3)| + |d(x_3)| \le C\, \Big( |g(0,x_3)| + |(\partial_{x_3}h_3^+)(0,x_3)|
  + \sum_\pm |(\partial_r h^\pm)(0,x_3)|\Big)
\]
with a constant independent of $x_3$. Furthermore, let
\[
v(x',x_3)=u(x',x_3)-h^+(0,x_3)-c(x_3)x_1-d(x_3)x_2.
\]
Since $\chi_1(\cdot,x_3)u(\cdot,x_3) \in W_\delta^2(K)^3$ and $v(0,x_3)=0$, it follows that
$\chi_1(\cdot,x_3)u(\cdot,x_3) \in V_\delta^2(K)^3$ (see \cite[Th.7.1.1]{kmr1}). With the notation
$v'=(v_1,v_2),\ f'=(f_1,f_2)$ we have
\[
-\Delta_{x'}v'(\cdot,x_3) + \nabla_{x'}p(\cdot,x_3)=F'(\cdot,x_3),\quad -\nabla_{x'}
  \cdot v'(\cdot,x_3)=G(\cdot,x_3),\quad -\Delta_{x'}v_3(\cdot,x_3)=F_3(\cdot,x_3)\quad\mbox{in }K
\]
and $v(\cdot,x_3)=H^\pm(\cdot,x_3)$ on $\gamma^\pm$, where
\begin{eqnarray*}
&& F'(x)=f'(x)+\partial_{x_3}^2 u'(x),\quad
F_3(x)=f_3(x)+\partial_{x_3}^2 u - \partial_{x_3}p,\\
&& G(x',x_3)=g(x',x_3)-g(0,x_3)+\partial_{x_3}\big( u_3(x',x_3)-u_3(0,x_3)\big),\\
&& H^\pm(r,x_3)=h^\pm(r,x_3) -h^\pm(0,x_3)-(\partial_r
h^\pm)(0,x_3)\, r
\end{eqnarray*}
Obviously, $\chi(\cdot,x_3)\, F(\cdot,x_3) \in W_\delta^1(K)^3\subset V_\delta^1(K)^3$ for almost all
$x_3\in M\cap \mbox{supp}\,\chi$. Furthermore, since $G(0,x_3)=0$ and $H^\pm(0,x_3)
= (\partial_r H^\pm)(0,x_3)=0$, we have $\chi(\cdot,x_3)\, G(\cdot,x_3) \in V_\delta^2(K)$ and
$\chi(\cdot,x_3)\, H^\pm(\cdot,x_3) \in V_\delta^{5/2}(\Gamma^\pm)^3$. Since in the strip
$0< \mbox{Re}\, \lambda \le 2-\delta$ there is only the eigenvalues $\lambda = 1$ of the pencil
$A(\lambda)$ with the corresponding eigenvector $(U,P)=(0,1)$ and without
generalized eigenvectors, we conclude that
\[
\zeta(\cdot,x_3) \big( v(\cdot,x_3),p(\cdot,x_3)-p(0,x_3)\big) \in  V_\delta^3(K)^3\times V_\delta^2(K)
\]
(see, e.g., \cite[Th.8.2.2]{kmr1}). From this we conclude that $\zeta(\cdot,x_3)\, u(\cdot,x_3)
\in  W_\delta^3(K)^3$ and $\zeta(\cdot,x_3)\, p(\cdot,x_3) \in  W_\delta^2(K)$. Analogously
to the proof of Lemma \ref{l12}, it follows that $\zeta\, (u,p) \in  W_\delta^3({\cal D})^3\times
W_\delta^2({\cal D})$. Thus, the theorem is proved for $l=3$. For $l>3$ the result holds
analogously to Theorem \ref{t4} by means of Lemma \ref{l9}. \hfill $\Box$\\

Furthermore, the following assertion holds (see the third part of the proof of Theorem \ref{t4}).

\begin{Co} \label{c6}
Let $(u,p)$ be a solution of problem {\em (\ref{Dirich})}, and let $\zeta,\eta$ be the same cut-off
functions as in Theorem {\em \ref{t3}}. Suppose that $\eta(u,p)\in {\cal H}\times L_2({\cal D})$,
$\eta \partial_{x_3}^j f\in W_\delta^{l-2}({\cal D})^2$, $\eta \partial_{x_3}^j g \in W_\delta^{l-1}
({\cal D})$, $\eta \partial_{x_3}^j h^\pm \in W_\delta^{l-1/2}(\Gamma^\pm)^3$ for
$j=0,1,\ldots,k$, where $l\ge 3$, $0<\delta<l-2$, $\delta$ is not integer, and $\theta<\pi$.
Suppose further that $\lambda=1$ is the only eigenvalue of the pencil $A(\lambda)$ in the strip
$0< \mbox{\em Re}\, \lambda \le l-1-\delta$ and that the compatibility conditions
$h^+|_M=h^-|_M$ and {\em (\ref{cc1})} are satisfied. Then $\zeta\partial_{x_3}^j (u,p) \in
W_\delta^{l}({\cal D})^3\times W_\delta^{l-1}({\cal D})$ for $j=0,1,\ldots,k$.
\end{Co}

\begin{Rem} \label{r2}
{\em The results of Theorem \ref{t6} and Corollary \ref{c6} can be extended to other boundary
conditions in the case (\ref{evend}). Then the traces of $h^\pm$, $\partial_r h^\pm$, $\phi^\pm$ and
$g$ on $M$ must be such that the system
\begin{eqnarray*}
&& S^\pm u|_M = h^\pm|_M, \quad S^\pm \big( \cos\mbox{$\frac \theta 2$}\, (\partial_{x_1}u)|_M
  \pm \sin\mbox{$\frac \theta 2$}\, (\partial_{x_2}u)|_M\big) = \partial_r h^\pm|_M,\quad
  N^\pm(u,p)|_M = \phi^\pm|_M ,\\
&& (\partial_{x_1}u_1)|_M + (\partial_{x_2}u_2)|_M + \partial_{x_3}u_3|_M = g|_M
\end{eqnarray*}
(with the unknowns $u_M, (\partial_{x_1}u)|_M, (\partial_{x_2}u)|_M$, and $p|_M$) is solvable.
For example, in the case of the Neumann problem ($d^+=d^-=3$) the boundary data $\phi^\pm$
must satisfy the condition
\[
n^+\cdot \phi^- = n^-\cdot \phi^+ \quad\mbox{on } M,
\]
while in the case $d^-=0$, $d^+=2$, the data $h^+$, $h^-$, $\phi^+$ and $g$ must satisfy the compatibility
conditions $h^-\cdot n^+ = h^+$ and
\[
\partial_r h^+\, \cos 2\theta - (2n^+\cos\theta\, + n^-)\, \partial_r h^- + 2\sin^2\theta\, (\phi_1^+\cos
\theta/2 + \phi_2^+\sin\theta/2) + \frac 12 (g+\partial_{x_3}h_3^-)\, \sin 2\theta=0
\]
on the edge $M$.}
\end{Rem}

\setcounter{equation}{0}
\setcounter{Th}{0}
\setcounter{Le}{0}
\setcounter{Co}{0}
\setcounter{Rem}{0}
\section{Green's matrix of the problems in half-space and dihedron}

In this section, we study Green's matrices of boundary value problems to Stokes system
in the half-space ${\Bbb R}_+^3=\{ x=(x_1,x_2,x_3)\in {\Bbb R}^3:\, x_3>0\}$ and in the
dihedron ${\cal D}$. For the problems in ${\Bbb R}_+^3$, we find explicit representations
of Green's matrix ${\cal G}(x,\xi)$.

\subsection{Green's matrix in the half-space}

We consider the Stokes system
\begin{equation} \label{Stokesspace}
-\Delta u + \nabla p = f,\quad -\nabla\cdot u =g
\end{equation}
in the half-space ${\Bbb R}_+^3$ with one of the following boundary conditions
\begin{itemize}
\item[(i)] $u(x) = 0$ (Dirichlet condition),
\item[(ii)] $u_1(x)=u_2(x)=-p+2\partial_{x_3}u_3(x)=0$,
\item[(iii)] $u_3(x)= \partial_{x_3}u_1(x)=\partial_{x_3} u_2(x) =0$ (free surface condition),
\item[(iv)] $\partial_{x_1}u_3(x)+\partial_{x_3}u_1(x)=\partial_{x_2}u_3(x)+\partial_{x_3}u_2(x)
  =-p+2\partial_{x_3}u_3(x)=0$ (Neumann condition)
\end{itemize}
on the plane $x_3=0$. By Green's matrix of this problem we mean a matrix ${\cal G}^+(x,\xi)=
\big( {\cal G}^+_{i,j}(x,\xi)\big)_{i,j=1}^4$ such that the vector $\vec{\cal G}^+_j=({\cal G}^+_{1,j},
{\cal G}^+_{2,j},{\cal G}^+_{3,j})^t$ and the function ${\cal G}^+_{4,j}$ satisfy the equations
\begin{equation} \label{greenf}
-\Delta_x \vec{\cal G}^+_{j}(x,\xi) + \nabla_x {\cal G}^+_{4,j}(x,\xi)= \delta(x-\xi)\
  \big( \delta_{1,j},\delta_{2,j},\delta_{3,j}\big)^t,\quad
  -\nabla_x \cdot \vec{\cal G}^+_{j}(x,\xi) = \delta_{4,j}\delta(x-\xi)
\end{equation}
for $x,\xi\in {\Bbb R}_+^3$, $j=1,2,3,4$, and the corresponding boundary condition for $x_3=0$,
$\xi\in {\Bbb R}_+^3$. Here $\delta_{i,j}$ denotes the Kronecker symbol and $(a_1,a_2,a_3)^t$ the
column vector with the components $a_1,a_2,a_3$. Furthermore, the vectors $\vec{\cal H}^+_i=({\cal G}^+_{i,1},
{\cal G}^+_{i,2},{\cal G}^+_{i,3})^t$ and the function ${\cal G}^+_{i,4}$ satisfy the equations
\begin{equation} \label{greenf1}
-\Delta_\xi \vec{\cal H}^+_{i}(x,\xi) + \nabla_\xi {\cal G}^+_{i,4}(x,\xi)= \delta(x-\xi)\
  \big( \delta_{i,1},\delta_{i,2},\delta_{i,3}\big)^t,\quad
  -\nabla_\xi\cdot \vec{\cal H}^+_{i}(x,\xi) = \delta_{i,4}\delta(x-\xi)
\end{equation}
for $x,\xi \in {\Bbb R}_+3$, $i=1,2,3,4$, and the corresponding boundary condition for $\xi_3=0$.
From (\ref{greenf1}) and Green's formula in the half-space (cf. (\ref{Green2})) it follows
that the solution $(u,p)$ of the boundary value problem is given by
\begin{eqnarray*}
u_i(x) & = & \int_{{\Bbb R}_+^3} \big( f(\xi)+ \nabla_\xi g(\xi)\big) \cdot \vec{\cal H}^+_i(x,\xi)\, d\xi
  + \int_{{\Bbb R}_+^3} g(\xi) \, {\cal G}^+_{i,4}(x,\xi)\, d\xi, \\
p(x) & = & -g(x)+\int_{{\Bbb R}_+^3} \big( f(\xi)+ \nabla_\xi g(\xi)\big) \cdot \vec{\cal H}^+_4(x,\xi)\, d\xi
  + \int_{{\Bbb R}_+^3} g(\xi) \, {\cal G}^+_{4,4}(x,\xi)\, d\xi.
\end{eqnarray*}
Note that in the case of boundary conditions (i) and (iii), there are the simpler formulas
\begin{eqnarray*}
u_i(x) & = & \int\limits_{{\Bbb R}_+^3} f(\xi)\cdot \vec{\cal H}^+_i(x,\xi)\, d\xi
  + \int\limits_{{\Bbb R}_+^3} g(\xi)\, {\cal G}^+_{i,4}(x,\xi)\, d\xi,\\
p(x) & = & \int\limits_{{\Bbb R}_+^3} f(\xi)\cdot \vec{\cal H}^+_4(x,\xi)\, d\xi
  + \int\limits_{{\Bbb R}_+^3} g(\xi)\, {\cal G}^+_{4,4}(x,\xi)\, d\xi.
\end{eqnarray*}
We denote by ${\cal G}(x,\xi)=\big( {\cal G}_{i,j}(x,\xi)\big)_{i,j=1}^4$ Green's matrix of  Stokes system
in ${\Bbb R}^3$, i.e., the matrix satisfying (\ref{greenf}) for
$x,\xi \in {\Bbb R}^3$, $j=1,2,3,4$. The components of ${\cal G}(x,\xi)$ are (see \cite{lad,mps})
\begin{eqnarray} \label{greenfsp1}
&& {\cal G}_{i,j}(x,\xi) = \frac{1}{8\pi}\Big( \frac{\delta_{i,j}}{|x-\xi|} +
  \frac{(x_i-\xi_i)(x_j-\xi_j)}{|x-\xi|^3}\Big) \ \mbox{ for } i,j=1,2,3, \\ \label{greenfsp2}
&& {\cal G}_{4,j}(x,\xi) = - {\cal G}_{j,4}(x,\xi) = \frac{1}{4\pi} \, \frac{x_k-\xi_k}{|x-\xi|^3}\ \mbox{ for }j=1,2,3,
  \quad {\cal G}_{4,4}(x,\xi) = - \delta(x-\xi).
\end{eqnarray}
Note that ${\cal G}_{i,j}(x,\xi)={\cal G}_{j,i}(\xi,x)$ for $i,j=1,2,3,4$. In \cite{mps} also the
Green matrices for the problems in the half-space with boundary conditions (i) and (iii) are given.
We calculate here the Green matrices for the problems with boundary conditions (ii) and (iv).

\begin{Le} \label{lg1}
The components of Green's matrix for the problem in ${\Bbb R}_+^3$ with boundary condition {\em (ii)} are
\[
{\cal G}^+_{i,j}(x,\xi) = {\cal G}_{i,j}(x,\xi) - (-1)^{\delta_{j,3}} {\cal G}_{i,j}(x,\xi^*) \
  \mbox{ for }i+j\le 7, \quad {\cal G}^+_{4,4}(x,\xi)= -\delta(x-\xi),
\]
where $\xi^*=(\xi_1,\xi_2,-\xi_3)$, and ${\cal G}_{i,j}$ is defined by {\em (\ref{greenfsp1}), (\ref{greenfsp2})}.
\end{Le}

\noindent P r o o f.
Let $F(\xi) = f(\xi)+\nabla_\xi g(\xi)$. Furthermore, we define $\tilde{F}(\xi)$ and $\tilde{g}(\xi)$ as
\[
\tilde{F}(\xi) = F(\xi), \ \tilde{g}(\xi)=g(\xi) \ \mbox{for }\xi_3>0,\quad
\tilde{F}_j(\xi) = - (-1)^{\delta_{j,3}} F_j(\xi^*), \ \tilde{g}(\xi) = - g(\xi^*) \ \mbox{ for }\xi_3<0.
\]
Then the functions
\begin{eqnarray*}
&& \hspace{-3em} u_i(x) = \sum_{j=1}^3 \int_{{\Bbb R}^3} \tilde{F}_j(\xi) \, {\cal G}_{i,j}(x,\xi)\, d\xi
  + \int_{{\Bbb R}^3} \tilde{g}(\xi)\, {\cal G}_{i,4}(x,\xi) \\
&&  = \sum_{j=1}^3 \int_{{\Bbb R}_+^3} F_j(\xi) \, \big( {\cal G}_{i,j}(x,\xi)-(-1)^{\delta_{j,3}}
  {\cal G}_{i,j}(x,\xi^*)\big) \, d\xi + \int_{{\Bbb R}^3} g(\xi)\, \big( {\cal G}_{i,4}(x,\xi)
  -{\cal G}_{i,4}(x,\xi^*)\big)\, d\xi,\\
&& \hspace{-3em} p(x)\ = -2g(x) + \sum_{j=1}^3 \int_{{\Bbb R}^3} \tilde{F}_j(\xi) \, {\cal G}_{4,j}(x,\xi)\, d\xi\\
&&  = -2g(x) + \sum_{j=1}^3 \int_{{\Bbb R}_+^3} F_j(\xi) \, \big( {\cal G}_{4,j}(x,\xi)-(-1)^{\delta_{j,3}}
  {\cal G}_{4,j}(x,\xi^*)\big) \, d\xi
\end{eqnarray*}
satisfy (\ref{Stokesspace}) for $x_3>0$. Furthermore, since ${\cal G}_{i,j}(x',0,\xi)=(-1)^{\delta_{j,3}}
{\cal G}_{i,j}(x',0,\xi^*)$ for $i=1,2$, we have $u_1(x',0)=u_2(x',0)=0$. We show that
$\Phi(x) = - p(x) + 2\partial_{x_3}u_3(x)$ vanishes for $x_3=0$. Using (\ref{greenfsp1}), (\ref{greenfsp2}),
we get
\[
\Phi(x)  = 2 g(x) -\frac{3}{4\pi} \Psi(x) - \frac{1}{2\pi} \frac{\partial}{\partial x_3} \int_{{\Bbb R}_+^3}
g(\xi) \frac{\partial}{\partial\xi_3} \Big( \frac{1}{|x-\xi|} + \frac{1}{|x-\xi^*|}\Big)\, d\xi,
\]
where
\begin{eqnarray*}
\Psi(x) & = & \sum_{j=1}^2 \ \int\limits_{{\Bbb R}_+^3} F_j(\xi) \Big(
  \frac{(x_j-\xi_j)(x_3-\xi_3)^2}{|x-\xi|^5}-\frac{(x_j-\xi_j)(x_3+\xi_3)^2}{|x-\xi^*|^5}\Big)\, d\xi\\
&&  + \int\limits_{{\Bbb R}_+^3} F_3(\xi)\, \Big( \frac{(x_3-\xi_3)^3}{|x-\xi|^5} +
  \frac{(x_3+\xi_3)^3}{|x-\xi^*|^5}\Big)\, d\xi
\end{eqnarray*}
Integrating by parts, we obtain
\begin{eqnarray*}
\Phi(x) & = & 2g(x)-\frac{3}{4\pi} \Psi(x) + \frac{1}{2\pi} \frac{\partial}{\partial x_3}\int_{{\Bbb R}_+^3}
  \frac{\partial g(\xi)}{\partial \xi_3}\, \Big( \frac{1}{|x-\xi|}+\frac{1}{|x-\xi^*|}\Big)\, d\xi \\
&&  + \frac{1}{\pi} \frac{\partial}{\partial x_3} \int_{{\Bbb R}^2} g(\xi',0)\, \frac 1{|x-(\xi',0)|}\, d\xi'.
\end{eqnarray*}
Here $\Psi(x)\to 0$ as $x_3\to 0$,
\[
\frac{\partial}{\partial x_3}\int_{{\Bbb R}_+^3} \frac{\partial g(\xi)}{\partial \xi_3}\,
\Big( \frac{1}{|x-\xi|}+\frac{1}{|x-\xi^*|}\Big)\, d\xi = -\int_{{\Bbb R}_+^3}
  \frac{\partial g(\xi)}{\partial \xi_3}\, \Big( \frac{x_3-\xi_3}{|x-\xi|^3}+\frac{x_3+\xi_3}{|x-\xi^*|^3}
  \Big)\, d\xi \to 0
\]
as $x_3\to 0$, and from the equality
\[
\frac{1}{2\pi} \int_{{\Bbb R}^2} \frac{x_3}{|x-(\xi',0)|^3}\, d\xi' = x_3\int_0^\infty
  \frac{\rho\, d\rho}{(\rho^2+x_3^2)^{3/2}} =1
\]
it follows that
\begin{eqnarray*}
&& \frac{1}{\pi} \frac{\partial}{\partial x_3} \int_{{\Bbb R}^2} g(\xi',0)\, \frac 1{|x-(\xi',0)|}\, d\xi'
  = - \frac{1}{\pi} \int_{{\Bbb R}^2} g(\xi',0)\, \frac{x_3}{|x-(\xi',0)|}\, d\xi' \\
&& = \frac 1\pi \int_{{\Bbb R}^2} \frac{(g(x',0)-g(\xi',0))x_3}{|x-(\xi',0)|^3}\, d\xi'-2g(x',0) \\
&& = \frac 1\pi \int_{{\Bbb R}^2} \frac{g(x',0)-g(x'+x_3 z',0)}{(1+|z'|^2)^{3/2}}\, dz' -2g(x',0)
  \to - 2 g(x',0) \ \mbox{ as }x_3\to 0.
\end{eqnarray*}
Consequently, $\Phi(x',0)=0$. This means, the vector function $(u,p)$ introduced above is a solution of
system (\ref{Stokesspace}) with boundary condition (ii). The result follows. \hfill  $\Box$

\begin{Le} \label{lg2}
The components of Green's matrix for the Neumann problem to the Stokes system in the half-space
${\Bbb R}_+^3$ are given by
\begin{eqnarray*}
&& {\cal G}^+_{i,j}(x,\xi) = {\cal G}_{i,j}(x,\xi)+{\cal G}_{i,j}(x,\xi^*)+\frac 1{4\pi}
  \frac{\partial^2}{\partial x_i\partial\xi_j}  \frac{x_3\xi_3}{|x-\xi^*|},\quad i,j=1,2,\\
&& {\cal G}^+_{i,3}(x,\xi) = {\cal G}_{i,3}(x,\xi)-{\cal G}_{i,3}(x,\xi^*)-\frac 1{4\pi}
  \frac{\partial}{\partial x_i}\Big( \frac{x_3}{|x-\xi^*|}+ \frac{x_3\xi_3(x_3+\xi_3)}{|x-\xi^*|^3}\Big)
  ,\quad i=1,2,\\
&& {\cal G}^+_{i,4}(x,\xi) = {\cal G}_{i,4}(x,\xi)+{\cal G}_{i,4}(x,\xi^*)+\frac 1{2\pi}
  \frac{\partial^2}{\partial x_i\partial\xi_3} \frac{x_3}{|x-\xi^*|}, \quad i=1,2, \\
&& {\cal G}^+_{3,3}(x,\xi) = {\cal G}_{3,3}(x,\xi)-{\cal G}_{3,3}(x,\xi^*) + \frac 1{4\pi}
  \Big( \frac{1}{|x-\xi^*|}+ \frac{(x_3+\xi_3)^2}{|x-\xi^*|^3}+\frac{x_3\xi_3(x_3+\xi_3)^2}{|x-\xi^*|^5}\Big),\\
&& {\cal G}^+_{3,4}(x,\xi) = {\cal G}_{3,4}(x,\xi)+{\cal G}_{3,4}(x,\xi^*)-\frac 1{2\pi} \frac{\partial}{\partial\xi_3}
  \Big( \frac{1}{|x-\xi^*|}+ \frac{x_3(x_3+\xi_3)}{|x-\xi^*|^3}\Big),\\
&& {\cal G}^+_{4,4}(x,\xi)=-\delta(x-\xi)+ \frac 1\pi \frac{\partial^2}{\partial x_3\partial\xi_3} \frac 1{|x-\xi|^*},
  \quad {\cal G}^+_{i,j}(x,\xi)={\cal G}^+_{j,i}(\xi,x) \ \mbox{ for }i,j=1,2,3,4.
\end{eqnarray*}
\end{Le}

\noindent P r o o f.
We define the functions $F$, $\tilde{F}$ and $\tilde{g}$ by $F(\xi)=f(\xi)+\nabla_\xi g(\xi)$,
\[
\tilde{F}(\xi) = F(\xi), \ \tilde{g}(\xi)=g(\xi) \ \mbox{for }\xi_3>0,\quad
\tilde{F}_j(\xi) =  (-1)^{\delta_{j,3}} F_j(\xi^*), \ \tilde{g}(\xi) =  g(\xi^*) \ \mbox{ for }\xi_3<0.
\]
Then the functions
\begin{eqnarray*}
&& \hspace{-2em} u_i(x) = \sum_{j=1}^3 \int_{{\Bbb R}^3} \tilde{F}_j(\xi) \, {\cal G}_{i,j}(x,\xi)\, d\xi
  + \int_{{\Bbb R}^3} \tilde{g}(\xi)\, {\cal G}_{i,4}(x,\xi) \\
&& \quad = \sum_{j=1}^3 \int_{{\Bbb R}_+^3} F_j(\xi) \, \big( {\cal G}_{i,j}(x,\xi)+(-1)^{\delta_{j,3}}
  {\cal G}_{i,j}(x,\xi^*)\big) \, d\xi + \int_{{\Bbb R}^3} g(\xi)\, \big( {\cal G}_{i,4}(x,\xi)
  +{\cal G}_{i,4}(x,\xi^*)\big)\, d\xi,\\
&& \hspace{-2em}p(x) = -2g(x) + \sum_{j=1}^3 \int_{{\Bbb R}^3} \tilde{F}_j(\xi) \, {\cal G}_{4,j}(x,\xi)\, d\xi\\
&& \quad =  -2g(x) + \sum_{j=1}^3 \int_{{\Bbb R}_+^3} F_j(\xi) \, \big( {\cal G}_{4,j}(x,\xi)+(-1)^{\delta_{j,3}}
  {\cal G}_{4,j}(x,\xi^*)\big) \, d\xi
\end{eqnarray*}
satisfy (\ref{Stokesspace}) for $x_3>0$ and the boundary conditions $\partial_{x_1}u_3(x)+\partial_{x_3}u_1(x)
=\partial_{x_2}u_3(x)+\partial_{x_3}u_2(x)=0$ on the plane $x_3=0$. We consider the function
\begin{eqnarray*}
-p(x) + 2\partial_{x_3} u_3(x) & = & - \frac{3}{4\pi} \sum_{j=1}^2 \int_{{\Bbb R}_+^3} F_j(\xi) \Big(
  \frac{(x_j-\xi_j)(x_3-\xi_3)^2}{|x-\xi|^5}+\frac{(x_j-\xi_j)(x_3+\xi_3)^2}{|x-\xi^*|^5}\Big)\, d\xi \\
&& -\frac{3}{4\pi} \int_{{\Bbb R}_+^3} F_3(\xi)\, \Big( \frac{(x_3-\xi_3)^3}{|x-\xi|^5} -
  \frac{(x_3+\xi_3)^3}{|x-\xi^*|^5}\Big)\, d\xi +2g(x) \\
&& -\frac{1}{2\pi} \frac{\partial}{\partial x_3} \int_{{\Bbb R}_+^3} g(\xi)\,
  \frac{\partial}{\partial\xi_3}\Big( \frac{1}{|x-\xi|}-\frac{1}{|x-\xi^*|}\Big)\, d\xi .
\end{eqnarray*}
The restriction to the plane $x_3=0$ is
\begin{eqnarray*}
\Phi(x') & = & - \frac{3}{2\pi} \sum_{j=1}^2 \int_{{\Bbb R}_+^3} F_j(\xi) \,
  \frac{(x_j-\xi_j)\xi_3^2}{|(x',0)-\xi|^5}\, d\xi
  +\frac{3}{2\pi} \int_{{\Bbb R}_+^3} F_3(\xi)\, \frac{\xi_3^3}{|(x',0)-\xi|^5} \, d\xi \\
&&  +2g(x',0) + \frac{1}{\pi} \int_{{\Bbb R}_+^3} \frac{\partial g(\xi)}{\partial\xi_3}\,
  \frac{\xi_3}{|(x',0)-\xi|^3}\,  d\xi .
\end{eqnarray*}
The vector $(v,q)$ with the potentials
\begin{eqnarray*}
&& v_i(x)=-\frac{1}{4\pi} \int_{{\Bbb R}^2} \frac{(x_i-y_i)x_3}{|x-y'|^3}\, \Phi(y')\, dy',\ \ i=1,2,\\
&&  v_3(x)=-\frac{1}{4\pi} \int_{{\Bbb R}^2} \Big( \frac 1{|x-y'|}+ \frac{x_3^2}{|x-y'|^3}\Big)\, \Phi(y')\, dy',\\
&& q(x) = \frac{1}{2\pi} \frac{\partial}{\partial x_3} \int_{{\Bbb R}^2} \frac{1}{|x-y'|}\, \Phi(y')\, dy'
\end{eqnarray*}
(here, for the sake of brevity, we wrote $y'$ instead of $(y',0)$) satisfies the equations
$-\Delta v + \nabla q=0,\ \nabla\cdot v=0$. Furthermore, from the equalities
\[
\int_{{\Bbb R}^2} \frac{(x_i-y_i)x_3^2}{|x-y'|^5}\, dy' =0 \ \mbox{ for }i=1,2, \quad
  \frac{3}{2\pi} \, \int_{{\Bbb R}^2} \frac{x_3^3}{|x-y'|^5}\, dy' = 1
\]
(see \cite[Appendix 1]{mps}) it follows that
\begin{eqnarray*}
-q+2\frac{\partial v_3}{\partial x_3}& = & \frac{3}{2\pi} \int_{{\Bbb R}^2} \frac{x_3^3\, \Phi(y')}{|x-y'|^5}
  \, dy'= \Phi(x') + \frac{3}{2\pi} \int_{{\Bbb R}^2} \frac{x_3^3\, \big( \Phi(y')-\Phi(x')\big)}{|x-y'|^5}
  \, dy'\\
& = & \Phi(x') + \frac{3}{2\pi} \int_{{\Bbb R}^2} \frac{\Phi(x'+x_3z')-\Phi(x')\big)}{(1+|z'|^2)^{5/2}}
  \, dz' \to \Phi(x') \ \mbox{ as } x_3\to 0
\end{eqnarray*}
and
\[
\frac{\partial v_i}{\partial x_3} + \frac{\partial v_3}{\partial x_i} = \frac{3}{2\pi} \int_{{\Bbb R}^2}
  \frac{(x_i-y_i)x_3^2}{|x-y'|^5}\, \Phi(y')\, dy' = -\frac{3}{2\pi} \int_{{\Bbb R}^2}
  \frac{z_i\big(\Phi(x'+   x_3 z')-\Phi(x')\big)}{(1+|z'|^2)^{5/2}}\, dz' \to 0
\]
as $x_3\to 0$, $i=1,2$. Consequently, the vector function $(u-v,p-q)$ is a solution of the system
(\ref{Stokesspace}) satisfying the Neumann condition (iv). It remains to represent $v$ and $q$ in terms
of $F$ and $g$. For $i=1,2$ we have
\begin{eqnarray*}
v_i(x) & = & -\frac{1}{8\pi^2} \sum_{j=1}^2 \int_{{\Bbb R}_+^3} F_j(\xi)\,
  \frac{\partial^2}{\partial x_i\partial\xi_j}\int_{{\Bbb R}^2} \frac{x_3\xi_3^2}{|x-y'|\, |y'-\xi|^3}
  \, dy'\, d\xi \\
&&  + \frac{1}{8\pi^2} \int_{{\Bbb R}_+^3} F_3(\xi)\, \frac{\partial}{\partial x_i} \int_{{\Bbb R}^2}
  \Big( \frac{2x_3\xi_3}{|x-y'|\, |y'-\xi|^3}-\frac{\partial}{\partial\xi_3}
  \frac{x_3\xi_3^2}{|x-y'|\, |y'-\xi|^3}\Big)\, dy'\, d\xi \\
&& -\frac{1}{2\pi} \int_{{\Bbb R}^2}\frac{(x_i-y_i)x_3}{|x-y'|^3}\, g(y')\, dy' +\frac{1}{4\pi^2}
  \int_{{\Bbb R}_+^3} \frac{\partial g(\xi)}{\partial \xi_3} \, \frac{\partial}{\partial x_i}\int_{{\Bbb R}^2}
  \frac{x_3\xi_3}{|x-y'|\, |y'-\xi|^3}\, dy'\, d\xi
\end{eqnarray*}
Using the equality
\[
\int_{{\Bbb R}^2} \frac{x_3\, dy'}{|x-y'|^3\, |y'-\xi|} = \frac{2\pi}{|x-\xi^*|}
\]
(see \cite[Appendix 1]{mps}), we obtain
\begin{eqnarray*}
v_i(x) & = & \frac{1}{4\pi} \int_{{\Bbb R}_+^3} F_3(\xi)\, \frac{\partial}{\partial x_i}
  \Big( \frac{x_3}{|x-\xi^*|}+ \frac{x_3\xi_3(x_3+\xi_3)}{|x-\xi^*|^3}\Big)\, d\xi \\
&& -\frac{1}{4\pi} \sum_{j=1}^2 \int_{{\Bbb R}_+^3} F_j(\xi)\, \frac{\partial^2}{\partial x_i\partial\xi_j}
  \frac{x_3\xi_3}{|x-\xi^*|}\, d\xi  -\frac{1}{2\pi}\int_{{\Bbb R}_+^3} g(\xi)\,
  \frac{\partial^2}{\partial x_i\partial\xi_3}\frac{x_3}{|x-\xi^*|} \, d\xi
\end{eqnarray*}
for $i=1,2$. Analogously,
\begin{eqnarray*}
\hspace{-2em}v_3(x) & = &  \frac{1}{8\pi^2} \sum_{j=1}^2 \int_{{\Bbb R}_+^3} F_j(\xi)\,
  \frac{\partial}{\partial\xi_j}\int_{{\Bbb R}^2} \Big( \frac{2\xi_3^2}{|x-y'|\, |y'-\xi|^3}
  - \frac{\partial}{\partial x_3} \frac{x_3\xi_3^2}{|x-y'|\, |y'-\xi|^3}\Big)  \, dy'\, d\xi \\
&&  - \frac{1}{8\pi^2} \int_{{\Bbb R}_+^3} F_3(\xi) \int_{{\Bbb R}^2}
  \Big( \frac{2}{|x-y'|}- \frac{\partial}{\partial x_3} \frac{x_3}{|x-y'|}\Big)\, \Big(
  \frac{2\xi_3}{|y'-\xi|^3}- \frac{\partial}{\partial \xi_3} \frac{\xi_3^2}{|y'-\xi|^3}\Big)\, dy'\, d\xi \\
&& -\frac{1}{2\pi} \int_{{\Bbb R}^2}\Big( \frac{1}{|x-y'|}+\frac{x_3^2}{|x-y'|^3}\Big)\, g(y')\, dy' \\
&&  -\frac{1}{4\pi^2} \int_{{\Bbb R}_+^3} \frac{\partial g(\xi)}{\partial \xi_3} \, \int_{{\Bbb R}^2}
  \Big( \frac{\xi_3}{|x-y'|\, |y'-\xi|^3}- \frac{\partial}{\partial \xi_3}
  \frac{x_3^2}{|x-y'|^3\, |y'-\xi|}\Big)\, dy'\, d\xi \\
& = & \frac{1}{4\pi} \sum_{j=1}^2 \int_{{\Bbb R}_+^3} F_j(\xi)\, \frac{\partial}{\partial\xi_j}
  \Big(  \frac{\xi_3}{|x-\xi^*|}+ \frac{x_3\xi_3(x_3+\xi_3)}{|x-\xi^*|^3}\Big) \, d\xi \\
&&  - \frac{1}{4\pi} \int_{{\Bbb R}_+^3} F_3(\xi)\, \Big( \frac{1}{|x-\xi^*|}+ \frac{(x_3+\xi_3)^2}{|x-\xi^*|^3}
  + \frac{x_3\xi_3(x_3+\xi_3)^2}{|x-\xi^*|^5}\Big)\, d\xi\\
&& +\frac{1}{2\pi}\int_{{\Bbb R}_+^3} g(\xi)\, \frac{\partial}{\partial\xi_3}
  \Big( \frac{1}{|x-\xi^*|}+\frac{x_3(x_3+\xi_3)}{|x-\xi^*|^3}\Big)\, d\xi
\end{eqnarray*}
and
\begin{eqnarray*}
\hspace{-2em}q(x) & = & \!\! -\frac{1}{4\pi^2} \sum_{j=1}^2 \frac{\partial}{\partial x_3}\int_{{\Bbb R}_+^3} F_j(\xi)\,
  \frac{\partial}{\partial\xi_j}\int_{{\Bbb R}^2} \frac{\xi_3^2}{|x-y'|\, |y'-\xi|^3}\, dy'\, d\xi \\
&& +  \frac 1{4\pi^2} \frac{\partial}{\partial x_3}\int_{{\Bbb R}_+^3} F_3(\xi)\,
  \int_{{\Bbb R}^2} \Big( \frac{2\xi_3}{|x-y'|\, |y'-\xi|^3} - \frac{\partial}{\partial\xi_3}
  \frac{\xi_3^2}{|x-y'|\, |y'-\xi|^3}\Big) \, d\xi\\
&& + \frac 1\pi \, \frac{\partial}{\partial x_3}\int_{{\Bbb R}^2} \frac{g(y')\, dy'}{|x-y'|}
  + \frac{1}{2\pi^2}\, \frac{\partial}{\partial x_3} \int_{{\Bbb R}_+^3}
  \frac{\partial g(\xi)}{\partial \xi_3}\, \int_{{\Bbb R}^2} \frac{\xi_3}{|x-y'|\, |y'-\xi|^3}\, dy'\, d\xi\\
& = &\!\! -\frac 1{2\pi} \sum_{j=1}^2 \int_{{\Bbb R}_+^3} F_j(\xi) \, \frac{\partial^2}{\partial x_3\partial\xi_j}
  \frac{\xi_3}{|x-\xi^*|}\, d\xi + \frac{1}{2\pi} \int_{{\Bbb R}_+^3} F_3(\xi) \, \frac{\partial}{\partial x_3}
  \Big( \frac{1}{|x-\xi^*|} +  \frac{\xi_3(x_3+\xi_3)}{|x-\xi^*|^3}\Big)\, d\xi\\
&& - \frac 1\pi \int_{{\Bbb R}_+^3} g(\xi) \, \frac{\partial^2}{\partial x_3\partial\xi_3}
  \frac{1}{|x-\xi^*|}\, d\xi
\end{eqnarray*}
Now the assertion of the lemma follows directly from the above obtained representation of the solution
$(u-v,p-q)$. \hfill $\Box$\\

As a consequence of the last two lemmas and the analogous results for the boundary conditions
(i) and (iii), we obtain the following theorem.

\begin{Th} \label{tg1}
{\em 1)} ${\cal G}^+_{i,j}(x,\xi)={\cal G}^+_{j,i}(\xi,x)$ for $i,j=1,2,3,4$.

{\em 2)} The components of Green's matrix $G(x,\xi)$ satisfy the estimate
\[
|\partial_x^\alpha\partial_\xi^\gamma {\cal G}^+_{i,j}(x,\xi)| \le c_{\alpha,\gamma} \,
  |x-\xi|^{-1-\delta_{i,4}-\delta_{j,4}-|\alpha|-|\gamma|},
\]
where the constants $c_{\alpha,\gamma}$ are independent of $x$ and $\xi$.

{\em 3)} There exist vector functions $\vec{P}_j(x,\xi)=(P_{1,j}(x,\xi),P_{2,j}(x,\xi),P_{3,j}(x,\xi))$,
$j=1,\ldots,4$, such that ${\cal G}^+_{4,j}(x,\xi) = - \nabla_x\cdot \vec{P}_j(x,\xi)$,
$P_{3,j}(x,\xi)|_{x_3=0}=0$, and
\begin{equation} \label{1tg1}
|\partial_x^\alpha\partial_\xi^\gamma \vec{P}_j(x,\xi)| \le c_{\alpha,\gamma} \,
  |x-\xi|^{-1-\delta_{j,4}-|\alpha|-|\gamma|}.
\end{equation}
\end{Th}

\noindent P r o o f.
Assertions 1) and 2) are obvious. The functions ${\cal G}^+_{4,j}$, $j=1,2$, are of the form
$\partial_{x_j}P_j(x,\xi)$ with a function $P_j$ satisfying (\ref{1tg1}). The function
${\cal G}^+_{4,3}$ can be written in the form
\begin{eqnarray*}
&&{\cal G}^+_{4,3}(x,\xi) = -\frac{1}{4\pi} \frac{\partial}{\partial x_3} \Big( \frac{1}{|x-\xi|}+
  \frac{1}{|x-\xi^*|}\Big) \\
&& =\frac{1}{4\pi} \sum_{j=1}^2 \frac{\partial}{\partial x_j} \Big(
  \frac{(x_j-\xi_j)(x_3-\xi_3)}{|x-\xi|^3} + \frac{(x_j-\xi_j)(x_3-\xi_3)}{|x-\xi^*|^3}\Big)
 + \frac{1}{4\pi}\frac{\partial}{\partial x_3} \Big( \frac{(x_3-\xi_3)^2}{|x-\xi|^3}
  + \frac{x_3^2-\xi_3^2}{|x-\xi^*|^3}\Big)
\end{eqnarray*}
in the case of boundary condition (ii) and in the form
\begin{eqnarray*}
&& {\cal G}^+_{4,3}(x,\xi) =  -\frac{1}{4\pi} \frac{\partial}{\partial x_3} \Big( \frac{1}{|x-\xi|}+
  \frac{1}{|x-\xi^*|} + 2 \frac{\xi_3(x_3+\xi_3)}{|x- \xi^*|^3}\Big) \\
&& =  \frac{1}{2\pi} \sum_{j=1}^2 \frac{\partial}{\partial x_j}  \frac{(x_j-\xi_j)(x_3+\xi_3)}{|x-\xi^*|^3}
  - \frac{1}{4\pi} \frac{\partial}{\partial x_3} \Big( \frac{1}{|x-\xi|} -\frac{1}{|x-\xi^*|}
  - 2\frac{x_3(x_3+\xi_3)}{|x-\xi^*|^3}\Big)
\end{eqnarray*}
in the case of Neumann condition (iv). For the function ${\cal G}^+_{4,4}$ there are the representations
\begin{eqnarray*}
&& {\cal G}^+_{4,4}(x,\xi) = -\delta(x-\xi) = \frac{1}{4\pi}\, \Delta_x\Big( \frac{1}{|x-\xi|}+
  \frac{1}{|x-\xi^*|}\Big) \\
&& =  \frac 1{4\pi} \sum_{j=1}^2 \frac{\partial^2}{\partial x_j^2} \Big( \frac{1}{|x-\xi|}+
  \frac{1}{|x-\xi^*|}\Big) - \frac{1}{4\pi} \frac{\partial}{\partial x_3}\Big( \frac{x_3-\xi_3}{|x-\xi|^3}+
  \frac{x_3+\xi_3}{|x-\xi^*|}\Big)
\end{eqnarray*}
in case (ii) and
\begin{eqnarray*}
&& {\cal G}^+_{4,4}(x,\xi) = -\delta(x-\xi) + \frac 1\pi \frac{\partial^2}{\partial x_3^2}
  \frac{1}{|x-\xi^*|}= \frac{1}{4\pi} \Delta_x\Big( \frac{1}{|x-\xi|}-3 \frac{1}{|x-\xi^*|}\Big)
  + \frac 1\pi \frac{\partial^2}{\partial x_3^2}\frac 1{|x-\xi^*|}\\
&& = \frac{1}{4\pi} \sum_{j=1}^2 \frac{\partial^2}{\partial x_j^2}\Big( \frac{1}{|x-\xi|}-\frac{3}{|x-\xi^*|}
  \Big) -\frac{1}{4\pi}\frac{\partial}{\partial x_3}\Big( \frac{x_3-\xi_3}{|x-\xi|^3}
  + \frac{x_3+\xi_3}{|x-\xi^*|^3}\Big)
\end{eqnarray*}
in case (iv). Similar representations hold for boundary conditions (i) and (iii). This proves assertion 3).
\hfill $\Box$

\subsection{Estimates of Green's matrix in the dihedron}

Let $G(x,\xi)=\big( G_{i,j}(x,\xi)\big)_{i,j=1}^4$ be Green's matrix for problem
(\ref{StokesD}), (\ref{bcD}), i.e. the vector functions $\vec{G}_j=( G_{1,j},G_{2,k},G_{3,k})$
and the function $G_{4,j}$ are solutions of the problem
\begin{eqnarray*}
&& -\Delta_x \vec{G}_j(x,\xi)+ \nabla_x  G_{4,j}(x,\xi)= \delta(x-\xi)\, (\delta_{1,j},\delta_{2,j},
  \delta_{3,j})^t ,\\
&& -\nabla_x\cdot \vec{G}_j(x,\xi) = \delta(x-\xi)\, \delta_{4,j} \quad
  \mbox{for }x,\xi\in {\cal D}, \nonumber \\
&& S^\pm \vec{G}_j(x,\xi)=0,\quad N^\pm(\partial_x)\, \big(\vec{G}_j(x,\xi),G_{4,j}(x,\xi)\big)=0
  \quad \mbox{for }x\in \Gamma^\pm,\ \xi\in {\cal D}
\end{eqnarray*}
More precisely, if ${\cal G}(x,\xi)$ denotes Green's matrix for Stokes system in ${\Bbb R}^3$
(see (\ref{greenfsp1}), (\ref{greenfsp2})) , then we define Green's matrix $G(x,\xi)=(G_{i,j}(x,\xi))_{i,j=1}^4$
of problem (\ref{StokesD}), (\ref{bcD}) by the formula
\begin{equation} \label{gr}
G(x,\xi) = \psi(x,\xi)\, {\cal G}(x,\xi) + R(x,\xi).
\end{equation}
Here $\psi(x,\xi)$ is a smooth function of $x$ and $\xi$ equal to 1 for small $|x-\xi|$ and to zero
for large $|x-\xi|$ is large and for $x$ near the edge $M$, and
the vector function  $\big( \vec{R}_j,R_{4,j}\big) =\big(R_{1,j},R_{2,j},R_{3,j},R_{4,j}\big)$ is
the uniquely determined weak solution of the problem
\begin{eqnarray*}
&& -\Delta_x \vec{R}_j(x,\xi) + \nabla_x R_{4,j}(x,\xi) = \ \delta(x-\xi)\, (\delta_{1,j},\delta_{2,j},\delta_{3,j})\\
&& \hspace{14em}+ \ \Delta_x \big(\psi(x,\xi)\vec{\cal G}_j(x,\xi)\big) -\nabla_x\big( \psi(x,\xi){\cal G}_{4,j}(x,\xi)\big),\\
&& -\nabla_x \vec{R}_j(x,\xi) = \delta_{4,j}\delta(x-\xi)+\nabla x\big( \psi(x,\xi)\vec{\cal G}_j(x,\xi)\big)
  \quad \mbox{for }x\in {\cal D},\ j=1,2,3,4, \\
&& S_j^\pm \vec{R}_j(x,\xi) = -S_j^\pm\big( \psi(x,\xi)\vec{\cal G}_j(x,\xi)\big), \\
&& N_j^\pm \big(\vec{R}_j(x,\xi),R_{4,j}(x,\xi)\big) = -N_j^\pm \big( \psi(x,\xi)\vec{\cal G}_j(x,\xi),
  \psi(x,\xi){\cal G}_{4,j}(x,\xi)\big) \quad \mbox{for }x\in \Gamma^\pm,
\end{eqnarray*}
$j=1,2,3,4.$ Since the right-hand sides of the above equations are smooth, have compact supports and vanish near the edge,
this problem has a unique weak solution $\big( \vec{R}_j(\cdot,\xi),R_{4,j}(\cdot,\xi)\big)$ for arbitrary
$\xi\in {\cal D}$. According to Theorem \ref{t4}, we have have even $\zeta \big(
\vec{R}_j(\cdot,\xi),R_{4,j}(\cdot,\xi)\big) \in W_\delta^l({\cal D})^3\times W_\delta^{l-1}({\cal D})$
with a certain $\delta<l-1$ for an arbitrary smooth function $\zeta$ with compact support.

\begin{Th} \label{t7}
{\em 1)} For $i,j=1,2,3,4$ there are the equalities
\begin{equation} \label{propG2}
G_{i,j}(tx,t\xi)=t^{-1-\delta_{i,4}-\delta_{j,4}} \, G_{i,j}(x,\xi) \quad \mbox{for arbitrary }t>0,\ x,\xi\in {\cal D}
\end{equation}
and
\begin{equation} \label{propG1}
G_{i,j}(x,\xi)=G_{j,i}(\xi,x)
\end{equation}

{\em 2)} Every solution $(u,p) \in C_0^\infty(\bar{\cal D})^4$ of equation {\em (\ref{StokesD})
satisfying the homogeneous boundary condition (\ref{bcD})} is given by the formulas
\begin{eqnarray} \label{repu}
u_i(x) & = & \int_{\cal D} \big( f(\xi)+\nabla_\xi g(\xi)\big)\cdot \vec{H}_i(x,\xi)\, d\xi
  + \int_{\cal D} g(\xi)\, G_{i,4}(x,\xi)\, d\xi,\quad i=1,2,3, \\ \label{repp}
p(x) & = & -g(x) + \int_{\cal D} \big( f(\xi)+\nabla_\xi g(\xi)\big)\cdot \vec{H}_4(x,\xi)\, d\xi
  + \int_{\cal D} g(\xi)\, G_{4,4}(x,\xi)\, d\xi,
\end{eqnarray}
where $\vec{H}_i$ denotes the vector function $(G_{i,1},G_{i,2},G_{i,3})^t$.

{\em 3)} The functions $G_{i,j}(x,\xi)$ are infinitely
differentiable with respect to $x,\xi \in \bar{\cal D}\backslash
M$, $x\not= \xi$. For $|x-\xi|<\min(|x'|,|\xi'|)$ there are the
estimates
\begin{equation} \label{1t7}
\big| \partial_x^\alpha\partial_\xi^\beta G_{i,j}(x,\xi)\big|\le
c\, |x-\xi|^{-1-\delta_{i,4}-\delta_{j,4}-|\alpha|-|\beta|}
\end{equation}
For $j=1,\ldots,4$ there is the representation $G_{4,j}(x,\xi) = - \nabla_x\cdot \vec{P}_j(x,\xi) + Q_j(x,\xi)$,
where $\vec{P}_j(x,\xi)\cdot n^\pm=0$ for $x\in \Gamma^\pm$, $\xi\in {\cal D}$, and
$\vec{P}_j$, $Q_j$ satisfy the estimates
\begin{equation} \label{2t7}
|\partial_x^\alpha\partial_\xi^\gamma \vec{P}_j(x,\xi)| \le c_{\alpha,\gamma} \,
  |x-\xi|^{-1-\delta_{i,4}-|\alpha|-|\gamma|}, \quad
|\partial_x^\alpha\partial_\xi^\gamma Q_j(x,\xi)| \le c_{\alpha,\gamma} \,
  |\xi'|^{-2-\delta_{i,4}-|\alpha|-|\gamma|}
\end{equation}
for $|x-\xi|<\min(|x'|,|\xi'|)$.
\end{Th}

\noindent P r o o f. 1) (\ref{propG2}) follows immediately from the definition of $G(x,\xi)$. We show that
Green's formula (\ref{Green2}) is valid for $u(x)=\vec{G}_i(x,y)$, $p(x)= G_{i,4}(x,y)$,
$v(x)=\vec{G}_j(x,z)$, $q(x)=G_{4,j}(x,z)$, where $y$ and $z$ are arbitrary points in ${\cal D}$. Let
$\zeta$ be a smooth function on the interval $(0,\infty)$, $\zeta(t)=1$ for $t<1/2$, $\zeta(t)=0$ for $t>1$,
and let $\varepsilon$ be a sufficiently small positive number. By Theorem \ref{t4}, the vector functions
$\eta \big(\vec{G}_j(\cdot,\xi),G_{4,j}(\cdot,\xi)\big)$ belong to $W_\delta^l({\cal D})^3\times
W_\delta^{l-1}({\cal D})$ with a certain $\delta<l-1$ for an arbitrary smooth function $\eta$ with
compact support equal to zero in a neighborhood of $\xi$. Consequently (\ref{Green2}) is valid for the
following vector functions $(u,p)$ and $(v,q)$.
\begin{itemize}
\item[(i)] $u(x)=\zeta(\frac{x-y}{\varepsilon})\, \vec{G}_i(x,y)$, $p(x)=\zeta(\frac{x-y}{\varepsilon})\,
  G_{4,i}(x,y)$, $v(x)=\big( 1-\zeta(\frac{x-z}{\varepsilon})\big)\, \vec{G}_j(x,z)$,\\
  $q(x)=\big( 1-\zeta(\frac{x-z}{\varepsilon})\big)\, G_{4,j}(x,z)$
\item[(ii)] $u(x)=\big( 1-\zeta(\frac{x-y}{\varepsilon})\big)\, \vec{G}_i(x,y)$,
  $p(x)=\big( 1-\zeta(\frac{x-y}{\varepsilon})\big)\, G_{4,i}(x,y)$,
  $v(x)=\zeta(\frac{x-z}{\varepsilon})\, \vec{G}_j(x,z)$,\\
  $q(x)=\zeta(\frac{x-z}{\varepsilon})\, G_{4,j}(x,z)$,
\item[(iii)] $u(x)=\big( 1-\zeta(\frac{x-y}{\varepsilon})\big)\, \vec{G}_i(x,y)$,
  $p(x)=\big( 1-\zeta(\frac{x-y}{\varepsilon})\big)\, G_{4,i}(x,y)$,\\
  $v(x)=\big( \zeta(\frac{x-z}R)-\zeta(\frac{x-z}{\varepsilon})\big)\, \vec{G}_j(x,z)$,
  $q(x)=\big( \zeta(\frac{x-z}R)-\zeta(\frac{x-z}{\varepsilon})\big)\, G_{4,j}(x,z)$,
\end{itemize}
where $R$ is an arbitrary positive number. From the last statement it follows that (\ref{Green2}) is also
valid for
\begin{itemize}
\item[(iv)] $u(x)=\big( 1-\zeta(\frac{x-y}{\varepsilon})\big)\, \vec{G}_i(x,y)$,
  $p(x)=\big( 1-\zeta(\frac{x-y}{\varepsilon})\big)\, G_{4,i}(x,y)$,
  $v(x)=\big( 1-\zeta(\frac{x-z}{\varepsilon})\big)\, \vec{G}_j(x,z)$,\\
  $p(x)=\big( 1-\zeta(\frac{x-z}{\varepsilon})\big)\, G_{4,j}(x,z)$
\end{itemize}
To see this , one has to show that all integrals in (\ref{Green2}) tend to zero as $R\to \infty$ if
\begin{eqnarray} \label{upvq1}
&& u(x)=\big( 1-\zeta(\frac{x-y}{\varepsilon})\big)\, \vec{G}_i(x,y), \ \
  p(x)=\big( 1-\zeta(\frac{x-y}{\varepsilon})\big)\, G_{4,i}(x,y), \\ \label{upvq2}
&& v(x)=\big( 1-\zeta(\frac{x-z}{R})\big)\, \vec{G}_j(x,z), \ \
  q(x)=\big( 1-\zeta(\frac{x-z}{R})\big)\, G_{4,j}(x,z).
\end{eqnarray}
For the integral
\[
\int_{\cal D} (-\Delta u - \nabla\nabla\cdot u+\nabla p)\cdot v\, dx
\]
this is evident. It vanishes for large $R$, since $(u,p)$ and $v$ have disjoint supports. Furthermore, for these
$(u,p)$ and $(v,q)$ and sufficiently large $R$ we have
\begin{eqnarray*}
&& \Big| \int_{\cal D} u\cdot (-\Delta v-\nabla\nabla\cdot v+\nabla q)\, dx\Big|^2 \\
&& \le c \Big( \int\limits_{{\cal D}_R(z)} |\vec{G}_i(x,y)|\, \big(
  |R^{-2} |\vec{G}_j(x,z)| + R^{-1}|\nabla \vec{G}_j(x,z)|+|R^{-1}|G_{4,j}(x,z)|\big)\, dx\Big)^2\\
&& \le c \int\limits_{{\cal D}_R(z)} |x|^{-2} |\vec{G}_i(x,y)|^2\, dx
   \int\limits_{{\cal D}_R(z)} \big( |x|^{-2} |\vec{G}_j(x,z)|^2 + |\nabla
  |\vec{G}_j(x,z)|^2 + |G_{4,j}(x,z)|^2\big)\, dx,
\end{eqnarray*}
where ${\cal D}_R(z)=\{ x: \, R/2<|x-z|<R\}$.
The right-hand side of the last inequality tends to zero as $R\to \infty$, since $G(x,\xi)$ can be written
in the form (\ref{gr}) and the norm in ${\cal H}$ is equivalent to (\ref{normH2}). Analogously, it can be
shown that the other integrals in (\ref{Green2}) tend to zero as $R\to \infty$ if $u,v,p,q$ are given by
(\ref{upvq1}), (\ref{upvq2}).

Finally, all integrals in (\ref{Green2}) vanish if
\begin{itemize}
\item[(v)] $u(x)=\zeta(\frac{x-y}{\varepsilon})\, \vec{G}_i(x,y)$, $p(x)=\zeta(\frac{x-y}{\varepsilon})\,
  G_{4,i}(x,y)$, $v(x)=\zeta(\frac{x-z}{\varepsilon})\, \vec{G}_j(x,z)$,
  $q(x)=\zeta(\frac{x-z}{\varepsilon})\, G_{4,j}(x,z)$,
\end{itemize}
and $\varepsilon$ is sufficiently small. Then the supports of $(u,p)$ and $(v,q)$ are disjoint.
From the validity of (\ref{Green2}) for (i), (ii), (iv), (v) it follows that this formula is applicable
to $u(x)=\vec{G}_i(x,y)$, $p(x)=G_{4,i}(x,y)$, $v(x)=\vec{G}_j(x,z)$, and $q(x)=G_{4,j}(x,z)$. This
implies $G_{i,j}(y,z) = G_{j,i}(z,y).$

2) follows immediately from (\ref{Green2}) and (\ref{propG1}).

3) Let ${\cal D}_+$ be the set of all $x\in {\cal D}$ such that dist$(x,\Gamma^+)< 2\,$dist$(x,\Gamma^-)$,
and let $t_0$ denote the distance of the set $\{ x\in {\cal D}_+:\, |x'|=1\}$ to $\Gamma^-$.
Furthermore, let $\zeta$ be a smooth function on $(0,\infty)$, $\zeta(t)=1$ for $t<t_0/2$, $\zeta(t)=0$
for $t>t_0$.

Suppose that $\xi\in {\cal D}_+$. Then the function $x\to \zeta(\frac{|x-\xi|}{|\xi'|})$ vanishes on $\Gamma^-$.
We write the Green matrix $G(x,\xi)$ in the form
\begin{equation} \label{3t7}
G(x,\xi) = \zeta\big( \frac{|x-\xi|}{|\xi'|}\big) \, {\cal G}^+(x,\xi) + R^+(x,\xi),
\end{equation}
where ${\cal G}^+(x,\xi)$ is the Green matrix of the problem in the half-space with the boundary conditions
$S^+ u =0$, $N^+(u,p)=0$ on the plane containing $\Gamma^+$. Then the vector functions
$\partial_\xi^\gamma \big(\vec{R}^+_j(\cdot,\xi), R^+_{4,j}(\cdot,\xi)\big)$ are the unique solutions
in ${\cal H}\times L_2({\cal D})$ of the problem
\begin{eqnarray*}
&& -\Delta_x \partial_\xi^\gamma \vec{R}_j^+(x,\xi)+ \nabla_x R^+_{4,j}(x,\xi) = \vec{\Phi}_j(x,\xi),
  \quad -\nabla\cdot \vec{G}^+_j(x,\xi) = \Psi_j(x,\xi)\ \mbox{ for }x\in {\cal D},\\
&& S^+ \partial_\xi^\gamma \vec{R}_j^+(x,\xi) =0, \ N^+\partial_\xi^\gamma \big(\vec{R}^+_j(x,\xi),
  R^+_{4,j}(x,\xi)\big) = \vec\Upsilon_j\ \mbox{ for }x\in \Gamma^+, \\
&& S^- \partial_\xi^\gamma \vec{R}_j^+(x,\xi) =0, \ N^-\partial_\xi^\gamma \big(\vec{R}^+_j(x,\xi),
  R^+_{4,j}(x,\xi)\big) = 0\ \mbox{ for }x\in\Gamma^-,
\end{eqnarray*}
where $\vec{\Phi}_j=-\Delta_x\partial_\xi^\gamma\big( \vec{G}_j - \zeta(\frac{|x-\xi|}{|\xi'|})\vec{\cal G}_j^+
\big) + \nabla_x\partial_\xi^\gamma\big( G_{4,j}-\zeta(\frac{|x-\xi|}{|\xi'|}){\cal G}_{4,j}^+\big)$,
$\Psi_j=-\Delta_x\partial_\xi^\gamma\big( \vec{G}_j - \zeta(\frac{|x-\xi|}{|\xi'|})\vec{\cal G}_j^+\big)$.
The functions $\vec{\Phi}_j$, $\Psi_j$ and $\vec{\Upsilon}_j$ are infinitely differentiable with respect to $x$
and vanish for $|x-\xi|<t_0|\xi'|/2$ and $|x-\xi|>t_0|\xi'|$.
For $\xi\in {\cal D}_+$, $\xi'|=1$ all derivatives $\partial_x^\alpha \vec{\Phi}_j(x,\xi)$,
$\partial_x^\alpha \Psi_j(x,\xi)$ and $\partial_x^\alpha \vec{\Upsilon}_j(x,\xi)$ are bounded by constants
independent of $x$ and $\xi$. Consequently, there exist constants $c_{\alpha,\gamma}$ such that
\[
\big| \partial_x^\alpha\partial_\xi^\gamma R_{i,j}^+(x,\xi)\big| \le c_{\alpha,\gamma}\quad\mbox{for }
  \xi\in {\cal D}_+, \ |\xi'|<1, \ 1/2 < |x'| < 2.
\]
Since the functions $R_{i,j}^+(x,\xi)$ (as well as $G_{i,j}(x,\xi)$ and ${\cal G}^+_{i,j}(x,\xi)$) are
positively homogeneous of degree $-1-\delta_{i,4}-\delta_{j,4}$, we conclude that
\begin{equation} \label{4t7}
\big| \partial_x^\alpha\partial_\xi^\gamma R_{i,j}^+(x,\xi)\big| \le c_{\alpha,\gamma}\,
  |\xi'|^{-1-\delta_{i,4}-\delta_{j,4}-|\alpha|-|\gamma|} \quad\mbox{for }
  \xi\in {\cal D}_+, \  |\xi'|/2 < |x'| < 2|\xi'|.
\end{equation}
Analogously, this estimates holds for $\xi \in {\cal D}_-=\{ x\in {\cal D}:\, \mbox{dist}(x,\Gamma^-)<
2\,\mbox{dist}(x,\Gamma^+)\}$, $|\xi'|/2 < |x'| < 2|\xi'|$. This proves (\ref{1t7}).

By Theorem \ref{tg1}, there exists a vector function $\vec{P}_j^+$ satisfying (\ref{1tg1}) such that
$-\nabla_x \cdot \vec{P}_j^+(x,\xi)={\cal G}_{4,j}^+(x,\xi)$ and  $\vec{P}_j^+(x,\xi)\cdot n^+=0$
for $x\in \Gamma^+$. Thus, one can write (\ref{3t7}) in the form
\[
G_{4,j}(x,\xi) = -\zeta\big(\frac{|x-\xi|}{|\xi'|}\big)\, \nabla_x \cdot \vec{P}_j^+(x,\xi) + R^+_{4,j}(x,\xi)
\]
for $\xi \in {\cal D}_+$, where $R^+_{4,j}$ satisfies (\ref{4t7}). Analogously, we obtain the representation
\[
G_{4,j}(x,\xi) = -\zeta\big(\frac{|x-\xi|}{|\xi'|}\big)\, \nabla_x \cdot \vec{P}_j^-(x,\xi) + R^-_{4,j}(x,\xi)
\]
for $\xi \in {\cal D}_-$, where $\vec{P}_j^-$, $R_{4,j}^-$ satisfy the estimates (\ref{1tg1}) and (\ref{4t7}),
respectively, and $\vec{P}_j^-(x,\xi)\cdot n^-=0$ for $x\in \Gamma^-$.
Let $\eta^+(\xi)$ be a function depending only on $\xi'/|\xi'|$, infinitely differentiable for $\xi'\not=0$
such that $\eta^+(\xi)=1$ for $\xi \in {\cal D}\backslash {\cal D}_-$, $\eta^+(\xi)=0$ for
$\xi \in {\cal D}\backslash{\cal D}_+$. Furthermore, let $\eta^-(\xi)=1-\eta^+(\xi)$. Then
\[
G_{4,j}(x,\xi) = - \zeta\big(\frac{|x-\xi|}{|\xi'|}\big)\, \sum_\pm \eta^\pm(\xi)\,
  \nabla_x \cdot \vec{P}_j^\pm(x,\xi) + \sum_\pm \eta^\pm(\xi)\, R_{4,j}^\pm(x,\xi) \\
= -\nabla_x \cdot \vec{P_j}(x,\xi) + Q_j(x,\xi),
\]
where
\[
\vec{P_j}(x,\xi)=\sum_\pm \eta^\pm(\xi)\vec{P}_j^\pm(x,\xi), \quad
Q_j(x,\xi) = \sum_\pm \eta^\pm(\xi)\Big( \big(1- \zeta\big(\frac{|x-\xi|}{|\xi'|}\big)\big)\, \nabla_x \cdot
 \vec{P}_j^\pm(x,\xi) +  R_{4,j}^\pm(x,\xi)\Big)
\]
Obviously $\vec{P}_j$ and $Q_j$ satisfy (\ref{2t7}). The proof is complete. \hfill $\Box$\\

Next we prove estimates of Green's matrix in the case $|x-\xi|>\min(|x'|,|\xi'|)$. Let $\lambda_1$
be the eigenvalue of the pencil $A(\lambda)$ with smallest positive real part, and let $\lambda_2$
be the eigenvalue with smallest real part greater than 1. We define
\begin{equation} \label{defmu}
\mu = \left\{ \begin{array}{ll} \mbox{Re}\, \lambda_1 & \mbox{if }d^+ +d^- \mbox{ is odd }\
  \mbox{ or }  \ d^+ + d^- \mbox{ is even and } \theta \ge \pi/m, \\
  \mbox{Re}\, \lambda_2 & \mbox{if } d^+ + d^- \mbox{ is even and } \theta < \pi/m, \end{array}\right.
\end{equation}
where $m=1$ if $d^+=d^-$, $m=2$ if $d^+\not= d^-$.

\begin{Le} \label{l13}
Let ${\cal B}$ be a ball with radius 1 and center at $x_0$, $\mbox{\em dist}(x_0,M)\le 4$. Furthermore,
let $\zeta,\eta$ be smooth functions with support in ${\cal B}$ such that $\eta=1$ in
a neighborhood of $\mbox{\em supp}\, \zeta$. If $\eta(u,p)\in {\cal H}\times L_2({\cal D})$,
$-\Delta u+\nabla p=0$, $\nabla\cdot u=0$ in ${\cal D}\cap {\cal B}$ and $S^\pm u =0,
N^\pm (u,p)=0$ on $\Gamma^\pm \cap {\cal B}$, then
\begin{eqnarray*}
&& \sup_{x\in {\cal D}} \Big(|x'|^{\max(|\alpha|-\mu+\varepsilon,0)}  \big| \zeta(x)\partial_{x'}^\alpha
  \partial_{x_3}^j u(x)\big|  + |x'|^{\max(|\alpha|+1-\mu+\varepsilon,0)} \,
  \big| \zeta(x) \partial_{x'}^\alpha\partial_{x_3}^j p(x)\big|\Big) \\
&&  \le c\, \big( \| \eta u\|_{\cal H}+ \| \eta p\|_{L_2({\cal D})}\big),
\end{eqnarray*}
where $\varepsilon$ is an arbitrary positive constant.
\end{Le}

\noindent P r o o f. Let $\varepsilon$ be such that $\mu-\varepsilon \in (k,k+1)$. Then $\delta=k+1-\mu+\varepsilon
\in (0,1)$. Furthermore, let $\chi$ be a function from $C_0^\infty({\cal B})$ such that $\zeta\chi=\zeta$
and $\eta\chi=\chi$. Since the strip $0<\mbox{Re}\, \lambda \le k+1-\delta = \mu-\varepsilon$ does not
contain eigenvalues of the pencil $A(\lambda)$, it follows from Theorem \ref{t4} and Corollary \ref{c6}
(see also Remark \ref{r2}) that $\partial_{x_3}^j (\chi u) \in W_\delta^{k+2}({\cal D})^3$,
$\partial_{x_3}^j (\chi p) \in W_\delta^{k+1}({\cal D})$ for $j=0,1,\ldots$.
Using Lemma \ref{l4a}, we even get $\partial_{x_3}^j (\chi u) \in W_{\delta+\nu}^{k+\nu+2}
({\cal D})^\ell$, $\partial_{x_3}^j (\chi p) \in W_{\delta+\nu}^{k+\nu+1}({\cal D})^\ell$, for
$\nu=0,1,\ldots$ and
\begin{equation} \label{1l13}
\|  \partial_{x_3}^j (\chi u)\|_{W_{\delta+\nu}^{k+\nu+2}({\cal D})^\ell}
   + \| \partial_{x_3}^j (\chi p) \|_{W_{\delta+\nu}^{k+\nu+1}({\cal D})^\ell}
   \le c\, \big(  \| \eta u\|_{{\cal H}}+ \| \eta p\|_{L_2({\cal D})}\big) .
\end{equation}
In particular, for $0\le |\alpha|\le k-1$ we have $\partial_{x'}^\alpha\partial_{x_3}^j (\chi p) \in
W_\delta^2({\cal D})$. Since $W_\delta^2(K)$ is continuously imbedded into $C(\bar{K})$, we conclude that
\[
\sup_{x'\in K,x_3\in{\Bbb R}} \big| \partial_{x'}^\alpha\partial_{x_3}^j(\chi p)(x',x_3)\big|
  \le c\, \sup_{x_3\in {\Bbb R}} \| \partial_{x'}^\alpha\partial_{x_3}^j (\chi p)(\cdot,x_3)
  \|_{W_\delta^2(K)}
\]
Furthermore, using the continuity of the imbedding $W_2^1(M)\subset C(M)$, we obtain
\[
\sup_{x_3\in {\Bbb R}} \| \partial_{x'}^\alpha\partial_{x_3}^j (\chi p)(\cdot,x_3)
  \|_{W_\delta^2(K)} \le c\, \Big( \| \partial_{x'}^\alpha\partial_{x_3}^j (\chi p)
  \|_{W_\delta^2({\cal D})} + \| \partial_{x'}^\alpha\partial_{x_3}^{j+1}(\chi p)
  \|_{W_\delta^2({\cal D})}\Big).
\]
This implies
\[
\sup_{x\in {\cal D}} \big| \partial_{x'}^\alpha\partial_{x_3}^j(\chi p)(x)\big|
  \le c \big(  \| \eta u\|_{{\cal H}}+ \| \eta p\|_{L_2({\cal D})}\big)
\]
If $|\alpha|\ge k$, then we conclude from (\ref{1l13}) that $\partial_{x'}^\alpha\partial_{x_3}^j(\chi p)
\in W_{\delta-k+1+|\alpha|}^2({\cal D})\subset V_{\delta-k+1+|\alpha|}^2({\cal D})$. Using Lemma \ref{l0a},
the continuity of the imbedding $W_2^1(M)\subset C(M)$ and (\ref{1l13}), we obtain
\begin{eqnarray*}
&& \sup_{x'\in K,x_3\in {\Bbb R}} |x'|^{\delta-k+|\alpha|}\, \big| \partial_{x'}^\alpha
  \partial_{x_3}^j(\chi p)(x',x_3)\big| \le c\, \sup_{x_3\in {\Bbb R}}
  \| \partial_{x'}^\alpha \partial_{x_3}^j(\chi p)(\cdot,x_3)\|_{V_{\delta-k+1+|\alpha|}^2(K)} \\
&& \le c\, \Big( \| \partial_{x'}^\alpha\partial_{x_3}^j (\chi p)
  \|_{V_{\delta-k+1+|\alpha|}^2{\cal D})} + \| \partial_{x'}^\alpha\partial_{x_3}^{j+1}(\chi p)
  \|_{V_{\delta-k+1+|\alpha|}^2{\cal D})}\Big)
  \le c\, \big(  \| \eta u\|_{{\cal H}}+ \| \eta p\|_{L_2({\cal D})}\big)
\end{eqnarray*}
for $|\alpha|\ge k$. Consequently,
\[
\sup_{x\in {\cal D}} |x'|^{\max(0,|\alpha|+1-\mu+\varepsilon)}
\big| \zeta(x)\partial_{x'}^\alpha\partial_{x_3}^j
  p(x)\big| \le c\, \big(  \| \eta u\|_{{\cal H}}+ \| \eta p\|_{L_2({\cal D})}\big).
\]
Analogously, it can be shown (cf. \cite[Le.2.9]{mr-01}) that
\[
\sup_{x\in {\cal D}} |x'|^{\max(0,|\alpha|-\mu+\varepsilon)}
\big| \zeta(x)\partial_{x'}^\alpha\partial_{x_3}^j
  u(x)\big| \le c\, \big(  \| \eta u\|_{{\cal H}}+ \| \eta p\|_{L_2({\cal D})}\big).
\]
The proof is complete. \hfill $\Box$\\

If $\lambda=0$ is not an eigenvalue of the pencil $A(\lambda)$ (i.e., if $d^+\cdot d^- =0$ or condition (\ref{erg1})
is satisfied), then the estimate in the last lemma can be improved for $|\alpha|<\mbox{Re}\, \lambda_1$.

\begin{Le} \label{l13a}
Let ${\cal B},\zeta,\eta$ be as in Lemma {\em \ref{l13}}, and let $\eta(u,p)\in {\cal H}\times L_2({\cal D})$
be such that $-\Delta u+\nabla p=0$, $\nabla\cdot u=0$ in ${\cal D}\cap {\cal B}$ and $S^\pm u =0,
N^\pm (u,p)=0$ on $\Gamma^\pm \cap {\cal B}$. If $\lambda=0$ is not an eigenvalue of the pencil $A(\lambda)$, then
\[
\sup_{x\in {\cal D}} |x'|^{|\alpha|-\mbox{\em \scriptsize Re}\, \lambda_1+\varepsilon}\Big(  \big| \zeta(x)\partial_{x'}^\alpha
  \partial_{x_3}^j u(x)\big|  + |x'| \, \big| \zeta(x) \partial_{x'}^\alpha\partial_{x_3}^j p(x)\big|\Big)
  \le c\, \big( \| \eta u\|_{\cal H}+ \| \eta p\|_{L_2({\cal D})}\big).
\]
\end{Le}

\noindent P r o o f.
Let $\delta=1-\mbox{Re}\, \lambda_1+\varepsilon$, and let $\chi$ be a function from
$C_0^\infty({\cal B})$ such that $\zeta\chi=\zeta$ and $\eta\chi=\chi$. By Lemma \ref{l4} and Theorem \ref{t4a},
we have
\[
\| \partial_{x_3}^j (\chi u)\|_{V_{\delta+k}^{k+2}({\cal D})} + \| \partial_{x_3}^j (\chi p)
  \|_{V_{\delta+k}^{k+1}({\cal D})} \le c\, \big( \| \eta u\|_{\cal H}+ \| \eta p\|_{L_2({\cal D})}\big).
\]
for arbitrary integer $k$.
Using this inequality, Lemma \ref{l0a} and the continuity of the imbedding $W_2^1(M)\subset C(M)$, we obtain
\begin{eqnarray*}
&& \sup_{(x',x_3)\in {\cal D}} |x'|^{\delta-1+|\alpha|} \, \big| \partial_{x'}^\alpha \partial_{x_3}^j
  (\chi u)(x',x_3)\big| \le c\, \sup_{x_3\in {\Bbb R}} \| \partial_{x'}^\alpha \partial_{x_3}^j
  (\chi u)(\cdot,x_3)\|_{V_{\delta+|\alpha|}^2({\cal D})^3} \\
&& \le c\, \Big( \| \partial_{x'}^\alpha \partial_{x_3}^j (\chi u)\|_{V_{\delta+|\alpha|}^2({\cal D})^3}
  + \| \partial_{x'}^\alpha \partial_{x_3}^{j+1} (\chi u)\|_{V_{\delta+|\alpha|}^2({\cal D})^3}\Big)
  \le c\, \big( \| \eta u\|_{\cal H}+ \| \eta p\|_{L_2({\cal D})}\big).
\end{eqnarray*}
Analogously,
\[
 \sup_{(x',x_3)\in {\cal D}} |x'|^{\delta+|\alpha|} \, \big| \partial_{x'}^\alpha \partial_{x_3}^j
  (\chi p)(x',x_3)\big| \le c\, \big( \| \eta u\|_{\cal H}+ \| \eta p\|_{L_2({\cal D})}\big).
\]
The result follows. \hfill $\Box$

\begin{Th} \label{t8}
For $|x-\xi|\ge \min(|x'|,|\xi'|)$ there is the estimate
\begin{equation} \label{2t8}
\big| \partial_{x'}^\alpha  \partial_{x_3}^\sigma\partial_{\xi'}^\beta\partial_{\xi_3}^\tau G_{i,j}(x,\xi)
  \big| \le c\, |x-\xi|^{-T-|\alpha|-|\beta|-\sigma-\tau} \, \Big( \frac{|x'|}{|x-\xi|}
  \Big)^{\sigma_{i,\alpha}} \, \Big( \frac{|\xi'|}{|x-\xi|}
  \Big)^{\sigma_{j,\beta}} ,
\end{equation}
where $T=1+\delta_{i,4}+\delta_{j,4}$, $\varepsilon$ is an arbitrarily small positive number, and
$\sigma_{i,\alpha}=\min(0,\mu-|\alpha|-\delta_{i,4}-\varepsilon)$.

In the case when $\lambda=0$ is not an eigenvalue of the pencil $A(\lambda)$, then estimate
{\em (\ref{2t8})} holds even with
$\sigma_{i,\alpha}=\mbox{\em Re}\, \lambda_1- |\alpha|-\delta_{i,4}-\varepsilon$ for
$|\alpha|<\mbox{\em Re}\, \lambda_1-\delta_{i,4}$.
\end{Th}

\noindent P r o o f. Due to (\ref{propG2}), it suffices to prove the estimate for $|x-\xi|=2$. Then, under
the assumption $\min(|x'|,|\xi'|)\le |x-\xi|$, we have $\max(|x'|,|\xi'|)\le 4$. Let ${\cal B}_x$,
${\cal B}_\xi$ be balls with centers $x$ and $\xi$, respectively, and radius 1. Furthermore, let
$\zeta$ and $\eta$ be infinitely differentiable functions with supports in ${\cal B}_x$ and
${\cal B}_\xi$ equal to one in neighborhoods of $x$ and $\xi$, respectively. By Lemmas \ref{l13} and
\ref{l13a}, we have
\begin{equation} \label{1t8}
\sum_{j=1}^4 |\xi'|^{-\sigma_{j,\beta}}
\big| \partial_{x'}^\alpha \partial_{x_3}^\sigma\partial_{\xi'}^\beta\partial_{\xi_3}^\tau G_{i,j}(x,\xi)\big|
\le c \Big( \sum_{j=1}^3 \| \eta(\cdot) \partial_{x'}^\alpha \partial_{x_3}^\sigma G_{i,j}(x,\cdot)
  \|_{\cal H} + \| \eta(\cdot)\partial_{x'}^\alpha \partial_{x_3}^\sigma G_{i,4}(x,\cdot)
  \|_{L_2({\cal D})}\Big)
\end{equation}
for $i=1,2,3,4$. Let $F$ and $g$ be smooth functions, and let
\begin{eqnarray*}
u_i(y) & = & \int_{\cal D} \eta(z)\, F(z)\cdot \vec{H}_i(y,z) \, dz + \int_{\cal D}
  \eta(z)\, g(z)\, G_{i,4}(y,z)\, dz, \quad i=1,2,3,\\
p(y) & = & -\eta(y)\, g(y) + \int_{\cal D} \eta(z)\, F(z)\cdot \vec{H}_4(y,z) \, dz
  + \int_{\cal D} \eta(z)\, g(z)\, G_{4,4}(y,z)\, dz.
\end{eqnarray*}
By (\ref{GreenD}) and (\ref{Green2}), the vector $(u,p)$ is a solution of the problem
\[
b(u,v)-\int_{\cal D} p\nabla\cdot v\, dx = \int_{\cal D} \eta(y)\, F(y)\cdot v(y)\, dy \
  \mbox{ for all } v\in V,\  \ - \nabla\cdot u= \eta g \ \mbox{ in }{\cal D},\ \ S^\pm u =0 \
  \mbox{ on }\Gamma^\pm.
\]
Since $\eta F$ vanishes in ${\cal B}_x$, we conclude from Lemma \ref{l13} that
\[
|x'|^{-\sigma_{1,\alpha}} \big| \partial_{x'}^\alpha\partial_{x_3}^\sigma u(x)\big|
  + |x'|^{-\sigma_{4,\alpha}} \big| \partial_{x'}^\alpha\partial_{x_3}^\sigma p(x)\big|
  \le c\, \big( \|\zeta u\|_{\cal H} + \|\zeta p\|_{L_2({\cal D})}\big)
  \le  c\, \big( \| F\|_{V^*} + \| g\|_{L_2({\cal D})}\big),
\]
where $c$ is independent of $x$ and $\xi$. Consequently, the functionals mapping $(F,g)$ to
\[
|x'|^{-\sigma_{i,\alpha}} \partial_{x'}^\alpha\partial_{x_3}^\sigma u_i(x)
 = |x'|^{-\sigma_{i,\alpha}} \int_{\cal D} \eta(z)\,\Big(
  \sum_{j=1}^3 F_j(z)\, \partial_{x'}^\alpha\partial_{x_3}^\sigma G_{i,j}(x,z)
  + g(z)\, \partial_{x'}^\alpha\partial_{x_3}^\sigma G_{i,4}(x,z)\Big)\, dz,
\]
$i=1,2,3$, and
\[
|x'|^{-\sigma_{4,\alpha}}\partial_{x'}^\alpha\partial_{x_3}^\sigma p(x)
 = |x'|^{-\sigma_{4,\alpha}}
  \int_{\cal D} \eta(z)\,\Big( \sum_{j=1}^3 F_j(z)\, \partial_{x'}^\alpha\partial_{x_3}^\sigma
  G_{4,j}(x,z) +  g(z)\, \partial_{x'}^\alpha\partial_{x_3}^\sigma G_{4,4}(x,z)\Big)\, dz
\]
can be extended to linear and continuous functionals on $V^* \times L_2({\cal D})$. The norms of these
functionals are bounded by constants independent of $x$. Therefore,
\[
\sum_{j=1}^3 \| \eta(\cdot)\partial_{x'}^\alpha\partial_{x_3}^\sigma G_{i,j}(x,\cdot)\|_{\cal H}
  + \| \eta(\cdot)\partial_{x'}^\alpha \partial_{x_3}^\sigma G_{i,4}(x,\cdot)\|_{L_2({\cal D})}
  \le c\,  |x'|^{\sigma_{i,\alpha}}
\]
for $i=1,2,3,4.$ From this and from (\ref{1t8}) we obtain the
assertion of the theorem. \hfill $\Box$

\setcounter{equation}{0} \setcounter{Th}{0} \setcounter{Le}{0}
\setcounter{Co}{0} \setcounter{Rem}{0}
\section{The boundary value problem in a polyhedral cone}

For every $j=1,\ldots,n$ let $d_j$ be one of the natural numbers 0,1,2,3.
We consider the boundary value problem
\begin{eqnarray} \label{Stokes}
&& -\Delta u + \nabla p = f,\quad -\nabla\cdot u = g\ \ \mbox{in }{\cal K},\\
\label{bccone} && S_j u = h_j\, ,\quad N_j(u,p)= \phi_j \quad \mbox{on }\Gamma_j,\ j=1,\ldots,n.
\end{eqnarray}
Here $S_j$ is defined as
\[
S_j u=u \ \mbox{if }d_j=0,\quad S_j u=u_n=u\cdot n \ \mbox{if } d_j=2,\quad
S_j u=u_\tau=u-u_n n \ \mbox{if }d_j=1.
\]
The operators $N_j$ are defined as
\begin{eqnarray*}
&& N_j(u,p)=-p+2\varepsilon_{n,n}(u) \ \mbox{if }d_j=1,\qquad N_j(u,p)=\varepsilon_{n,\tau}(u) \
 \mbox{if } d_j=2, \\
&& N_j(u,p)=-pn+2\varepsilon_n(u) \ \mbox{if }d_j=3.
\end{eqnarray*}
In the case $d_j=0$ the condition $N_j(u,p)=\phi_j$ does not appear in (\ref{bccone}), whereas the
condition $S_ju=h_j$ does not appear if $d_j=3$.

\subsection{Weighted Sobolev spaces}

For an arbitrary point $x\in {\cal K}$ let $\rho(x)=|x|$ be the distance to the vertex of the cone,
 $r_j(x)$ the distance to the edge $M_j$, and $r(x)$ the distance to the set ${\cal S}=\{ 0\}
 \cup M_1\cup\cdots\cup M_n$ of all singular boundary points. Obviously, there exist positive constants
$c_1,c_2$ such that
\[
c_1 \, \rho(x)\prod_{j=1}^n \frac{r_j(x)}{\rho(x)} \le r(x) \le c_2 \, \rho(x)\prod_{j=1}^n
  \frac{r_j(x)}{\rho(x)}\quad\mbox{for all }x\in {\cal D}.
\]
Let $l$ be a nonnegative integer, $\beta\in {\Bbb R}$, $\delta=(\delta_1,\ldots,\delta_n)\in {\Bbb R}^n$,
By $V_{\beta,\delta}^{l}({\cal K})$ we denote the set of all functions with finite norm
\[
\| u\|_{V_{\beta,\delta}^{l}({\cal K})} = \Big( \int_{\cal K}
  \sum_{|\alpha|\le l} \rho^{2(\beta-l+|\alpha|)} \ \prod_{k=1}^n
  \big(\frac{r_k}{\rho}\big)^{2(\delta_k-l+|\alpha|)} \ |\partial_x^\alpha u|^2\, dx\Big)^{1/2}.
\]
The norm in the weighted Sobolev space $W_{\beta,\delta}^{l}({\cal K})$, where $\delta_k>-1$ for $k=1,\ldots,n$,
is defined as
\[
\| u\|_{W_{\beta,\delta}^{l}({\cal K})} = \Big( \int_{\cal K}
  \sum_{|\alpha|\le l} \rho^{2(\beta-l+|\alpha|)} \ \prod_{j=1}^n
  \big(\frac{r_j}{\rho}\big)^{2\delta_j} \ |\partial_x^\alpha u|^2\, dx\Big)^{1/2}.
\]
Furthermore, we introduce the following notation. If $d$ is real number,
then $W_{\beta,d}^{l,s}({\cal K})$ denotes the above introduced spaces with $\delta=(d,\ldots,d)$. If
$\delta=(\delta_1,\ldots,\delta_n)$ and $d$ is a real number, then we define
$W_{\beta,\delta+d}^{l,s}({\cal K})=W_{\beta,\delta'}^{l,s}({\cal K})$, where
$\delta'=(\delta_1+d,\ldots,\delta_n+d)$.
The trace space for $V_{\beta,\delta}^{l}({\cal K})$ and $W_{\beta,\delta}^{l}({\cal K})$ on the side
$\Gamma_j$ of ${\cal K}$ are denoted by $V_{\beta,\delta}^{l-1/2}(\Gamma_j)$ and
$W_{\beta,\delta}^{l-1/2}(\Gamma_j)$, respectively.

Note that $W_{\beta+1,\delta'}^{l+1}({\cal K})$ is continuously imbedded into
$W_{\beta,\delta}^{l}({\cal K})$ if $\delta'_j\le \delta_j+1$ for $j=1,\ldots,n$ (see
\cite[Le.4.1]{mr-01}). Obviously, $V_{\beta,\delta}^{l}({\cal K}) \subset W_{\beta,\delta}^{l}({\cal K})$.
If $\delta_k>l-1$ for $k=1,\ldots,n$, then $V_{\beta,\delta}^{l}({\cal K})=W_{\beta,\delta}^{l}({\cal K})$.
Moreover, if $u\in W_{\beta,\delta}^{l}({\cal K})$, $\delta_k>-1$, $\delta_k$ not integer for $k=1,\ldots,n$,
then for the inclusion $u\in V_{\beta,\delta}^{l}({\cal K})$ it is necessary and sufficient that
\[
(\partial_x^\alpha u)|_{M_k} =0 \quad\mbox{for }|\alpha|\le l-\delta_k-1.
\]
(cf. Lemma \ref{l0}). Using Lemma \ref{l11}, one can prove the following assertion
analogously to \cite[Le.4.2]{mr-01}.

\begin{Le} \label{l14}
Let $h_j \in W_{\beta,\delta}^{3/2}(\Gamma_j)^{3-d_j}$ and $\phi_j \in W_{\beta,\delta}^{1/2}
(\Gamma_j)^{d_j}$, $j=1,\ldots,n$, be given, where $\delta_j>0$ for all $j$. For
every $j=1,\ldots,n$ let $\Gamma_{j_+}$ and $\Gamma_{j_-}$ be the sides of ${\cal K}$ adjacent to the
edge $M_j$. We suppose that the functions $h_j$ satisfy the compatibility conditions
\begin{equation} \label{cc2}
\big( h_{j_+}|_{M_j},h_{j_-}|_{M_j}\big) \in R(T_j),
\end{equation}
where $R(T_j)$ is the range of the operator $T_j=(S_{j_+},S_{j_-})$ (cf. condition {\em (\ref{cc})}).
Then there exists a vector function $u\in W_{\beta,\delta}^{2}({\cal K})^3$ such that $S_j u=h_j$,
$N_j (u,0)=\phi_j$ on $\Gamma_j$, $j=1,\ldots,n$ and
\[
\| u\|_{W_{\beta,\delta}^{2}({\cal K})^3} \le
  c\, \sum_{j=1}^n \Big( \| h_j\|_{W_{\beta,\delta}^{3/2}(\Gamma_j)^{3-d_j}}
  + \|\phi_j \|_{W_{\beta,\delta}^{1/2}(\Gamma_j)^{d_j}}\Big)
\]
with a constant $c$ independent of $h_j$ and $\phi_j$.
\end{Le}

\subsection{Operator pencils generated by the boundary value problem}

We introduce the following operator pencils ${\mathfrak A}$ and $A_j$.

1. Let $\Gamma_{k_\pm}$ be the sides of ${\cal K}$ adjacent to the edge $M_k$, and let $\theta_k$ be the
angle at the edge $M_k$. By $A_k(\lambda)$ we denote the operator pencil introduced in Section 2.9, where
$S^\pm=S_{k_\pm}$ and $N^\pm=N_{k_\pm}$. Furthermore, let $\lambda_1^{(k)}$ denote the eigenvalue with smallest
positive real part of this pencil, while $\lambda_2^{(k)}$ is the eigenvalue with smallest real part greater
than 1. Finally, we define
\begin{equation} \label{defmuk}
\mu_k = \left\{ \begin{array}{ll} \mbox{Re}\, \lambda_1^{(k)} & \mbox{if }d_{k_+} +d_{k_-} \mbox{ is odd }\
  \mbox{ or }  \ d_{k_+} + d_{k_-} \mbox{ is even and } \theta_k \ge \pi/m_k, \\
  \mbox{Re}\, \lambda_2^{(k)} & \mbox{if } d_{k_+} + d_{k_-} \mbox{ is even and } \theta_k < \pi/m_k, \end{array}\right.
\end{equation}
where $m_k=1$ if $d_{k_+} = d_{k_-}$, $m_k=2$ if $d_{k_+} \not= d_{k_-}$. \\

2. Let $\rho=|x|$, $\omega=x/|x|$, $V_\Omega = \{ u\in W^1(\Omega)^3:\, S_j u=0$ on $\gamma_j$ for
$j=1,\ldots,n\}$, and
\[
a\Big( \Big( \begin{array}{c} u \\ p\end{array}\Big),\Big( \begin{array}{c} v \\ q\end{array}\Big);\lambda\Big)
  = \frac{1}{\log 2}\, \int\limits_{\substack{{\cal K}\\ 1<|x|<2}} \Big( 2\sum_{i,j=1}^3 \varepsilon_{i,j}(U)
  \cdot \varepsilon_{i,j}(V) - P\nabla\cdot V - (\nabla\cdot U)\, Q\Big)\, dx,
\]
where $U=\rho^\lambda u(\omega)$, $V=\rho^{-1-\lambda} v(\omega)$, $P=\rho^{\lambda-1} p(\omega)$,
$Q=\rho^{-2-\lambda}q(\omega)$, $u,v \in V_\Omega$, $p,q\in L_2(\Omega)$, and $\lambda \in {\Bbb C}$.
The bilinear form $a(\cdot,\cdot;\lambda)$ generates the linear and continuous operator
\[
{\mathfrak A}(\lambda):\, V_\Omega \times L_2(\Omega) \to V_\Omega^*\times L_2(\Omega)
\]
by
\[
\int_\Omega {\mathfrak A}(\lambda)\Big( \begin{array}{c} u \\ p\end{array}\Big) \cdot
  \Big( \begin{array}{c} v \\ q\end{array}\Big)\, d\omega
  = a\Big( \Big( \begin{array}{c} u \\ p\end{array}\Big),\Big( \begin{array}{c} v \\ q\end{array}\Big)
  ;\lambda\Big), \quad   u,v\in V_\Omega ,\ p,q \in L_2(\Omega).
\]
As is known, the spectrum of the pencil ${\mathfrak A}(\lambda)$ consists of isolated points, the eigenvalues
of this pencil. Detailed information on the spectrum can be found in \cite[Sec.5,6]{kmr2}.

Let $V_\delta^l(\Omega)$ and $W_{\delta}^l(\Omega)$ be the weighted Sobolev spaces with the norms
\begin{eqnarray*}
&&\| u\|_{V_{\delta}^l(\Omega)} = \Big( \int\limits_{\substack{{\cal K}\\ 1<|x|<2}} \sum_{|\alpha|\le l}
  \prod_{j=1}^n r_j^{2(\delta_j-l+|\alpha|)} \, |\partial_x^\alpha u(x)|^2\, dx\Big)^{1/2},\\
&&\| u\|_{W_{\delta}^l(\Omega)} = \Big( \int\limits_{\substack{{\cal K}\\ 1<|x|<2}} \sum_{|\alpha|\le l}
  \prod_{j=1}^n r_j^{2\delta_j} \, |\partial_x^\alpha u(x)|^2\, dx\Big)^{1/2}
\end{eqnarray*}
where the function $u$ is extended by $u(x)=u(x/|x|)$ to ${\cal K}$. The corresponding trace spaces on the
side $\gamma_j$ are denoted by $V_{\delta}^{l-1/2}(\gamma_j)$ and $W_{\delta}^{l-1/2}(\gamma_j)$, respectively.
We consider the restriction of the operator ${\mathfrak A}(\lambda)$ to the space $W_\delta^2(\Omega)^3
\times W_{\delta}^1(\Omega)$. By ${\mathfrak A}_{\delta}(\lambda)$, we denote the operator
\begin{eqnarray*}
W_{\delta}^2(\Omega)^3 \times W_{\delta}^1(\Omega) \ni (u,p) & \to &
   \big( f,g,\{h_j\},\{\phi_j\} \big) \\
&&\in W_{\delta}^0(\Omega)^3 \times W_{\delta}^1(\Omega)
   \times \prod_{j=1}^n W_{\delta}^{3/2}(\gamma_j)^{3-d_j} \times \prod_{j=1}^n
   W_{\delta}^{1/2}(\gamma_j)^{d_j}
\end{eqnarray*}
where $f(\omega)= \rho^{2-\lambda}(-\Delta U+ \nabla P)$, $g(\omega)=-\rho^{1-\lambda}\nabla\cdot U$,
$h_j=S_j u$, $\phi_j = \rho^{1-\lambda}N_j(U,P)$, and $U,P$ are as above. It can be proved that the spectra of
the pencils ${\mathfrak A}(\lambda)$ and ${\mathfrak A}_{\delta} (\lambda)$ coincide if
$\max(0,1-\mu_k)<\delta_k<1$ for $k=1,\ldots,n$. Furthermore, there exist positive constants $N$ and $\varepsilon$ such that for
all $\lambda$ in the set $\{ \lambda\in {\Bbb C}:\ |\lambda|>N, \ |\mbox{Re}\,
\lambda|<\varepsilon |\mbox{Im}\, \lambda\}$ the operator ${\mathfrak A}_{\delta}(\lambda)$ is an
isomorphism onto the subset of all $\big( f,g,\{h_j\},\{\phi_j\} \big)$ satisfying the
compatibility conditions $(h_{j_+},h_{j_-}) \in R(T_j)$ on the corners $M_j\cap S^2$ of $\Omega$. For every
$\lambda$ in this set und every solution $(u,p)\in W_{\delta}^2(\Omega)^3 \times
W_{\delta}^1(\Omega)$ of the equation ${\mathfrak A}_{\delta}(\lambda)\, (u,p)
=\big( f,g,\{h_j\},\{\phi_j\} \big)$ the estimate
\begin{eqnarray} \label{estimate} \nonumber
&& \hspace{-4em}\sum_{j=0}^2 |\lambda|^{2-j}\, \| u\|_{W_{\delta}^j(\Omega)^3} +
  \sum_{j=0}^1 |\lambda|^{1-j}\, \| p\|_{W_{\delta}^j(\Omega)} \le c\, \Big(
  \| f\|_{W_{\delta}^0(\Omega)^3} + \sum_{j=0}^1 |\lambda|^{1-j}\, \| g\|_{W_{\delta}^j(\Omega)} \\
&& \hspace{-4em} + \sum_{j=1}^n \Big( \| h_j \|_{W_{\delta}^{3/2}(\gamma_j)^{3-d_j}}
  + |\lambda|^{3/2}\, \| h_j\|_{W_{\delta}^0(\gamma_j)^{3-d_j}}+ \| \phi_j
  \|_{W_{\delta}^{1/2}(\gamma_j)^{d_j}} + |\lambda|^{1/2}\, \| \phi_j\|_{W_{\delta}^0(\gamma_j)^{d_j}}
  \Big)\Big)
\end{eqnarray}
holds with a constant $c$  independent of $(u,p)$ and $\lambda$.
For the proof we refer to \cite[Th.3.2]{mr-01}.

\subsection{Solvability of the boundary value problem}

The following theorem can be proved in a standard way (see, e.g.,
\cite[Ch.6]{kmr1}) using the estimate (\ref{estimate}).

\begin{Th} \label{t9}
{\em 1)} Suppose that there are no eigenvalues of the pencil ${\mathfrak A}$ on the line
$\mbox{\em Re}\, \lambda = -\beta+1/2$ and that the components of $\delta$ satisfy the inequalities
$\max(1-\mu_k,0) < \delta_k < 1$. Then the boundary value problem {\em (\ref{Stokes}), (\ref{bccone})}
is uniquely solvable in $W_{\beta,\delta}^2({\cal K})^3\times W_{\beta,\delta}^1({\cal K})$
for arbitrary $f \in W_{\beta,\delta}^{0}({\cal K})^3$, $g \in W_{\beta,\delta}^{0}({\cal K})$,
$h_j \in W_{\beta,\delta}^{3/2}(\Gamma_j)^{3-d_j}$ satisfying {\em (\ref{cc2})}, and
$\phi_j \in W_{\beta,\delta}^{3/2}(\Gamma_j)^{d_j}$.

{\em 2)} Let $(u,p)\in W_{\beta,\delta}^2({\cal K})^3\times W_{\beta,\delta}^1({\cal K})$
be a solution of the boundary value problem {\em (\ref{Stokes}), (\ref{bccone})}, where $f \in
W_{\beta',\delta'}^{0}({\cal K})^\ell$, $g\in W_{\beta',\delta'}^1({\cal K})$ $h_j \in
W_{\beta',\delta'}^{3/2}(\Gamma_j)^{3-d_j}$, and $\phi_j \in W_{\beta',\delta'}^{1/2}
(\Gamma_k)^{d_j}$. Suppose that the components of $\delta$ and $\delta'$ satisfy the
inequality $\max(1-\mu_k,0)<\delta'_k\le \delta_k<1$. If there are no eigenvalues of the pencil
${\mathfrak A}$ on the lines $\mbox{\em Re}\, \lambda = -\beta+1/2$ and $\mbox{\em Re}\,
\lambda =-\beta'+1/2$, then
\begin{equation} \label{zerlegung}
(u,p) =  \sum_{\nu=1}^N \sum_{j=1}^{I_\nu} \sum_{s=0}^{\kappa_{\nu,j}-1}
  c_{\nu,j,s} \sum_{\sigma=0}^s \frac{1}{\sigma!}\ (\log \rho)^\sigma\, \big( \rho^{\lambda_\nu}
  u^{(\nu,j,s-\sigma)}(\omega), \rho^{\lambda_\nu-1} p^{(\nu,j,s-\sigma)}(\omega)\big)+ (w,q),
\end{equation}
where $(w,q) \in W_{\beta',\delta'}^2({\cal K})^3\times W_{\beta',\delta'}^1({\cal K})$
is a solution of problem {\em (\ref{Stokes})--(\ref{bccone})}, $\lambda_\nu$ are the eigenvalues
of the pencil ${\mathfrak A}$ between the lines $\mbox{\em Re}\, \lambda = -\beta+1/2$ and
$\mbox{\em Re}\, \lambda =-\beta'+1/2$ and $\big( u^{(\nu,j,s)},p^{(\nu,j,s)}\big)$ are eigenvectors
and generalized eigenvectors corresponding to the eigenvalue $\lambda_\nu$.
\end{Th}

Furthermore, analogously to \cite[Le.4.3]{mr-01}, the following assertion holds.

\begin{Le} \label{l15}
Let $(u,p) \in W_{\beta,\delta}^2({\cal K})^3 \times W_{\beta,\delta}^1({\cal K})$ be a solution of problem
{\em (\ref{Stokes}), (\ref{bccone})}. We assume that $(\rho\partial_\rho)^\nu f\in W_{\beta,\delta}^0
({\cal K})$, $(\rho\partial_\rho)^\nu g \in W_{\beta,\delta}^1({\cal K})$,
$(\rho\partial_\rho)^\nu h_j\in W_{\beta,\delta}^{3/2}(\Gamma_j)^{3-d_j}$, and $(\rho\partial_\rho)^\nu
\phi_j\in W_{\beta,\delta}^{1/2}(\Gamma_j)$ for $\nu=0,\ldots,l$, $j=1,\ldots,n$. If
$\max(1-\mu_k,0)<\delta_k<1$ for $j=1,\ldots,n$ and the line $\mbox{\em Re}\, \lambda=-\beta+1/2$ is free of
eigenvalues of the pencil ${\mathfrak A}(\lambda)$, then $(\rho\partial_\rho)^\nu (u,p)\in
W_{\beta,\delta}^2({\cal K})^3 \times W_{\beta,\delta}^1({\cal K})$ for $\nu=0,\ldots,l$.
\end{Le}

\subsection{Existence of weak solutions}

Let $V_\beta=\{ u\in W_{\beta,0}^1({\cal K})^3:\ S_j u=0$ on $\Gamma_j$ for $j=1,\ldots,n\}$, and let the
operator ${\cal A}_\beta$ be defined as the mapping
\[
V_\beta \times W_{\beta,0}^0({\cal K}) \ni (u,p)\to (F,g)\in V_{-\beta}^* \times W_{\beta,0}^0({\cal K}),
\]
where
\[
F(v) = b(u,v)-\int_{\cal K}p\, \nabla\cdot v \, dx \ \mbox{ for all }v\in V_{-\beta}^*\quad\mbox{and}\quad
  g=-\nabla\cdot u.
\]

\begin{Le} \label{l16}
For arbitrary $u\in V_\beta$, $p\in V_\beta^0({\cal K})$, $(F,g)={\cal A}_\beta (u,p)$ there is the estimate
\[
\| u\|_{W_{\beta,0}^1({\cal K})^3} + \| p\|_{W_{\beta,0}^0({\cal K})} \le c\, \Big(
  \| F\|_{V_{-\beta}^*} + \| p\|_{W_{\beta,0}^0({\cal K})} +
  \| u\|_{W_{\beta-1,0}^0({\cal K})^3} + \| p\|_{W_{\beta-1,0}^{-1}({\cal K})} \Big)
\]
with a constant $c$ independent of $u$ and $p$. Here $W_{\beta-1,0}^{-1}({\cal K})$
denotes the dual space of $W_{1-\beta,0}^1({\cal K})$.
\end{Le}

The proof of this lemma proceeds analogously to \cite[Le.4.4]{mr-01}.

\begin{Th} \label{t10}
The operator ${\cal A}_\beta$ is an isomorphism if there are no eigenvalues of the pencil
${\mathfrak A}(\lambda)$ on the line $\mbox{\em Re}\, \lambda=-\beta-1/2$.
\end{Th}

\noindent P r o o f. We show first that
\begin{equation} \label{1t10}
\| u\|_{W_{\beta,0}^1({\cal K})^3} + \| p\|_{W_{\beta,0}^0({\cal K})} \le c\, \Big(
  \| F\|_{V_{-\beta}^*} +  \| g\|_{W_{\beta,0}^0({\cal K})}\Big)
\end{equation}
for all $u \in V_\beta$, $p\in W_{\beta,0}^0({\cal K})$, $(F,g)={\cal A}_\beta (u,p)$. Let $u\in
V_\beta \subset W_{\beta-1,\varepsilon-1}^0({\cal K})^3$, $p\in W_{\beta,0}^0({\cal K})$,
$w\in W_{1-\beta,1-\varepsilon}^0({\cal K})^3$, and $\psi \in W_{1-\beta,1-\varepsilon}^1({\cal K})$
with sufficiently small positive $\varepsilon$. By Theorem \ref{t9}, there exists a solution $(v,q) \in
W_{1-\beta,1-\varepsilon}^2({\cal K})^3 \times W_{1-\beta,1-\varepsilon}^1({\cal K})$ of the problem
\[
- \Delta v -\nabla\nabla\cdot v + \nabla q =w,\ \-\nabla\cdot v=\psi\ \mbox{ in }{\cal K},
  \quad S_j v=0,\ N_j(v,q)=0 \ \mbox{ on }\Gamma_j,\ j=1,\ldots,n
\]
satisfying the estimate
\[
\|v\|_{W_{1-\beta,1-\varepsilon}^2({\cal })^3} + \| p\|_{W_{1-\beta,1-\varepsilon}^0({\cal K})}
   \le c\, \Big( \| w\|_{W_{1-\beta,1-\varepsilon}^0({\cal K})^3}
   + \|\psi \|_{W_{1-\beta,1-\varepsilon}^1({\cal K})}\Big).
\]
From Green's formula it follows that
\[
\int_{\cal K} u\cdot w\, dx + \int_{\cal K} p\psi\, dx = b(u,v) -\int_{\cal K} p\, \nabla\cdot v \, dx
  - \int_{\cal K} (\nabla\cdot u)\, q\, dx = F(v) + \int_{\cal K} g\, q\, dx.
\]
Hence,
\begin{eqnarray*}
&& \Big| \int_{\cal K} u\cdot w\, dx + \int_{\cal K} p\psi\, dx \Big| \le
  \| {\cal A}_\beta (u,p)\|_{V_{-\beta}^* \times W_{\beta,0}^0({\cal K})}
  \, \Big( \| v\|_{W_{-\beta,0}^1({\cal K})^3} + \| q\|_{W_{-\beta,0}^0({\cal K})}\Big) \\
&& \le c\, \| {\cal A}_\beta (u,p)\|_{V_{-\beta}^* \times W_{\beta,0}^0({\cal K})} \, \Big( \|
  v\|_{W_{1-\beta,1-\varepsilon}^2({\cal K})^3}+\| q \|_{W_{1-\beta,1-\varepsilon}^0({\cal K}))} \Big) \\
&& \le c\, \| {\cal A}_\beta (u,p)\|_{V_{-\beta}^*\times W_{\beta,0}^0({\cal K})} \, \Big(
  \| w\|_{W_{1-\beta,1-\varepsilon}^0({\cal K})^3}+\|\psi \|_{W_{1-\beta,1-\varepsilon}^1({\cal K})}\Big).
\end{eqnarray*}
Setting $w=\rho^{2(\beta-1)}\prod (r_j/\rho)^{2(\varepsilon-1)}u$ and $\psi=0$, we obtain
\[
\| u\|^2_{W_{\beta-1,\varepsilon-1}^0({\cal K})^3} \le c\, \| {\cal A}_\beta (u,p)
  \|_{V_{-\beta}^*\times W_{\beta,0}^0({\cal K})}\, \| u\|_{W_{\beta-1,\varepsilon-1}^0({\cal K})^3}
\]
and, therefore,
\[
\| u\|_{W_{\beta-1,0}^0({\cal K})^3} \le c\, \| u\|_{W_{\beta-1,\varepsilon-1}^0({\cal K})^3}
  \le c\, \| {\cal A}_\beta (u,p)\|_{V_{-\beta}^*\times W_{\beta,0}^0({\cal K})}.
\]
Analogously, for $w=0$ and arbitrary $\psi \in W_{1-\beta,0}^1({\cal K})$ we get
\[
\Big| \int_{\cal K} p\psi\, dx\Big| \le c\, \| {\cal A}_\beta (u,p)
  \|_{V_{-\beta}^*\times W_{\beta,0}^0({\cal K})} \, \| \psi\|_{W_{1-\beta,1-\varepsilon}^1({\cal K})}
  \le c\, \| {\cal A}_\beta (u,p) \|_{V_{-\beta}^*\times W_{\beta,0}^0({\cal K})} \,
  \| \psi\|_{W_{1-\beta,0}^1({\cal K})}
\]
what implies the inequality
\[
\| p\|_{W_{\beta-1,0}^{-1}({\cal K})}\le c\, \| {\cal A}_\beta (u,p)
  \|_{V_{-\beta}^*\times W_{\beta,0}^0({\cal K})}
\]
Using Lemma \ref{l16}, we arrive at (\ref{1t10}).

From (\ref{1t10}) it follows that ${\cal A}_\beta$ is injective and its range is closed.
By Theorem \ref{t9}, the range of ${\cal A}_\beta$ contains the set $W_{\beta-1,\varepsilon-1}^0
({\cal K})^3 \times W_{\beta-1,\varepsilon-1}^1({\cal K})$ which is dense in $V_{-\beta}^* \times
W_{\beta,0}^0({\cal K})$. This proves the theorem. \hfill $\Box$ \\

Let $F\in V_{-\beta}^*$, $g\in W_{\beta,0}^0({\cal K})$, and let $h_j \in W_{\beta,0}^{1/2}(\Gamma_j)$
be such that there exists a vector function $w\in W_{\beta,0}^1({\cal K})^3$ satisfying the
boundary condition $S_jw=h_j$ on $\Gamma_j$. By a weak solution $(u,p)\in W_{\beta,0}^1({\cal K})
\times W_{\beta,0}^0({\cal K})$ of problem (\ref{Stokes}), (\ref{bccone}) we mean a pair $(u,p)$
satisfying
\begin{eqnarray} \label{ws1}
&& b(u,v) - \int_{\cal K} p\, \nabla\cdot v\, dx = F(v) \quad\mbox{for all }v\in V_{-\beta} \\
\label{ws2} && -\nabla\cdot u=g \ \mbox{ in }{\cal K},\quad S_j u=h_j\ \mbox{ on }\Gamma_j,\ j=1,\ldots,n.
\end{eqnarray}
If $g\in W_{\beta+1,\delta}^1({\cal K})$ with $\delta_j<1$, $j=1,\ldots,n$, and the functional $F$ has the form
\begin{equation} \label{functF}
F(v) = \int_{\cal K} (f+\nabla g)\, v\, dx+ \sum_{j=1}^n
\int_{\Gamma_j} \phi_j \cdot v\, dx,
\end{equation}
with $f\in W_{\beta+1,\delta}^0({\cal K})$, $\phi_j \in W_{\beta+1,\delta}^{1/2}(\Gamma_j)$,
then $(u,p)$ is a strong solution of problem (\ref{Stokes}), (\ref{bccone}). Theorem
\ref{t10} ensures the existence and uniqueness of weak solutions provided the line $\mbox{Re}\, \lambda
=-\beta-1/2$ does not contain eigenvalues of the pencil ${\mathfrak A}(\lambda)$.

\subsection{Regularity of weak solutions}

Analogously to \cite[Th.4.4]{mr-01}, the following result holds.
The proof is essentially based on Theorem \ref{t4}.

\begin{Th} \label{t11}
Let $(u,p)\in W_{\beta-l+1,0}^1({\cal K})^3 \times W_{\beta-l+1,0}^0({\cal K})$ be a solution of problem
{\em (\ref{ws1}), (\ref{ws2})}. Suppose that the functional $F\in V_{-\beta+l-1}^*$ has the form
{\em (\ref{functF})}, where
\[
f\in W_{\beta,\delta}^{l-2}({\cal K})^3, \ \ g\in W_{\beta,\delta}^{l-1}({\cal K}),\ \
  \phi_j\in W_{\beta,\delta}^{l-3/2}(\Gamma_j)^{d_j},
\]
$l\ge 2$, $\delta_k$ is not integer, $\max(l-1-\mbox{\em Re}\, \lambda_1^{(k)},0)<\delta_k<l-1$.
Suppose further that the vector functions $h_j\in W_{\beta,\delta}^{l-1/2}(\Gamma_j)^{d_j}$
satisfy the compatibility condition {\em (\ref{cc2})} on the edges $M_j$. Then
$u\in W_{\beta,\delta}^l ({\cal K})^3$ and $p\in W_{\beta,\delta}^{l-1}({\cal K})$.
\end{Th}

\begin{Co} \label{c7}
Let the assumptions of Theorem {\em \ref{t11}} be valid. Additionally, we assume that $(\rho\partial_\rho)^\nu
f\in W_{\beta,\delta}^{l-2}({\cal K})^3$, $(\rho\partial_\rho)^\nu g\in W_{\beta,\delta}^{l-1}({\cal K})$,
$(\rho\partial_\rho)^\nu h_j\in W_{\beta,\delta}^{l-1/2}(\Gamma_j)^{3-d_j}$, $(\rho\partial_\rho)^\nu
\phi_j\in W_{\beta,\delta}^{l-3/2}(\Gamma_j)^{d_j}$ for $\nu=1,\ldots,N$ and $j=1,\ldots,n$.
Then $(\rho\partial_\rho)^\nu u\in W_{\beta,\delta}^l({\cal K})^3$ and $(\rho\partial_\rho)^\nu p
\in W_{\beta,\delta}^{l-1}({\cal K})$ for $\nu=1,\ldots,N$.
\end{Co}

{\it Proof}: It suffices to prove the assertion for $N=1$. For $N>1$ it can proved by induction.
Let first $\delta_k>l-2$ and, therefore, $\max(1-\mbox{Re}\, \lambda_1^{(k)},0)<\delta_k-l+2<1$
for $k=1,\ldots,n$. Then from Theorem \ref{t11} and Lemma \ref{l15} it follows that $\rho\partial_\rho u\in
W_{\beta-l+2,\delta-l+2}^2({\cal K})^3$ and $\rho\partial_\rho p\in
W_{\beta-l+2,\delta-l+2}^1({\cal K})$. Using Theorem \ref{t11}, the equations
\begin{equation} \label{1c7}
-\Delta (\rho\partial_\rho u) + \nabla(\rho\partial_\rho p) = (\rho\partial_\rho+2)f -\nabla p
  \in W_{\beta,\delta}^{l-2}({\cal K})^3,\quad
-\nabla\cdot (\rho\partial_\rho u)= (\rho\partial_\rho+1)\, g
  \in W_{\beta,\delta}^{l-1}({\cal K})
\end{equation}
and analogous equations for $S_j(\rho\partial_\rho u)$ and $N_j(\rho\partial_\rho u,\rho\partial_\rho p)$,
we obtain $\rho\partial_\rho u\in W_{\beta,\delta}^l({\cal K})^3$ and $\rho\partial_\rho p
\in W_{\beta,\delta}^{l-1}({\cal K})$.

Now let $\delta_k<l-2$ for $k=1,\ldots,n$. By Theorem \ref{t11}, we have $(u,p) \in W_{\beta,\delta}^l
({\cal K})^3\times W_{\beta,\delta}^{l-1}({\cal K})$ and, consequently, $\rho\partial_\rho (u,p)
\in W_{\beta-1,\delta}^{l-1}({\cal K})^3\times W_{\beta-1,\delta}^{l-2}({\cal K})$. Using again
Theorem \ref{t11} and (\ref{1c7}), we obtain $\rho\partial_\rho (u,p)\in W_{\beta,\delta}^l
({\cal K})^3\times W_{\beta,\delta}^{l-1}({\cal K})$.

Finally, let $\delta_k<l-2$ for some, but not all, $k$. Then let $\psi_1,\ldots,\psi_n$
be smooth functions on $\overline{\Omega}$ such that $\psi_j\ge 0$, $\psi_j=1$ near $M_j\cap S^2$, and $\sum
\psi_j =1$. We extend $\psi_j$ to ${\cal K}$ by the equality $\psi_j(x)=\psi_j(x/|x|)$. Then
$\partial_x^\alpha \psi_j(x) \le c\, |x|^{-|\alpha|}$. Consequently, the assumptions of the
corollary are satisfied for $(\psi_j u,\psi_j p)$, and from what has been shown above it follows that
$\rho \partial_\rho (\psi_j u,\psi_j p)\in W_{\beta,\delta}^l({\cal K})^3 \times
W_{\beta,\delta}^{l-1}({\cal K})$ for $j=1,\ldots,n$. This completes the proof. \rule{1ex}{1ex}

\begin{Rem} \label{r6}
{\em If $d_{k_+}+d_{k_-}$ is even and, moreover, $\theta_k<\pi$ for $d_{k_+}=d_{k_-}$,
$\theta_k <\pi/2$ for $d_{k_+}\not=d_{k_-}$, then the number $\mbox{Re}\, \lambda_1^{(k)}$
in the condition on $\delta_k$ of Theorem \ref{t11} and Corollary \ref{c7}
is equal to 1 and can be replaced by $\mu_k$. However, if $\delta_k<l-2$, then $h_{k_\pm}$,
$\phi_{k_\pm}$ and $g$ must satisfy certain additional compatibility condition on the edge
$M_k$ (cf. Theorem \ref{t6} and Remark \ref{r2}).}
\end{Rem}

In the case when $\lambda=0$ is not an eigenvalue of the pencils $A_k(\lambda)$, we obtain
a regularity result analogous to Corollary \ref{c7} in the class of the weighted spaces
$V_{\beta,\delta}^l$. In the following corollary, let $\zeta_k,\eta_k$ be smooth cut-off
functions on $\bar{\Omega}$ equal to one near the corner
$\bar{\Omega}\cap M_k$ such that $\eta_k=1$ in a neighborhood of supp$\, \zeta_k$ and
$\eta_k=0$ in a neighborhood of the corners $\bar{\Omega}\cap M_j$, $j\not=k$. We extend
$\zeta_k$ and $\eta_k$ by
\[
\zeta_k(x)=\zeta_k(x/|x|),\quad  \eta_k(x)=\eta_k(x/|x|)
\]
to functions on ${\cal K}$.

\begin{Co} \label{c8}
Let $(u,p)\in W_{\beta-l+1,0}^1({\cal K})^3 \times W_{\beta-l+1,0}^0({\cal K})$
be a solution of problem {\em (\ref{ws1}), (\ref{ws2})}, where $l\ge 2$, $F\in V_{-\beta+l-1}^*$ has the form
{\em (\ref{functF})},
$
\eta_k (\rho\partial_\rho)^\nu f\in V_{\beta,\delta}^{l-2}({\cal K})^3, \ \
\eta_k (\rho\partial_\rho)^\nu g\in V_{\beta,\delta}^{l-1}({\cal K}),\ \
\eta_k (\rho\partial_\rho)^\nu h_j\in V_{\beta,\delta}^{l-1/2}(\Gamma_j)^{3-d_j}, \ \
\eta_k (\rho\partial_\rho)^\nu \phi_j\in V_{\beta,\delta}^{l-3/2}(\Gamma_j)^{d_j},
$
for $\nu=1,\ldots,N$.
Suppose that $\delta_k$ is positive and not integer and that the strip $0\le \mbox{\em Re}\, \lambda\le l-1-\delta_k$
does not contain eigenvalues of the pencil $A_k(\lambda)$.
Then $\zeta_k (\rho\partial_\rho)^\nu u\in V_{\beta,\delta}^l({\cal K})^3$ and $\zeta_k (\rho\partial_\rho)^\nu p
\in V_{\beta,\delta}^{l-1}({\cal K})$ for $\nu=1,\ldots,N$.
\end{Co}

\noindent P r o o f.
From Corollary \ref{c7} it follows that $\zeta_k (\rho\partial_\rho)^\nu u\in W_{\beta,\delta}^l({\cal K})^3$
and $\zeta_k (\rho\partial_\rho)^\nu p\in W_{\beta,\delta}^{l-1}({\cal K})$ for $\nu=1,\ldots,N$.
Moreover, due to Lemma \ref{l12b}, we have $\chi\zeta_k (\rho\partial_\rho)^\nu u\in V_{\beta,\delta}^l({\cal K})^3$
and $\chi\zeta_k (\rho\partial_\rho)^\nu p\in V_{\beta,\delta}^{l-1}({\cal K})$ for $\nu=1,\ldots,N$ and
for every smooth function $\chi$ with compact support vanishing near the vertex of the cone. Therefore,
\[
\partial_x^\alpha \zeta_k(\rho\partial_\rho)^\nu u =0 \ \mbox{on }M_k \ \mbox{for }|\alpha|< l-1-\delta_1, \quad
\partial_x^\alpha \zeta_k(\rho\partial_\rho)^\nu p =0 \ \mbox{on }M_k \ \mbox{for }|\alpha|< l-2-\delta_1.
\]
This implies $\zeta_k (\rho\partial_\rho)^\nu u\in V_{\beta,\delta}^l({\cal K})^3$ and $\zeta_k (\rho\partial_\rho)^\nu p
\in V_{\beta,\delta}^{l-1}({\cal K})$ for $\nu=1,\ldots,N$. \hfill $\Box$\\

The following theorem can be proved analogously to \cite[Cor.4.3]{mr-01} by means of the the second part
of Theorem \ref{t9}.

\begin{Th} \label{t12}
Let $(u,p)\in W_{\beta,0}^1({\cal K})^3 \times W_{\beta,0}^0({\cal K})$ be solution
of problem {\em (\ref{ws1}), (\ref{ws2})}, where $g\in W_{\beta,0}^0({\cal K})\cap
W_{\beta',0}^0({\cal K})$, $h_j\in W_{\beta,0}^{1-1/2}(\Gamma_j) \cap W_{\beta',0}^{1-1/2}(\Gamma_j)$,
and $F\in V_{-\beta}^*\cap V_{-\beta'}^*$. If  there are no eigenvalues of the pencil ${\mathfrak A}
(\lambda)$ on the lines $\mbox{\em Re}\, \lambda=-\beta-1/2$ and $\mbox{\em Re}\, \lambda=-\beta'-1/2$,
then $(u,p)$ admits the decomposition
{\em (\ref{zerlegung})}, where $w\in W_{\beta',0}^1({\cal K})^3$, $q\in W_{\beta',0}^0
({\cal K})$, and $\lambda_\nu$ are the eigenvalues of ${\mathfrak A}(\lambda)$ between the lines
$\mbox{\em Re}\, \lambda=-\beta-1/2$ and $\mbox{\em Re}\, \lambda=-\beta'-1/2$.
\end{Th}

\subsection{Estimates of Green's matrix}

A matrix $G(x,\xi)=\big( G_{i,j}(x,\xi)\big)_{i,j=1}^4$ is called
Green's matrix for problem (\ref{Stokes}), (\ref{bccone}) if
\begin{eqnarray}
&& -\Delta_x \vec{G}_j(x,\xi)+ \nabla_x G_{4,j}(x,\xi)= \delta(x-\xi)\,
  (\delta_{1,j},\delta_{2,j},\delta_{3,j})^t \quad \mbox{for }x,\xi\in {\cal K}, \label{g1}\\ \label{g2}
&& -\nabla_x\cdot \vec{G}_j(x,\xi) = \delta_{4,j}\, \delta(x-\xi) \quad \mbox{for }x,\xi\in {\cal K},\\
&&  S_k \vec{G}_j(x,\xi)=0,\quad N_k(\partial_x)\, \big(\vec{G}_j(x,\xi),G_{4,j}(x,\xi)\big)=0
  \quad \mbox{for }x\in \Gamma_k,\ \xi\in {\cal K},\label{g3}
\end{eqnarray}
$k=1,\ldots,n$. Here $\vec{G}_j$ denotes the vector with the components $G_{1,j},G_{2,j},G_{3,j}$.

\begin{Th} \label{t13}
{\em 1)} Suppose that the line $\mbox{\em Re}\, \lambda=-\beta-1/2$ is free of eigenvalues of the pencil
${\mathfrak A}(\lambda)$. Then there exists a unique Green matrix $G(x,\xi)$ such that the function
$x\to \zeta(|x-\xi|/r(\xi))\, G_{i,j}(x,\xi)$ belongs to $W_{\beta,0}^1({\cal K})$ for $i=1,2,3$ and to
$W_{\beta,0}^0({\cal K})$ for $i=4$, where $\zeta$ is an arbitrary smooth function on $(0,\infty)$ equal
to one in $(1,\infty)$ and to zero in $(0,\frac 12)$.

{\em 2)} There are the equalities
\begin{equation} \label{1t13}
G_{i,j}(tx,t\xi) = t^{-T} G_{i,j}(x,\xi)  \ \mbox{ for }x,\xi\in {\cal K},\ t>0,
\end{equation}
where $T=1+\delta_{i,4}+\delta_{j,4}$.

{\em 3)} The functions $G_{i,j}(x,\xi)$ are infinitely differentiable with respect to $x,\xi \in
\bar{\cal K} \backslash {\cal S}$, $x\not= \xi$, and satisfy the following estimates
for $|x|/2<|\xi|<2|x|$.
\begin{eqnarray*}
&& \hspace{-4em}\big| \partial_x^\alpha\partial_\xi^\gamma G_{i,j}(x,\xi)\big|\le c\, |x-\xi|^{-T-|\alpha|-|\gamma|}
  \ \mbox{ if } |x-\xi|<\min(r(x),r(\xi)),\\
&& \hspace{-4em}\big| \partial_{x}^\alpha \partial_{\xi}^\gamma G_{i,j}(x,\xi) \big| \le c\, |x-\xi|^{-T-|\alpha|-|\gamma|}\,
  \Big( \frac{r(x)}{|x-\xi|}\Big)^{\sigma_{i,\alpha}(x)} \
  \Big( \frac{r(\xi)}{|x-\xi|}\Big)^{\sigma_{j,\gamma}(\xi)},\ |x-\xi|>\min(r(x),r(\xi)),
\end{eqnarray*}
where $\sigma_{i,\alpha}(x)=\min(0,\mu_x-|\alpha|-\delta_{i,4}-\varepsilon)$, $\mu_x=\mu_{k(x)}$, and $k(x)$
is the smallest integer $k$ such that $r(x)=r_k(x)$. In the case when $\lambda=0$ is not an eigenvalue
of the pencil $A_{k(x)}(\lambda)$, one can even put $\sigma_{i,\alpha}(x)=\mbox{\em Re}\, \lambda_1^{(k(x))}-
\delta_ {i,4}-|\alpha|-\varepsilon$ for $|\alpha|< \mbox{\em Re}\, \lambda_1^{(k(x))}- \delta_ {i,4}$.

Furthermore, for $j=1,\ldots,4$ there is the representation $G_{4,j}(x,\xi) = - \nabla_x\cdot \vec{P}_j(x,\xi)
+ Q_j(x,\xi)$, where $\vec{P}_j(x,\xi)\cdot n$ for $x\in \Gamma_k$, $k=1,\ldots,n$, $\xi\in {\cal D}$, and
the inequalities
\begin{equation} \label{2t13}
|\partial_x^\alpha\partial_\xi^\gamma \vec{P}_j(x,\xi)| \le c_{\alpha,\gamma} \,
  |x-\xi|^{-1-\delta_{j,4}-|\alpha|-|\gamma|}, \quad
|\partial_x^\alpha\partial_\xi^\gamma Q_j(x,\xi)| \le c_{\alpha,\gamma} \,
  r(\xi)^{-2-\delta_{j,4}-|\alpha|-|\gamma|}
\end{equation}
are satisfied for $|x-\xi|<\min(r(x),r(\xi))$.

{\em 4)} The functions $\xi\to \zeta(|x-\xi|/r(x))\, G_{i,j}(x,\xi)$ belong to
$W_{-\beta,0}^1({\cal K})$ for $j=1,2,3$ and to $W_{-\beta,0}^0({\cal K})$ for $j=4$.
The vector functions $\vec{H}_i =(G_{i,1},G_{i,2},G_{i,3})^t$ and the functions
$G_{i,4}$, $i=1,2,3,4$, are solutions of the problems
\begin{eqnarray*}
&& -\Delta_\xi \vec{H}_i(x,\xi)+ \nabla_\xi G_{i,4}(x,\xi)= \delta(x-\xi)\,
   (\delta_{i,1},\delta_{i,2},\delta_{i,3})^t \quad \mbox{for }x,\xi\in {\cal K},\\
&&  -\nabla_\xi\cdot \vec{H}_i(x,\xi) = \delta_{i,4}\, \delta(x-\xi) \quad \mbox{for }x,\xi\in {\cal K},\\
&& S_k \vec{H}_i(x,\xi)=0,\quad N_k(\partial_\xi)\,  \big(\vec{H}_i(x,\xi),G_{i,4}(x,\xi)\big)=0
  \quad \mbox{for }x\in {\cal K},\ \xi\in \Gamma_k,\ k=1,\ldots,n.
\end{eqnarray*}
This means that every solution $(u,p) \in C_0^\infty(\bar{\cal K})^4$ of equation {\em (\ref{Stokes})}
satisfying the homogeneous boundary condition {\em (\ref{bccone})} is given by the formulas
\begin{eqnarray} \label{3t13}
u_i(x) & = & \int_{\cal K} \big( f(\xi)+\nabla_\xi g(\xi)\big)\cdot \vec{H}_i(x,\xi)\, d\xi
  + \int_{\cal K} g(\xi)\, G_{i,4}(x,\xi)\, d\xi,\quad i=1,2,3, \\ \label{4t13}
p(x) & = & -g(x) + \int_{\cal K} \big( f(\xi)+\nabla_\xi g(\xi)\big)\cdot \vec{H}_4(x,\xi)\, d\xi
  + \int_{\cal K} g(\xi)\, G_{4,4}(x,\xi)\, d\xi.
\end{eqnarray}
\end{Th}

\noindent P r o o f.
1) We define Green's matrix by the formula
\[
G(x,\xi) = \big( 1 - \zeta(\frac{|x-\xi|}{r(\xi)})\big)\, {\cal G}(x,\xi) +R(x,\xi),
\]
where ${\cal G}(x,\xi)$ is Green's matrix of Stokes system in ${\Bbb R}^3$. The existence of a matrix
$R(x,\xi)$, where $R_{i,j}(\cdot,\xi) \in W_{\beta,0}^1({\cal K})$, $i=1,2,3$, and
$R_{4,j}(\cdot,\xi) \in W_{\beta,0}^0({\cal K})$ are such that $G(x,\xi)$
satisfies (\ref{g1})--(\ref{g3}) follows from Theorem \ref{t10}.

2) Eq. (\ref{1t13}) follows directly from the definition of $G(x,\xi)$.

3) The smoothness of $G(x,\xi)$ for $x\not=\xi$, $x,\xi \in \bar{\cal K}\backslash {\cal S}$
follows from the manner of its construction. The estimates of Green's matrix and the representation
for the components $G_{4,j}$ can be proved analogously to the third part of Theorem \ref{t7}
using the estimates for Green's matrix in a dihedron.

4) The last assertion can be proved analogously to \cite[Th.2.1]{mp79}.  \hfill $\Box$\\

As a consequence of items 3) and 4) of the last theorem, the following assertion holds.

\begin{Co} \label{c8a}
For $i=1,\ldots,4$ there is the representation $G_{i,4}(x,\xi) = - \nabla_\xi\cdot \vec{\cal P}_i(x,\xi)
+ {\cal Q}_i(x,\xi)$, where $\vec{\cal P}_i(x,\xi)\cdot n$ for $\xi\in \Gamma_k$, $x\in {\cal D}$, and
$\vec{\cal P}_i$ and ${\cal Q}_i$ satisfy the estimates
\[
|\partial_x^\alpha\partial_\xi^\gamma \vec{\cal P}_i(x,\xi)| \le c_{\alpha,\gamma} \,
  |x-\xi|^{-1-\delta_{i,4}-|\alpha|-|\gamma|}, \quad
  |\partial_x^\alpha\partial_\xi^\gamma {\cal Q}_i(x,\xi)| \le c_{\alpha,\gamma} \,
  r(\xi)^{-2-\delta_{i,4}-|\alpha|-|\gamma|}
\]
for $|x-\xi|<\min(r(x),r(\xi))$.
\end{Co}

Finally, we estimate Green's matrix and its derivatives in the cases $|x|>2|\xi|$ and $|\xi|>2|x|$.
For this we need the following lemma.

\begin{Le} \label{l17}
{\em 1)} If $u\in V_{\beta,\delta}^l({\cal K})$, $\rho\partial_\rho u\in V_{\beta,\delta}^l({\cal K})$,
$l\ge 2$, then there exists a constant $c$ independent of $u$ and $x$ such that
\begin{equation} \label{1l17}
\rho^{\beta-l+3/2} \prod_{j=1}^n \Big( \frac{r_j}{\rho}\Big)^{\max(\delta_j-l+1,0)} |u(x)| \le  c\, \Big(
 \| u\|_{W_{\beta,\delta}^l({\cal K})} + \|\rho\partial_\rho u\|_{W_{\beta,\delta}^l({\cal K}))}\Big)
\end{equation}

{\em 2)} If $u\in W_{\beta,\delta}^l({\cal K})$, $\rho\partial_\rho u\in W_{\beta,\delta}^l({\cal K})$,
$l\ge 2$, $\delta_k\not=l-1$ for $k=1,\ldots,n$, then there is the estimate
\begin{equation} \label{2l17}
\rho^{\beta-l+3/2} \prod_{k=1}^n \Big( \frac{r_k}{\rho}\Big)^{\max(\delta_k-l+1,0)} |u(x)| \le  c\, \Big(
 \| u\|_{W_{\beta,\delta}^l({\cal K})} + \|\rho\partial_\rho u\|_{W_{\beta,\delta}^l({\cal K}))}\Big).
\end{equation}
\end{Le}

\noindent P r o o f.
1) Applying the estimate
\[
\sup_{0<\rho<\infty} |v(\rho)|^2 \le c\, \int_0^\infty\big( |v(\rho)|^2 + |\rho v'(\rho)|^2\big)\, \frac{d\rho}\rho
\]
(which follows immediately from Sobolev's lemma) to the function $\rho^{\beta-l+3/2}u(\rho,\omega)$,
we obtain
\begin{equation} \label{3l17}
\sup_{0<\rho<\infty} \rho^{2(\beta-l)+3} \, \big| u(\rho,\omega)\big|^2 \le
  c\, \int_0^\infty \rho^{2(\beta-l+1)} \big( |u(\rho,\omega)|^2+ |\rho\partial_\rho u(\rho,\omega)|^2\big)\, d\rho.
\end{equation}
Furthermore, by Lemma \ref{l0a}, we have
\begin{equation} \label{4l17}
\sup_{\omega\in \Omega} \prod_{k=1}^n r_k(\omega)^{\delta_k-l+1} \, \big| \rho\partial_\rho u(\rho,\omega)\big|
  \le c\, \big\| \rho\partial_\rho u(\rho,\cdot)\big\|_{V_\delta^l(\Omega)}.
\end{equation}
Since the norm in $V_{\beta,\delta}^l({\cal K})$ is equivalent to
\[
\| u\| = \Big( \int_0^\infty \rho^{2(\beta-l+1)} \sum_{k=0}^l \big\| (\rho\partial_\rho)^k u(\rho,\cdot)
  \big\|^2_{V_\delta^{l-k}(\Omega)}\, d\rho\Big)^{1/2},
\]
The last inequality together with (\ref{3l17}) implies (\ref{1l17}).

2) If $\delta_k>l-1$ for all $k$, then $W_{\beta,\delta}^l({\cal K})=V_{\beta,\delta}^l({\cal K})$ and
(\ref{2l17}) follows from (\ref{1l17}). If $\delta_k<l-1$ for a certain $k$, then any function
$v\in W_\delta^l(\Omega)$ is continuous at the corner $M_k\cap\bar{\Omega}$ and the supremum of $v$ in
a neighborhood of this corner can be estimated by the $W_\delta^l$ norm of $v$. Therefore, instead of
(\ref{4l17}), we have
\[
\sup_{\omega\in \Omega} \prod_{k=1}^n r_k(\omega)^{\max(\delta_k-l+1,0)} \, \big| \rho\partial_\rho u(\rho,\omega)\big|
  \le c\, \big\| \rho\partial_\rho u(\rho,\cdot)\big\|_{W_\delta^l(\Omega)}.
\]
This together with (\ref{3l17}) implies (\ref{1l17}).\hfill $\Box$

\begin{Th} \label{t15}
Let $G(x,\xi)$ be Green's matrix introduced in Theorem {\em \ref{t13}}. Furthermore, let $\Lambda_-<\mbox{\em Re}\,
\lambda<\Lambda_+$ be the widest strip in the complex plane which is free of eigenvalues of the pencil
${\mathfrak A}(\lambda)$ and which contains the line $\mbox{\em Re}\, \lambda=-\beta-1/2$. Then for
$|x|>2|\xi|$ there is the estimate
\[
\big| \partial_x^\alpha\partial_\xi^\gamma G_{i,j}(x,\xi)\big|  \le  c\,
  |x|^{\Lambda_- -\delta_{i,4} -|\alpha|+\varepsilon}\ |\xi|^{-\Lambda_- -1 -\delta_{j,4}
  -|\gamma|-\varepsilon}\,
  \prod_{k=1}^n \Big( \frac{r_k(x)}{|x|}\Big)^{\sigma_{k,i,\alpha}}
  \prod_{k=1}^n \Big( \frac{r_k(\xi)}{|\xi|}\Big)^{\sigma_{k,j,\gamma}}\, ,
\]
where $\varepsilon$ is an arbitrarily small positive number and $\sigma_{k,i,\alpha}
=\min(0,\mu_k-|\alpha|-\delta_{i,4}-\varepsilon)$. In the case when $\lambda=0$ is not an eigenvalue
of the pencil $A_k(\lambda)$ and $|\alpha|< \mbox{\em Re}\, \lambda_1^{(k)}-\delta_{i,4}$, we can even
put $\sigma_{k,i,\alpha}=\mbox{\em Re}\, \lambda_1^{(k)}-|\alpha|-\delta_{i,4}-\varepsilon$.
Analogously,
\[
\big| \partial_x^\alpha\partial_\xi^\gamma G_{i,j}(x,\xi)\big| \le
c\, |x|^{\Lambda_+ -\delta_{i,4} -|\alpha|-\varepsilon}\ |\xi|^{-\Lambda_+ -1 -\delta_{j,4}
  -|\gamma|+\varepsilon}\,
  \prod_{k=1}^n \Big( \frac{r_k(x)}{|x|}\Big)^{\sigma_{k,i,\alpha}}
  \prod_{k=1}^n \Big( \frac{r_k(\xi)}{|\xi|}\Big)^{\sigma_{k,j,\gamma}}
\]
for $|\xi|> 2|x|$.
\end{Th}

\noindent P r o o f. Suppose that $|x|=1$. We denote by $\zeta$ and $\eta$
smooth functions on $\bar{\cal K}$ such that $\zeta(\xi)=1$ for $|\xi|<1/2$, $\eta=1$ in a neighborhood
of supp$\,\zeta$, and $\eta(\xi)=0$ for $|\xi|>3/4$. Furthermore, let $l$ be an integer,
$l>\max \mu_k+1$, $l\ge 3$. By Theorem \ref{t13}, we have
\begin{eqnarray*}
&& \hspace{-3em}  \eta(\xi)\big(-\Delta_\xi \partial_x^\alpha\vec{H}_i(x,\xi)+
  \nabla_\xi \partial_x^\alpha G_{i,4}(x,\xi)\big) = 0, \quad
  \eta(\xi)\, \nabla_\xi \cdot \vec{H}_i(x,\xi)=0 \  \mbox{ for }\xi\in{\cal K}, \\
&& \hspace{-3em} \eta(\xi) S_k \partial_x^\alpha\vec{H}_i(x,\xi)=0,\quad \eta(\xi) N_k(\partial_\xi)\,
  \big( \partial_x^\alpha\vec{H}_i(x,\xi), \partial_x^\alpha G_{i,4}(x,\xi)\big)=0
  \quad \mbox{for } \xi\in \Gamma_k,\ \ k=1,\ldots,n,
\end{eqnarray*}
for $i=1,2,3,4$. Since $\eta(\cdot)\partial_x^\alpha \vec{H}_i \in W_{-\beta,0}^1({\cal K})^3$ and
$\eta(\cdot) \partial_x^\alpha G_{i,4}(x,\cdot) \in W_{-\beta,0}^0({\cal K})$, we conclude from
Corollary \ref{c7} (see also Remark \ref{r6}) and Theorem \ref{t12} that
\[
\zeta(\cdot)\, (|\xi|\partial_{|\xi|})^j \partial_\xi^\gamma \partial_x^\alpha \vec{H}_i(x,\cdot) \in
  W^l_{\beta'+|\gamma|,\delta+|\gamma|}({\cal K})^3, \ \
  \zeta(\cdot)\, (|\xi|\partial_{|\xi|})^j \partial_\xi^\gamma \partial_x^\alpha \vec{G}_{i,4}(x,\cdot)
  \in W^0_{\beta'+|\gamma|,\delta+|\gamma|}({\cal K})
\]
for $j=0,1,\ldots$, where $\beta'=l+\Lambda_- +\varepsilon-1/2$, $\delta_k=l-1-\mu_k+\varepsilon$.
If $\lambda=0$ is not an eigenvalue of the pencil $A_k(\lambda)$, then by Corollary \ref{c8}, we have also
$\zeta(\cdot)\, \zeta_k(\cdot)\, (|\xi|\partial_{|\xi|})^j \partial_\xi^\gamma \partial_x^\alpha \vec{H}_i(x,\cdot)
  \in  V^l_{\beta'+|\gamma|,\delta'+|\gamma|}({\cal K})^3$ and
$\zeta(\cdot)\, \zeta_k(\cdot)\, (|\xi|\partial_{|\xi|})^j \partial_\xi^\gamma \partial_x^\alpha
  \vec{G}_{i,4}(x,\cdot) \in V^0_{\beta'+|\gamma|,\delta'+|\gamma|}({\cal K}),$
where $\delta'_k= l-1-\mbox{Re}\, \lambda_1^{(k)}+\varepsilon$ (here $\zeta_k$ is the cut-off function introduced
before Corollary \ref{c8}).  Using Lemma \ref{l17}, we obtain
\begin{eqnarray} \label{1t15}
&& |\xi|^{\Lambda_- +1+\delta_{j,4}+|\gamma|+\varepsilon} \prod_{k=1}^n \Big( \frac{r_k(\xi)}{|\xi|}
  \Big)^{-\sigma_{k,j,\gamma}} \, \big| \partial_x^\alpha\partial_\xi^\gamma
  G_{i,j}(x,\xi)\big| \nonumber \\
&& \le c\, \Big( \| \eta(\cdot) \partial_x^\alpha \vec{H}_i(x,\cdot)\|_{W_{-\beta,0}^1({\cal K})^3}+
  \| \eta(\cdot) \partial_x^\alpha G_{i,4}(x,\cdot)\|_{W_{-\beta,0}^0({\cal K})}\Big)
\end{eqnarray}
for $i,j=1,2,3,4$, and $|\xi|<1/2$, where $c$ is independent of $x$ and $\xi$. By Theorem \ref{t10}, the problem
\begin{eqnarray*}
&& b(u,v)-\int_{\cal K} p\, \nabla\cdot v = \int_{\cal K} \eta(y)\, F(y)\cdot v(y)\, dy \quad
  \mbox{for all } v\in V_{-\beta} \\
&& -\nabla \cdot u = \eta g\ \mbox{ in }{\cal K},\quad S_k u = 0 \ \mbox{ on }\Gamma_k
\end{eqnarray*}
has a unique solution $(u,p)\in V_\beta \times W_{\beta,0}^0({\cal K})$ which van be written in the form
\begin{eqnarray*}
&& u_i(y) = \int_{\cal K} \eta(z)\,\big( F(z)\cdot \vec{H}_i(y,z) + g(z)\, G_{i,4}(y,z)\big)\,
  dz, \ i=1,2,3, \\
&& p(y) = -\eta(y)\, g(y)+ \int_{\cal K} \eta(z)\,\big( F(z)\cdot \vec{H}_4(y,z) + g(z)\,
  G_{4,4}(y,z)\big)\, dz.
\end{eqnarray*}
Let $\chi_1$ and $\chi_2$ be smooth cut-off function, $\chi_2=1$ in a neighborhood of $x$,
$\chi_1=1$ in a neighborhood of $\chi_2$, $\chi_1(y)=0$ for $|x-y|>1/4$. Since $\chi_1$ and $\eta$ have
disjunct supports, we have
\[
\chi_1(-\Delta u+\nabla p)=0, \ \chi_1\nabla\cdot u=0, \ \chi_1 S_k u=0, \ \chi_1 N_k(u,p)=0.
\]
Consequently, from Corollary \ref{c7} (see also Remark \ref{r6}) and Theorem \ref{t12} it follows that
\[
\chi_2 (\rho\partial_\rho)^j \partial_x^\alpha u\in W_{\beta'',\delta+|\alpha|}^l({\cal K})^3,\quad
\chi_2 (\rho\partial_\rho)^j \partial_x^\alpha p\in W_{\beta'',\delta+|\alpha|}^{l-1}({\cal K})^3
\]
for arbitrary integer $j$ and for arbitrary $\beta''$.
In the case when $\lambda=0$ is not an eigenvalue of the pencil $A_k(\lambda)$, then by Corollary \ref{c8},
\[
\chi_2\zeta_k (\rho\partial_\rho)^j \partial_x^\alpha u\in V_{\beta'',\delta'+|\alpha|}^l({\cal K})^3,\quad
\chi_2\zeta_k (\rho\partial_\rho)^j \partial_x^\alpha p\in V_{\beta'',\delta'+|\alpha|}^{l-1}({\cal K})^3.
\]
Thus, by Lemma \ref{l17},
\[
\prod_{k=1}^n r_k(x)^{-\sigma_{k,1,\alpha}}\, |\partial_x^\alpha u(x)|+
  \prod_{k=1}^n r_k(x)^{-\sigma_{k,4,\alpha}}\, |\partial_x^\alpha p(x)|
 \le c\, \big( \| F\|_{V_{-\beta}^*} + \| g\|_{W_{\beta,0}^0({\cal K})}\big).
\]
This means that the functionals
\begin{eqnarray*}
&& V_{-\beta}^* \times W_{-\beta,0}^0({\cal K}) \ni (F,g) \to \prod_{k=1}^n r_k(x)^{-\sigma_{k,i,\alpha}}
  \partial_x^\alpha u_i(x) \\ && \qquad = \prod_{k=1}^n r_k(x)^{-\sigma_{k,i,\alpha}}
  \int_{\cal K} \eta(z)\,\big( F(z)\cdot \partial_x^\alpha \vec{H}_i(x,z) + g(z)\,
  \partial_x^\alpha G_{i,4}(x,z)\big)\, dz,
\end{eqnarray*}
$i=1,2,3$, and
\begin{eqnarray*}
&& V_{-\beta}^* \times W_{-\beta,0}^0({\cal K}) \ni (F,g) \to \prod_{k=1}^n r_k(x)^{-\sigma_{k,4,\alpha}}
  \partial_x^\alpha p(x) \\ && \qquad  = \prod_{k=1}^n r_k(x)^{-\sigma_{k,4,\alpha}}
  \int_{\cal K} \eta(z)\,\big( F(z)\cdot \partial_x^\alpha \vec{H}_4(x,z) + g(z)\,
  \partial_x^\alpha G_{4,4}(x,z)\big)\, dz,
\end{eqnarray*}
are continuous and their norms are bounded by constants independent of $x$. Consequently,
\[
\| \eta(\cdot)\partial_x^\alpha \vec{H}_i(x,\cdot)\|_{V_{-\beta}} + \| \eta(\cdot)\partial_x^\alpha
  G_{i,4}(x,\cdot)\|_{W_{-\beta,0}^0({\cal K})} \le c\, \prod_{k=1}^n r_k(x)^{\sigma_{k,i,\alpha}}
\]
for $i,1,2,3$ and
\[
\| \eta(\cdot)\partial_x^\alpha \vec{H}_4(x,\cdot)\|_{V_{-\beta}} + \| \eta(\cdot)\partial_x^\alpha
  G_{4,4}(x,\cdot)\|_{W_{-\beta,0}^0({\cal K})} \le c\, \prod_{k=1}^n r_k(x)^{\sigma_{k,4,\alpha}}.
\]
Combining the last two inequalities with (\ref{1t15}), we obtain the assertion of the theorem for
$|x|=1$, $|\xi|<1/2$. Since
$G_{i,j}(x,\xi)$ is positively homogeneous of degree $-1-\delta_{i,4}-\delta_{j,4}$, the estimate holds also
for arbitrary $x,\xi$, $|\xi|<|x|/2$. The proof for the case $|\xi|>2|x|$ proceeds analogously.
\hfill $\Box$

\begin{Rem} \label{r7}
{\em The estimates in Theorems \ref{t13}, \ref{t15} can be improved by means of Theorem \ref{t8} and
Corollary \ref{c7} if the direction of the derivative is tangential to the edges. In particular, we have
\[
|\partial_\rho G_{i,j}(x,\xi)| \le c\, |x-\xi|^{-2-\delta_{i,4}-\delta_{j,4}}\,
  \Big( \frac{r(x)}{|x-\xi|}\Big)^{\min(0,\mu_x-\delta_{i,4}-\varepsilon)} \
  \Big( \frac{r(\xi)}{|x-\xi|}\Big)^{\min(0,\mu_\xi-\delta_{j,4}-\varepsilon)}
\]
if $|x|/2<|\xi|<2|x|$, $|x-\xi|>\min(r(x),r(\xi))$,
\[
\big| \partial_\rho G_{i,j}(x,\xi)\big| \le c\,
  |x|^{\Lambda_- -1-\delta_{i,4}+\varepsilon}\ |\xi|^{-\Lambda_- -1 -\delta_{j,4} -\varepsilon} \
 \prod_{k=1}^n \Big( \frac{r_k(x)}{|x|}\Big)^{\sigma_{k,i,0}}
  \prod_{k=1}^n \Big( \frac{r_k(\xi)}{|\xi|}\Big)^{\sigma_{k,j,0}}
\]
if $|x|>2|\xi|$ and
\[
\big| \partial_\rho G_{i,j}(x,\xi)\big| \le c\, |x|^{\Lambda_+ -1-\delta_{i,4}-\varepsilon}\
  |\xi|^{-\Lambda_+ -1 -\delta_{j,4}+\varepsilon}
  \prod_{k=1}^n \Big( \frac{r_k(x)}{|x|}\Big)^{\sigma_{k,i,0}}
  \prod_{k=1}^n \Big( \frac{r_k(\xi)}{|\xi|}\Big)^{\sigma_{k,j,0}}
\]
if $|\xi|> 2|x|$, where $\sigma_{k,i,0}=\min(0,\mu_k-\delta_{i,4}-\varepsilon)$.}
\end{Rem}

\end{document}